# Isotopetronics - New Direction of Nanoscience.

Vladimir G. Plekhanov

Computer Science College, Erika Street 7a, Tallinn, 10416, ESTONIA.

**Abstract.** Isotopetronics is a new branch of the nanoscience. The present paper is devoted to brief description of the main parts of isotopetronics: human health, optical fiber, neutron transmutation doping semiconductors, isotope memory of information, solid - state processor for quantum computers as well as fundamental task of theoretical physics, for example, the enigma of the mass.

**0. Preface.**

**1. Elementary excitations of isotope - mixed crystals.**

1.1. Introduction.
1.2. Phonons.
1.3. Excitons.

**2. Methods of the preparation of low dimensional structures.**

2.1. Molecular beam epitaxy (MBE) and metal - organic chemical vapour deposition (MOCVD).
2.2. Nanolitography and etching technologies.
2.3. Techniques for characterization of nanostructures.
2.4. Nuclear technology.

**3. Electron excitations in low - dimensional structures.**

3.1. Wave - like properties of electrons.
3.2. Dimensionality and density of states.
3.3. Electron in quantum dot.
3.4. Excitons in nanostructures.
3.4.1 Excitons in quantum wells.
3.4.2. Excitons in quantum wires.
3.4.3. Excitons in quantum dots.
3.5. Biexcitons in quantum dots.

**4. Applications of low - dimensional structures.**

**5. Conclusion and outlook.**

**0. Preface.**

    The experience of the past shows that throughout constant technology improvement electronics (optoelectronics) has become more reliable, faster, more powerful, and less expensive by reducing the dimensions of integrated circuits. These advantages are the driver for the development of modern microelectronics. The long - term goal of this development will lead to nanoelectronics. Advancing to the nanoscale is not just a step toward miniaturization, but requires the introduction and consideration of many additional phenomena. At the nanoscale, most phenomena and processes are dominated by quantum physics and they exhibit unique behavior. Nanotechnology includes the integration of manmade nanostructures into larger material components and systems (see, e.g. [11]). Importantly, within these larger scale systems, the active elements of the system will remain at nanoscale.

    Low - dimensional structures have become one of the most active research in nanoscience and nanotechnology. Quantum wells, quantum wires and quantum dots structures produced in the main by epitaxial growth techniques (mainly molecular beam epitaxy (MBE) and metal -organic chemical vapour deposition (MOCVD) and their various variations such as chemical beam epitaxy (CBE), atomic layer epitaxy (ALE) etc. (see, e.g. [1 -6])). MBE and MOCVD are of considerable technological  interest since they are used as active components in modern devides. These devices are high -

electron - mobility transistors, diodes and lasers, as well as quantum dots from quantum computations and communications perspectives.

The seminal works of Esaki and Tsu [7] and others on the semiconductor superlattice stimulated a vast international research effort to understand the fabrication and electronic properties of superlattice, quantum wells, quantum wires and quantum dots (see, for example. [8 - 11]). The dimensional scale of such samples is between 10 and 100 nm are subject of nanoscience - is a broad and interdisciplinary field of emerging research and development. Nanoscience and nanotechnology are concerned with materials, structures, and systems whose components exhibit novel and significantly modified physical, chemical properties due to their nanoscale sizes. New direction of nanoscience is isotopetronics, who is studied the more low - dimensional size, as a rule the sizes of the sample of isotopetronics compare to the atomic size. Nuclear technology - neutron irradiation [12] - is very useful method to preparing low - dimensional structure: quantum wells, quantum wires and quantum dots [13]. A principal goal of isotopetronics as new directions of the nanotechnology is to control and exploit their new properties in structures and devices at atomic, molecular and supramolecular levels. The minituarization required by modern electronics is one of the driving forces for isotopetronics - new direction of nanotechnology (see, also [14]).

Modern nanoscience and nanotechnology is fertile ground for teaching, as it brings together the quantum theory of materials, novel physics in the electronic and optical properties of solids, the engineering of small structures and the design of high performance electronic, photonic and optoelectronic systems. The treatments attempts to be introductory, comprehensive and phenomenological in the main. The new physics described in this review comes from one important consideration - length scale (see, also [9, 10, 15, 16]) especially in mesoscopic physics. As we all know mesoscopic physics deals with structures which have a size between the macroscopic and the microscopic or atomic one. These structures are also called mesoscopic systems, or nanostructures [10] in a more colloquial way since their size usually ranges from a few nanometers to about 100 nm. The electrons in such mesoscopic systems show their wavelike properties [15, 16] and therefore their behavior is markedly dependent on the geometry of the samples. In this case, the states of the electrons are wave - like and somewhat similar to electromagnetic waves (see, e.g. [72]).

As was saying above for the description of the behavior of electrons in solids it is very convenient to define a series of characteristic lengths. If the dimension of the solids in which the electron embedded is of the order of, or smaller than these characteristic lengths ($\lambda_B$ de Broglie wavelength, or a$_{ex}$ - exciton radius, etc.) the material might show new properties, which in general are more interesting than the corresponding ones in macroscopic materials. On the contrary, a mesoscopic system approaches its macroscopic limit if its size is several times its characteristic length.

As was mentioned above when the dimensions of the solid get reduced to a size comparable with, or smaller $\lambda_B$, then the particles behave wavelike and quantum mechanics should be used. Let us suppose that we have an electron confined within a box of dimensions L$_x$, L$_y$, L$_z$. If the characteristic length is , we can have the following situations:

1) $l \langle$ L$_x$, L$_y$, L$_z$. In this case the electron behaves as in regular 3D bulk semiconductor (insulator).

2) $l \rangle L_x$ and $L_x \langle\langle L_y, L_z$. In this situation we have a 2D semiconductor perpendicular to the x - axis. This mesoscopic system is also called a quantum well (for details see chapter 3).

3) $l \rangle L_x, L_y$ and $L_x, L_y \langle\langle L_z$. This case corresponds to a 1D semiconductor or quantum wire, located along the z - axis.

4) $l \rangle\rangle L_x, L_y, L_z$. In this case it is said that we have a 0D or a quantum dot [8. 9].

In general, we say in mesoscopic physics that a solid, very often a crystal, is of reduced dimensionality if at least one of its dimensions $L_i$ is smaller than the characteristic length. For instance, if $L_x$ and $L_y$ are smaller than *l* we a crystal of dimensionality equal to one. We could also have the case that *l* is comparable, or a little larger, than one of the dimensions of the solid but much smaller than the other two. Then we have a quasi 2D system, which in practice is a very thin film, but not thin enough to show quantum size effect (for details see chapter 3).

This review is organized into five chapters. In chapter 1 I review the present status of elementary excitations in solids. Preparation methods of low - dimensional structures describe in chapter 2. Chapter 3 deals with physics of low - dimensional structure. In this chapter of the most frequently structures - quantum dots - revised. The applications of low - dimensional structures was done in chapter four. Conclusion and outlook was described in the last chapters of our review.

**1. Elementary excitations of isotope - mixed crystals.**

1.1 Introduction.

The modern view of solid state physics is based on the presentation of elementary excitations, having mass, quasiimpuls, electrical charge and so on (see, e.g. [17]). According to this presentation the elementary excitations of the non - metallic materials are electrons (holes), excitons (polaritons [18]) and phonons [19]. The last one are the elementary excitations of the crystal lattice, the dynamics of which is described in harmonic approximation (see e.g. [20]). As is well - known, the base of such view on solid are the multiparticle approach. In such view, the quasiparticles of solid are ideal gas, which described the behavior of the system, e.g. noninteracting electrons. We should add such approach to consider the theory of elementary excitations as suitable model for the application of the common methods of the quantum mechanics for the solution solid state physics task. In this part of our review will be briefly consider not only the manifestations of the isotope effect in different solids, but also will bring the new accurate results, showing the quantitative changes of different characteristics of phonons and electrons (excitons) in solid with isotopical substitution (see, also [21]). By the way, isotopic effect become more pronounced when we dealing with solids. For example, on substitution of H with D the change in energy of the electron transition in solid state (e.g. LiH ) is two orders of magnitude larger than in atomic hydrogen (see, e.g. [22]). In the use of an elementary excitations to describe the complicated motion of many particles has turned out to be an extraordinary useful device in contemporary physics, and it is view of a solid which we describe in this part of review,

The basic Hamiltonian which our model of the solid is of the form [21]

$$H = H_{ion} + H_{electron} + H_{electron-ion} \qquad (1)$$

where

$$H_{ion} = \sum_i \frac{p_i^2}{2m} + \frac{1}{2}\sum_{i \neq j} V(R_i - R_j), \qquad (2)$$

$$H_{electron} = \sum_i \frac{p_i^2}{2m} + \frac{1}{2}\sum_{i \neq j} \frac{e^2}{|r_i - r_j|}, \qquad (3)$$

$$H_{electron-ion} = \sum_{i,j} v(r_i - R_j). \qquad (4)$$

$H_{ion}$ describes a collection of ions (of a single species) which interact through a potential $V(R_i - R_j)$ which depends only on the distance between ions. By ion we mean a nucleus plus the closed - shell, or core, electrons, that is, those electrons which are essentially unchanged when the atoms are brought together to make a solid. $H_{electron}$ presents the valence electrons (the electrons outside the last closed shell), which are assumed to interact via a Coulomb interaction. Finally, $H_{electron-ion}$ describes the interaction between the electrons (excitons) and the ions, which is again assumed to be represented by a suitable chosen potential.

In adopting (1) as our basic Hamiltonian, we have already made a number of approximation in a treatment of a solid. Thus, in general the interaction between ions is not well - represented by a potential $V(R)$, when the coupling between the closed - shell electrons on different ions begins to play an important role (see, e.g. [23, 24]). Again, in using a potential to represent electron - ion interaction, we have neglected the fact that the ions possess a structure (the core electrons); again, when the Pauli principle plays an important role in the interaction between the valence electrons, that interaction may no longer be represented by a simple potential. It is desirable to consider the validity of these approximations in detail (for detail see, e.g. [24]). In general one studies only selected parts of the Hamiltonian (91). Thus, for example, the band theory of solids is based upon the model Hamiltonian [23, 24].

$$H_B = \sum_i \frac{p_i^2}{2m} + \sum_{i,j} v(r_i - R_{j0}) + V_H(r_i), \qquad (5)$$

where the $R_{j0}$ represents the fixed equilibrium positions of the ions and the potential $V_H$ describes the (periodic) Hartree potential of the electrons. One studies the motion of a single electron in the periodic field of the ions and the Hartree potential, and takes the Pauli principle into account in the assignment of one - electron states. In so doing one neglects aspects other than the Hartree potential of the interaction between electrons. On the other hand, where one is primarily interested in understanding the interaction between electrons in metals, it is useful to consider only (3), replacing the effect of the ion cores by a uniform distribution of positive charge [25]. In this way one can approximate the role that electron interaction plays without having present the additional complications introduced by the periodic ion potential. Of course one wants finally to keep both the periodic ion potential and the electron interactions, and to include as well the effects associated with departure of the ions from the equilibrium positions, since only in this way does not arrive at a generally adequate description of the solid. Usually for the elementary excitations in solids by first considering various different parts of the Hamiltonian (1) and then taking into account the remaining terms which act to couple different excitations.

## 1.2. Phonons.

The simplest kind of motion in solids is the vibrations of atoms around the equilibrium point. The interaction of the crystal forming particles with the one another at the move of the one atom entanglements neighbor atoms [20]. The analysis of this kind motion shows that the elementary form of motion is the wave of the atom displacement. As is well - known that the quantization of the vibrations of the crystal lattice and after introduction of the normal coordinates, the Hamiltonian of our task will be have the following relation (see, e.g. [26])

$$H(Q,P) = \sum_{i,q} \left[ -\frac{\hbar^2}{2} \frac{\partial^2}{\partial Q^2(\vec{q})} + \frac{1}{2}\omega_j^2 Q_j^2(\vec{q}) \right] \qquad (6).$$

In this relation, the sum, where every addend means the Hamiltonian of linear harmonic oscillator with coordinate $Q_j(\vec{q})$, the frequency $\omega_j(\vec{q})$ and the mass, which equals a unit. If the Hamiltonian system consists of the sum, where every addend depends on the coordinate and conjugate its quasiimpuls, then according to quantum mechanics [49] the wave function of the system equals the product of wave functions of every appropriate addend and the energy is equal to the sum of assigned energies. Any separate term of the Hamiltonian (96) corresponds, as indicate above, the linear oscillator

$$-\frac{\hbar^2}{2}\frac{\partial^2 \Psi}{\partial Q^2} + \frac{1}{2}\omega^2 Q^2 \Psi = \varepsilon \Psi. \qquad (7)$$

Solving last equation and finding the eigenvalues and eigenfunctions and then expressing explicitly the frequency, we will obtain for model with two atoms in primitive cell (with masses $M_1$ and $M_2$; $M_1 \rangle M_2$) the following equation

$$\omega^2 = C\left(\frac{M_1 + M_2}{M_1 M_2}\right) \pm \left[C^2\left(\frac{M_1 + M_2}{M_1 M_2}\right) - \frac{4C^2}{M_1 M_2}\sin^2\frac{ka}{2}\right]$$

or

$$\omega_\pm^2 = A \pm [A^2 - B\sin^2\frac{ka}{2}]^{1/2}. \qquad (8)$$

There are now two solutions for $\omega^2$, providing two distinctly separate groups of vibrational modes. The first group, associated with $\omega_-^2$, contains the acoustic modes. The second group arises with $\omega_+^2$ and contains the optical modes; these correspond to the movement of the different atom sorts in opposite directions (e.g. NaCl - structures), it is contra motion whereas the acoustic behavior is motion in unison

For small ka we have from (8) two roots:

$$\omega^2 \simeq 2C(\frac{1}{M_1} + \frac{1}{M_2}) \qquad (8')$$

and

$$\omega^2 \simeq \frac{C}{2(M_1 + M_2)}K^2 a^2. \qquad (9)$$

Taking into account that $K_{max} = \pm \pi/a$, where a is a period of the crystal lattice, i.e. $K_{max}$ respond the border of the first Brillouin zone (see also Fig. 1)

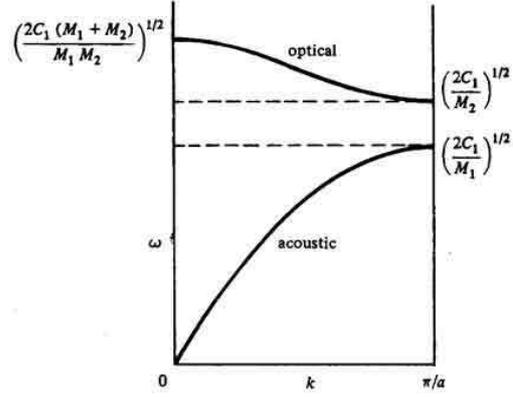

**Fig. 1. Optical and acoustic modes. The optical modes lie at higher frequencies and show less dispersion than the acoustic modes (for details see text).**

$\omega^2 = \frac{2C}{M_1}$ and $\omega^2 = \frac{2C}{M_2}$ (10)

As it is clear, formula (8') describes the optical branch of vibrations whereas (9) - acoustical branch of vibrations. Usually the last formula is written in this way

$\omega = \sqrt{\frac{\alpha}{M}}$, (11)

where $\alpha$ is so - called the force constant. Here, M is the mass of vibrated atom (ion). From the preceding relation it is clear that, as in molecular physics [21], in solids the isotope effect directly manifests in vibration spectrum, which depends on the symmetry [28] measures either in IR - absorption or in Raman scattering of light. Before analyzing Raman scattering spectra of different solids we briefly consider the classical approximation of the mechanism of Raman effect [29, 30].

Historically, Raman scattering denotes inelastic scattering of light by molecular vibrations or by optical phonons in solids. In a macroscopic picture, the Raman effect in crystals is explained in terms of the modulation of polarizability by the quasi - particle under consideration. The assumption that the polarization depends linearly upon the electric field strength [31] is a good approximation and is invariably used when discussing the scattering of light by crystal excited by lasers. However, the approximation is not valid for large strength such as can be obtained from pulsed lasers [32]. The polarization may then be expressed as

$P = \alpha E + \frac{1}{2}\beta E^2 + \frac{1}{6}\gamma E^3 + \frac{1}{24}\delta E^4 + $............, (12)

where $\beta$, the first hyperpolarizability coefficient, plays an important part for large values of E, since it responsible for the phenomenon of optical harmonic generation using Q - switched lasers. Isolated atoms have $\beta = 0$, since, like $\mu$ the dipole moment, it arises from interactions between atoms. A simplified theory of Rayleigh scattering, the Raman effect, harmonic generation and hyper Raman scattering is obtained by setting (see, e.g. [32])

$E = E_0 \cos\omega_0 t$, (13)
$\alpha = \alpha_0 + (\frac{\partial \alpha}{\partial Q})Q$, (14)
$\beta = \beta_0 + (\frac{\partial \beta}{\partial Q})Q$, (15)
$Q = Q_0 + \cos\omega_v t$. (16).

Here Q is a normal coordinate, $\omega_v$ is the corresponding vibrational frequency and $\omega_0$ is the laser frequency. After that we have

$$P = \alpha_0 E_0 \cos\omega_v t + \tfrac{1}{2}(\tfrac{\partial\alpha}{\partial Q})Q_0 E_0 \cos\omega_0 t \cos\omega_v t + \tfrac{1}{2}\beta_0 E_0^2 \cos^2\omega_0 t +$$
$$\tfrac{1}{2}(\tfrac{\partial\beta}{\partial Q})Q_0 E_0^2 \cos^2\omega_0 t \cos\omega_v t. \quad (17)$$

Then, after small algebra, we obtain

$$P = \alpha_0 E_0^2 \cos\omega_v t + \tfrac{1}{2}(\tfrac{\partial\alpha}{\partial Q})Q_0 E_0 \cos(\omega_0 - \omega_v)t + \cos(\omega_0 + \omega_v)t + \tfrac{1}{2}\beta_0 E_0^2 + \tfrac{\beta_0}{4}E_0^2 \cos 2\omega_0 t +$$
$$+ \tfrac{1}{2}Q_0 E_0^2 (\tfrac{\partial\beta}{\partial Q})\cos(2\omega_0 - \omega_v)t + \cos(2\omega_0 - \omega_v)t. \quad (18).$$

In last relation the first term describes the Rayleigh scattering, second - Raman scattering, third - d.c. polarization, fourth - frequency doubling and the last - hyper Raman effect. Thus the hyper Raman effect is observed with large electric field strength in the vicinity of twice the frequency of the exciting line with separations corresponding to the vibrational frequencies. $\alpha$ and $\beta$ are actually tensors and $\beta$ components $\beta_{\alpha\beta\gamma}$ which are symmetrical suffixes [33].

Semiconducting crystals (C, Si, Ge, $\alpha$ - Sn) with diamond - type structure present ideal objects for studying the isotope effect by the Raman light - scattering method. At present time this is facilitated by the availability of high - quality crystals grown from isotopically enriched materials (see, e.g [34] and references therein). In this part our understanding of first - order Raman light scattering spectra in isotopically mixed elementary and compound (CuCl, GaN, GaAs) semiconductors having a zinc blende structure is described. Isotope effect in light scattering spectra in Ge crystals was first investigated in the paper by Agekyan et al. [35]. A more detailed study of Raman light scattering spectra in isotopically mixed Ge crystals has been performed by Cardona and coworkers [34].

It is known that materials having a diamond structure are characterized by the triply degenerate phonon states in the $\Gamma$ point of the Brillouin zone ($\vec{k} = 0$). These phonons are active in the Raman scattering spectra, but not in the IR absorption one [28]. Figure 2[a] demonstrates the dependence of the shape and position of the first - order line of optical phonons in germanium crystal on the isotope composition at liquid nitrogen temperature (LNT) [36].

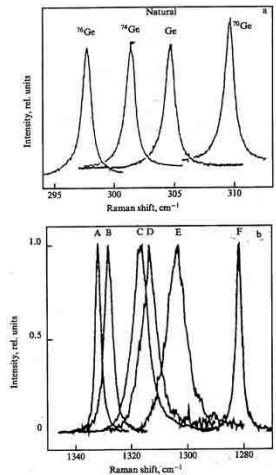

**Fig. 2. a - First - order Raman scattering spectra Ge with different isotope contents [34] and b - First - order Raman scattering in isotopically mixed diamond crystals $^{12}C_x^{13}C_{1-x}$. The peaks A, B, C, D, E and F correspond to x = 0.989; 0.90;**

**0.60; 0.50; 0.30 and 0.001 (after [38]).**

The coordinate of the center of the scattering line is proportional to the square root of the reduced mass of the unit cell, i.e $\sqrt{M}$. It is precisely this dependence that is expected in the harmonic approximation. An additional frequency shift of the line is observed for the natural and enriched germanium specimens and is equal, as shown in Ref. [34] to 0.34 ± 0.04 and 1.06 ± 0.04 cm$^{-1}$, respectively (see, e.g. Fig. 7 in Ch. 4 of Ref. [37]).

First - order Raman light - scattering spectrum in diamond crystals also includes one line with maximum at $\omega_{LTO}(\Gamma)$ = 1332.5 cm$^{-1}$. In Fig. 2$^b$ the first - order scattering spectrum in diamond crystals with different isotope concentration is shown [38]. As shown below, the maximum and the width of the first - order scattering line in isotopically - mixed diamond crystals are nonlinearly dependent on the concentration of isotopes x. The maximum shift of this line is 52.3 cm$^{-1}$, corresponding to the two limiting values of x = 0 and x = 1. Analogous structures of first - order light scattering spectra and their dependence on isotope composition has by now been observed many times, not only in elementary Si, and $\alpha$ - Sn, but also in compound CuCl and GaN semiconductors (for more details see reviews [34, 39]). Already short list of data shows a large dependence of the structure of first - order light - scattering spectra in diamond as compared to other crystals (Si, Ge). This is the subject detailed discussion in [40].

Second - order Raman spectra in natural and isotopically mixed diamond have been studied by Chrenko [41] and Hass et al. [42]. Second - order Raman spectra in a number of synthetic diamond crystals with different isotope composition shown in Fig. 3 are measured wit resolution (~ 4 cm$^{-1}$) worse than for first - order scattering spectra. The authors of cited work explain this fact by the weak signal in the measurement of second - order Raman scattering spectra. It is appropriate to note that the results obtained in [42] for natural diamond ($C_{13_C}$ = 1.1%), agree well with the preceding comprehensive studies of Raman light - scattering spectra in natural diamond [43].

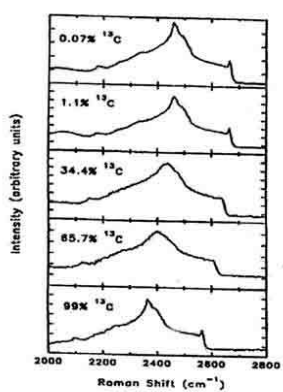

**Fig. 3. Second - order Raman scattering spectra in synthetic diamond with different isotope concentration at room temperature (after [42]).**

As is clearly seen from Fig. 3 the structure of second - order light scattering "follows" the concentration of the $^{13}$C isotope. It is necessary to add that in the paper by Chrenko [41] one observes a distinct small narrow peak above the high - frequency edge of LO phonons and the concentration of $^{13}$C x = 68%. Note is passing that second - order spectra in isotopically mixed diamond crystals were measured in the work by Chrenko [41] with a better resolution than the spectra shown in Fig. 3. Second - order Raman light scattering spectra and IR absorption spectra in crystals of natural and isotopically enriched $^{70}$Ge can be found in [39].

A comprehensive interpretation of the whole structure of second - order Raman light - scattering spectra in pure LIH (LiD) crystals is given in [22, 40]. Leaving this question, let us now analyze the behavior of the highest frequency peak after the substitution of hydrogen for deuterium (see, also [44]).

Absorption behavior of an IR - active phonon in mixed crystals with a change in the concentrations of the components can be classified into two main types: one and two - mode (see, e.g. the review [45]). Single - mode behavior means that one always has a band in the spectrum with a maximum gradually drifting from one endpoint to another. Two - mode behavior is defined by the presence, in the spectrum, of two bands characteristic of each components lead not only to changes in the frequencies of their maxima, but mainly to a redistribution of their intensities. In principle, one and the same system can show different types of behavior at opposite ends [46]. The described classification is qualitative and is rarely realized in its pure form (see, also [46]). The most important necessary condition for the two - mode behavior of phonons (as well as of electrons [47]) is considered to be the appearance of the localized vibration in the localized defect limit. In the review [45] a simple qualitative criterion for determining the type of the IR absorption behavior in crystals with an NaCl structure type has been proposed (see also [47]). Since the square of the TO ($\Gamma$) phonon frequency is proportional to the reduced mass of the unit cell M, the shift caused by the defect is equal to

$$\Delta = \omega_{TO}^2 (1 - \frac{\overline{M}}{M'}). \qquad (19)$$

This quantity is compared in [45] with the width of the optical band of phonons which, neglecting acoustical branches and using the parabolic dispersion approximation, is written as

$$W = \omega_{TO}^2 (\frac{\varepsilon_0 - \varepsilon_\infty}{\varepsilon_0 + \varepsilon_\infty}). \qquad (20)$$

A local or gap vibrations appears, provided the condition $|\Delta| > (1/2)W$ is fulfilled. As mentioned, however, in [45] in order for the two peaks to exist up to concentrations on the order of ~ 0.5, a stronger condition $|\Delta| > W$ has to met. Substituting the numerical values from Tables 1 and 2 of [40] into formulas (19) and (20) shows that for LiH (LiD) there holds (since $\Delta = 0.44\omega_{TO}^2$ and $W = 0.58\omega_{TO}^2$) the following relation:

$$|\Delta| > (1/2)W. \qquad (21)$$

Thereby it follows that at small concentrations the local vibration should be observed. This conclusion is in perfect agreement with earlier described experimental data [44]. As to the second theoretical relation $\Delta > W$, one can see from the above discussion that for LiH (LiD) crystals the opposite relation, i.e. $W > \Delta$, is observed [20].

Following the results of [48], in Fig. 4 we show the second - order Raman scattering spectra in mixed LiH$_x$D$_{1-x}$ crystals at room temperature. In addition to what has been said on Raman scattering spectra at high concentration [48], we note that as the concentration grows further (x > 0.15) one observes in the spectra a decreasing intensity in the maximum of 2LO ($\Gamma$) phonons in LiD crystal with a simultaneous growth in intensity of the highest frequency peak in mixed LiH$_x$D$_{1-x}$ crystals. The nature of the latter is in the renormalization of LO($\Gamma$) vibrations in mixed crystal [49].

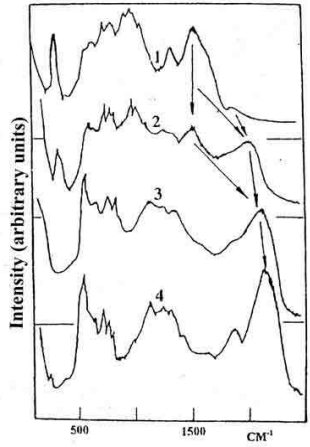

**Fig. 4. Second - order Raman scattering spectra in the isotopically mixed crystals LiH$_x$D$_{1-x}$ at room temperature. 1 - x = 0; 2 - 0.42; 3 - 0.76; 4 - 1. The arrows point out a shift of LO(Γ) phonons in the mixed crystals (after [40]).**

Comparison of the structure of Raman scattering spectra (curves 1 and 2 in Fig. 4) allows us, therefore, to conclude that in the concentration range of 0.1 < x < 0.45 the Raman scattering spectra simultaneously contain peaks of the LO(Γ) phonon of pure LiD and the LO(Γ) phonon of the mixed LiH$_x$D$_{1-x}$ crystal. For further concentration growth (x > 0.45) one could mention two effects in the Raman scattering spectra of mixed crystals. The first is related to an essential reconstruction of the acoustooptical part of the spectrum. This straightforwardly follows from a comparison of the structure of curves 1 -3 in Fig. 4. The second effect originates from a further shift of the highest frequency peak toward still higher frequencies, related to the excitation of LO(Γ) phonons. The limit of this shift is the spectral location of the highest frequency peak in LiH. Finishing our description of the Raman scattering spectra, it is necessary to note that a resonance intensity growth of the highest frequency peak is observed at x > 0.15 in all mixed crystals (for more details see [50]).

Once more reason of the discrepancy between theory and results of the experiment may be connected with not taking into account in theory the change of the force-constant at the isotope substitution of the smaller in size D by H ion [51]. We should stress once more that among the various possible isotope substitution, by far the most important in vibrational spectroscopy is the substitution of hydrogen by deuterium. As is well-known, in the limit of the Born-Oppenheimer approximation the force-constant calculated at the minimum of the total energy depends upon the electronic structure and not upon the mass of the atoms. It is usually assumed that the theoretical values of the phonon frequencies depend upon the force-constants determined at the minimum of the adiabatic potential energy surface. This leads to a theoretical ratio $\omega(H)/\omega(D)$ of the phonon frequencies that always exceed the experimental data. Very often anharmonicity has been proposed to be responsible for lower value of this ratio. In isotope effect two different species of the same atom will have different vibrational frequencies only because of the difference in isotopic masses. The ratio p of the optical phonon frequencies for LiH and LiD crystals is given in harmonic approximation by:

$$p = \frac{\omega(H)}{\omega(D)} = \sqrt{\frac{M(LiD)}{M(LiH)}} \simeq \sqrt{2} \qquad (22)$$

while the experimental value (which includes anharmonic effects) is 1.396 ÷ 1.288 (see Table1 in Ref. [51]). In this Table there are the experimental and theoretical values of p according to formula (22), as well as the deviation $\delta = \frac{P_{Theory} - p_{exp}}{p_{theory}}$ of these values from theoretical ones. Using the least squares method it was found the empirical formula of $\ln(\delta\%) \sim f(\ln[\frac{\partial E}{\partial M}])$ which is depicted on Fig. 5.

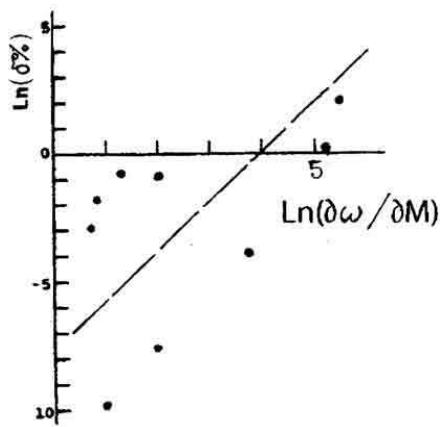

**Fig. 5. The dependence of $\ln(\delta\%) \sim f[\ln(\frac{\partial \omega}{\partial M})]$: points are experimental values and continuous line - calculation on the formulas (23) (after [51]).**

As can be seen the indicated dependence has in the first approximation a linear character:

$\ln(\delta\%) = -7.5 + 2\ln(\frac{\partial E}{\partial M})$.  (23)

From the results of Fig. 5, it can be concluded that only hydrogen compounds (and its isotope analog - deuterium) need to take into account the force-constant changes in isotope effect. It is also seen that for semiconductor compounds (on Fig. 5 - points, which is below of Ox line) the isotope effect has only the changes of the isotope mass (for details see [51]).

Thus, the experimental results presented in this section provide, therefore, evidence of, first, strong scattering potential (most importantly, for optical phonons) and, second, of the insufficiency of CPA model for a consistent description of these results [42].

1.3. Electronic excitations.

Isotopic substitution only affects the wavefunction of phonons; therefore, the energy values of electron levels in the Schrödinger equation ought to have remained the same. This, however, is not so, since isotopic substitution modifies not only the phonon spectrum, but also the constant of electron-phonon interaction (see above). It is for this reason that the energy values of purely electron transition in molecules of hydride and deuteride are found to be different [52]. This effect is even more prominent when we are dealing with a solid [53]. Intercomparison of absorption spectra for thin films of LiH and LiD at room temperature revealed that the longwave maximum (as we know now, the exciton peak [54]) moves 64.5 meV towards the shorter wavelengths when H is replaced with D. For obvious reasons this fundamental result could not then receive consistent and comprehensive interpretation, which does not be little its importance even today. As will be shown below, this effect becomes even more pronounced at low temperatures (see, also [39]).

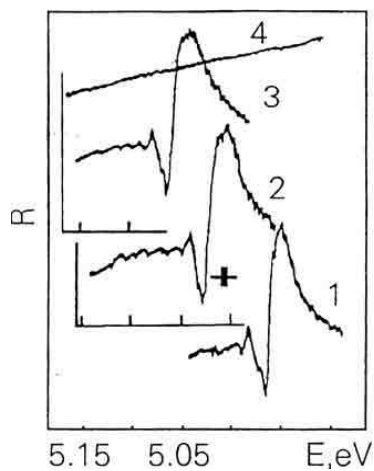

**Fig. 6. Mirror reflection spectra of crystals: 1 - LiH; 2 - LiH$_x$D$_{1-x}$; 3 - LiD; at 4.2 K. 4 - source of light without crystal. Spectral resolution of the instrument is indicated on the diagram (after [55]).**

The mirror reflection spectra of mixed and pure LiD crystals cleaved in liquid helium are presented in Fig. 6. For comparison, on the same diagram we have also plotted the reflection spectrum of LiH crystals with clean surface. All spectra have been measured with the same apparatus under the same conditions. As the deuterium concentration increases, the long-wave maximum broadens and shifts towards the shorter wavelengths. As can clearly be seen in Fig. 6, all spectra exhibit a similar long-wave structure. This circumstance allows us to attribute this structure to the excitation of the ground (Is) and the first excited (2s) exciton states. The energy values of exciton maxima for pure and mixed crystals at 2 K are presented in Table 17 of Ref. [21]. The binding energies of excitons $E_b$, calculated by the hydrogen-like formula, and the energies of interband transitions $E_g$ are also given in Table 17 of ref. [21].

Going back to Fig. 6, it is hard to miss the growth of $\Delta_{12}$, [55], which in the hydrogen-like model causes an increase of the exciton Rydberg with the replacement of isotopes (see Fig. 90 in [39]). When hydrogen is completely replaced with deuterium, the exciton Rydberg (in the Wannier-Mott model) increases by 20% from 40 to 50 meV, whereas $E_g$ exhibits a 2% increase, and at 2 ÷ 4.2 K is $\Delta E_g = 103$ meV. This quantity depends on the temperature, and at room temperature is 73 meV, which agrees well enough with $\Delta E_g = 64.5$ meV as found in the paper of Kapustinsky et al. Isotopic substitution of the light isotope ($^{32}$S) by the heavy one ($^{34}$S) in CdS crystals [56] reduces the exciton Rydberg, which was then attributed to the tentative contribution from the adjacent electron bands (see also [21]), which, however, are not present in LiH. The single-mode nature of exciton reflection spectra of mixed crystals LiH$_x$D$_{1-x}$ agrees qualitatively with the results obtained with the virtual crystal model (see e.g. Elliott et al. [45]; Onodera and Toyozawa [58]), being at the same time its extreme realization, since the difference between ionization potentials ($\Delta\zeta$) for this compound is zero. According to the virtual crystal model, $\Delta \zeta = 0$ implies that $\Delta E_g = 0$, which is in contradiction with the experimental results for $LiH_xD_{1-x}$ crystals. The change in $E_g$ caused by isotopic substitution has been observed for many broad-gap and narrow-gap semiconductor compounds.

All of these results are documented in Table 22 of Ref.[39], where the variation of $E_g$, $E_b$, are shown at the isotope effect. We should highlighted here that the most prominent isotope effect is observed in LiH crystals, where the dependence of $E_b = f(C_H)$ is also observed and investigated. To end this section, let us note that $E_g$ decreases by 97 cm$^{-1}$ when $^7Li$ is replaced with $^6Li$.

Further we will briefly discuss of the variation of the electronic gap ($E_g$) of semiconducting crystals with its isotopic composition. In the last time the whole raw of semiconducting crystals were grown. These crystals are diamond , copper halides , germanium , silicon, CdS and GaAs . All numerated crystals show the dependence of the electronic gap on the isotope masses (see, reviews [34, 39]).

Before we complete the analysis of these results we should note that before these investigations, studies were carried out on the isotopic effect on exciton states for a whole range of crystals by Kreingol'd and coworkers (see, also [37]). First, the following are the classic crystals $Cu_2O$ [58, 59] with the substitution $^{16}O \rightarrow {}^{18}O$ and $^{63}Cu \rightarrow {}^{65}Cu$. Moreover, there have been some detailed investigations of the isotopic effect on ZnO crystals , where $E_g$ was seen to increase by 55 cm$^{-1}$ ($^{16}O \rightarrow {}^{18}O$) and 12 cm$^{-1}$ ( at $^{64}Zn \rightarrow {}^{68}Zn$) [60, 61]. In [56] it was shown that the substitution of a heavy $^{34}S$ isotope for a light $^{32}S$ isotope in CdS crystals resulted in a decrease in the exciton Rydberg constant ($E_b$), which was explained tentatively by the contribution from the nearest electron energy bands, which however are absent in LiH crystals.

More detailed investigations of the exciton reflectance spectrum in CdS crystals were done by Zhang et al. [62]. Zhang et al. studied only the effects of Cd substitutions, and were able to explain the observed shifts in the band gap energies, together with the overall temperature dependence of the band gap energies in terms of a two-oscillator model provided that they interpreted the energy shifts of the bound excitons and n = 1 polaritons as a function of average S mass reported as was noted above, earlier by Kreingol'd et al. [56] as shifts in the band gap energies. However, Kreingol'd et al. [56] had interpreted these shifts as resulting from isotopic shifts of the free exciton binding energies (see, also [55]), and not the band gap energies, based on their observation of different energy shifts of features which they identified as the n = 2 free exciton states (for details see [56]). The observations and interpretations, according Meyer at al. [63], presented by Kreingol'd et al. [56] are difficult to understand, since on the one hand a significant band gap shift as a function of the S mass is expected [62], whereas it is difficult to understand the origin of the relatively huge change in the free exciton binding energies which they claimed. Meyer et al. [230] reexamine the optical spectra of CdS as function of average S mass, using samples grown with natural Cd and either natural S (∼ 95% $^{32}S$), or highly enriched (99% $^{34}S$). These author observed shifts of the bound excitons and the n = 1 free exciton edges consistent with those reported by Kreingol'd et al. [56], but, contrary to their results, Meyer et al. observed essentially identical shifts of the free exciton excited states, as seen in both reflection and luminescence spectroscopy.

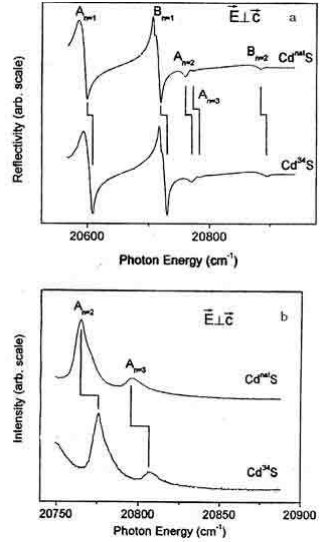

**Fig. 7. a** - Reflection spectra in the A and B excitonic polaritons region of Cd$^{nat}$S and Cd$^{34}$S at 1.3K with incident light in the $\vec{E} \perp \vec{C}$. **The broken vertical lines connecting peaks indicate measured enrgy shifts reported in Table 18. In this polarization, the n = 2 and 3 excited states of the A exciton, and the n = 2 excited state of the B exciton, can be observed. b** - Polarized photoluminescence spectra in the region of the A$_{n=2}$ and A$_{n=3}$ **free exciton recombination lines of Cd$^{nat}$S and Cd$^{34}$S taken at 1.3K with the $\vec{E} \perp \vec{C}$. The broken vertical lines connecting peaks indicate measured enrgy shifts reported in Table 18 (after [63]).**

The reflectivity and photoluminescence spectra i polarized light ($\vec{E} \perp \vec{C}$) over the A and B exciton energy regions for the two samples depicted on the Fig. 7. For the $\vec{E} \perp \vec{C}$ polarization used in Fig. 7 both A and B excitons have allowed transitions, and therefore reflectivity signatures. Fig. 7 also reveals both reflectivity signatures of the n = 2 and 3 states of the A exciton as well that of the n = 2 state of the B exciton.

In Table 18 Meyer et al. summarized the energy differences ΔE = E (Cd$^{34}$S) - E (Cd$^{nat}$S), of a large number of bound exciton and free exciton transitions, measured using photoluminescence, absorption, and reflectivity spectroscopy, in CdS made from natural S (Cd$^{nat}$S, 95% $^{32}$S) and from highly isotopically enriched $^{34}$S (Cd$^{34}$S, 99% $^{34}$S) (see, also [21]). As we can see, all of the observed shifts are consistent with a single value, 10.8±0.2 cm$^{-1}$. Several of the donor bound exciton photoluminescence transitions, which in paper [63] can be measured with high accuracy, reveal shifts which differ from each other by more than the relevant uncertainties, although all agree with the 10.8±0.2 cm$^{-1}$ average shift. These small differences in the shift energies for donor bound exciton transitions may reflect a small isotopic dependence of the donor binding energy in CdS. This value of 10.8±0.2 cm$^{-1}$ shift agrees well with the value of 11.8 cm$^{-1}$ reported early by Kreingol'd et al. [56] for the B$_{n=1}$ transition, particularly when one takes into account the fact that enriched $^{32}$S was used in that earlier study, whereas Meyer et al. have used natural S in place of an isotopically enriched Cd$^{32}$S (for details see [63]).

Authors [63] conclude that all of the observed shifts (see Table 18 of Ref. [21]) arise predominantly from an isotopic dependence of the band gap energies, and that the contribution from any isotopic dependence of the free exciton binding energies is much smaller. On the basis of the observed temperature dependencies of the excitonic transitions energies, together with a simple two-oscillator model, Zhang et al. [62] earlier calculated such a difference, predicting a shift with the S isotopic mass of 950 $\mu$eV/amu for the A exciton and 724 $\mu$eV/amu for the B exciton. Reflectivity and photoluminescence study of $^{nat}Cd^{32}S$ and $^{nat}Cd^{34}S$ performed by Kreingol'd et al. [56] shows that for anion isotope substitution the ground state (n = 1) energies of both A and B excitons have a positive energy shifts with rate of $\partial E/\partial M_S$ = 740 $\mu$eV/amu. Results of Meyer et al. [63] are consistent with a shift of ~710 $\mu$eV/amu for both A and B excitons. Finally, it is interesting to note that the shift of the exciton energies with Cd mass is 56 $\mu$eV/amu [62], an order of magnitude less than found for the S mass.

The present knowledge of the electronic band structure of Si stems from experimental observation of electronic transitions in transmission, reflectivity, or cyclotron resonance, on the one hand, and theoretical calculations, e.g. those based on pseudopotential or $\vec{k} \cdot \vec{p}$ methods (for details see [64, 21] and references therein). In this manner it has been established that the fundamental, indirect band gap of Si occurs between the $\Gamma_8^+$ valence band maximum and the $\Delta_0$ conduction band minima along (100).

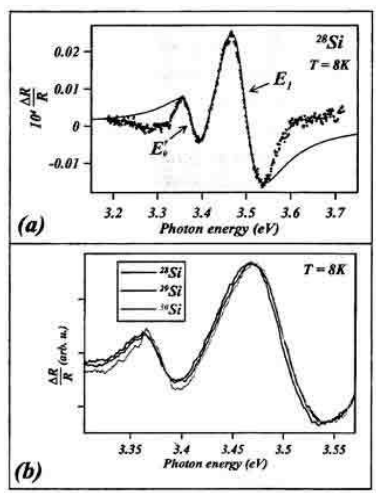

**Fig. 8. a -Signatures of the $E_0$' and $E_1$ excitonic band gaps of $^{28}$Si observed (dots) in photomodulated reflectivity. The solid line is a theoretical fit using the excitonic line shape. b - Photomodulated reflectivity spectra of isotopically enriched Si exhibiting isotopic shifts of the $E_0$' and $E_1$ gaps (after [67]).**

Recently, Lastras-Martinez et al. [65] performed ellipsometric measurements on isotopically enriched $^{28}$Si and $^{30}$Si and deduced the isotopic dependence of $E_1$ from the analysis of the data in reciprocal (Fourier inverse) space. However, these measurements did not resolve (see, also [60]) the nearly degenerate $E'_0$ and $E_1$ transitions and the isotopic shift was assigned solely to the stronger $E_1$ transitions (see, however, Fig. 8). We should add that in papers [67] very recently was studied the dependence of indirect band gap in Si on the isotopic mass. Photoluminescence and wavelength-modulated transmission spectra displaying phonon assisted indirect excitonic transitions in isotopically enriched $^{28}$Si, $^{29}$Si, $^{30}$Si as well as in natural Si have yielded the isotopic gap $E_{gx}$ which equals 1213.8±1.2 meV. This is purely electronic value in the absence of electron-phonon interaction and volume changes associated with anharmonicity (for details see [67] and below).

Returning to Fig. 8, we can see that the spectrum contains two characteric signatures, attributed to the excitonic transitions across the E'$_0$ and E$_1$ gaps. Isotopic dependence of the E'$_0$ and E$_1$ is displayed in Fig. 8, where the photomodulated reflectivity spectra of $^{28}$Si, $^{29}$Si, and $^{30}$Si are shown for the spectral range $3.3 \leq E \leq 3.58$ eV. The E'$_0$ and E$_1$ excitonic band gaps determined in paper [67] from the line-shape analysis. Linear least-squares fit yielded the corresponding isotopic dependences E'$_0$ = (3.4468 - 03378 M$^{-1/2}$) eV and E$_1$ = (3.6120 - 0.6821 M$^{-1/2}$) eV. In concluding, we should note that the spin-orbit interaction depends in Ge in contrast to that in Si [67].

As is well known ago, the fundamental energy gap in silicon, germanium, and diamond is indirect (see, e.g. [21] and references therein). While the conduction band minima in Si and diamond are located at the $\Delta$ point along <100>, with $\Delta_6$ symmetry, those of germanium with L$_6^+$ symmetry occur at the <111> zone boundaries [67]. The onset of the absorption edge corresponds to optical transition from the $\Gamma_8^+$ valence band maximum to the L$_6^+$ conduction band minima in Ge, and the $\Delta_6$ in Si and diamond; for wavector conservation, these indirect transitions require the emission or absorption of the relevant phonons. In Si and C, transverse acoustic (TA), longitudinal acoustic (LA), transverse optic (TO), or longitudinal optic (LO) phonons of $\Delta$ symmetry must be simultaneously emitted or absorbed. In Ge (see, also above), the wavector conserving phonons are TA, LA, TO or LO phonons with L symmetry. At low temperatures, these indirect transitions are assisted by phonon emission. In this case we should expect at low temperatures four excitonic derivative signatures at photon energies E$_{gx}$ + $\hbar\omega_{\vec{q},j}$ in modulated transmission experiments and in photoluminescence at the photon energies E$_{gx}$ - $\hbar\omega_{\vec{q},j}$. Here E$_{gx}$ is the excitonic band gap and j corresponds to a wave vector preserving phonon.

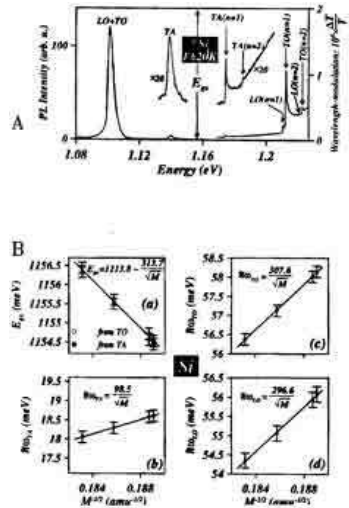

**Fig. 9. A - Photoluminescence (PL) and wavelength - modulated transmission (WMT) spectra of isotopically enriched $^{30}$Si recorded at 20K ; B - The excitonic indirect band gap and the associated phonon energies as a function of M (after [67]).**

In Fig. 9 - A the photoluminescence and wavelength - modulated spectra of $^{30}$Si (M = 2.81 amu) are displayed; the labels n = 1 and 2 designate the ground and the first excited states of the indirect TA and and TO excitons. From the energies of the photoluminescence and wavelength-modulated excitonic signatures in all isotopic specimens (see [67]) cited authors deduce $E_{gx}$ as well as the energies of the participating TO, LO and TA phonons, shown in Fig. 9 - B as function of $M^{-1/2}$. The excitonic band gap data are fitted well with expression $E_{gx}(M) = E_{gx}(\infty) - CM^{-1/2}$, yielding $E_{gx}(\infty) = (1213.8\pm1.2)$ meV and $C = (313.7 \pm5.3)$ meV/amu. A linear fit in M can be made over small range of available masses (see, Fig. 9 - B) with a slope $(\partial E_{gx}/\partial M)_{P,T}$ 1.01$\pm$0.04 meV/amu, which agrees with the results of bound exciton photoluminescence of Karaiskaj et al. [66]. The experiments in papers [67] also indicate that separation of the n = 2 and n = 1 excitons is isotope mass independent, implying, according these authors, the excitonic binding energy is independent on isotope mass within experimental error. In concluding this part we should note that recent high - resolution spectroscopic studies of excitonic and impurity transition in high - quality samples of isotopically enriched Si have discovered the broadening of bound exciton emission (absorption) lines connected with isotope - induced disorder as well as the depend of their binding energy on the isotope mass [66, 67]. The last effect was early observed on the bound excitons in diamond [68, 69], and earlier on the free excitons [70] in $LiH_xD_{1-x}$ mixed crystals (see, e.g. [71] and references therein).

**2. Methods of the preparation of low dimensional structures.**

2.1. Molecular beam epitaxy (MBE) and metal - organic chemical vapour deposition (MOCVD).

During the 1980s. two newer epitaxial techniques were introduced which are in widespread use today for the preparation of III - V (II - VI) semiconductor multilayers for both physics studies and device fabrication. These techniques are MBE and MOCVD [1, 2, 3, 73, 74]. With this it is possible to control both the chemical composition and the level of doping down to thickness that approach an atomic monolayer, and certainly within less than one nanometer ($10^{-9}$ m). MBE and MOVCVD thus approach the absolute limits of control for preparing layered semiconductor structures [1, 2].

Fabrication of a crystal layer upon wafer of a compatible crystal makes it possible to obtain very well - controlled growth regimes and to produce high - quality crystal with the desired crystalline orientation at temperatures typically well below the melting point of the substrate. The schematic heart of the MBE process is shown in Fig. 10, while the photograph of such equipment can be found in [75]. MBE is a conceptually very simple crystal growth technique. An MBE machine consists of a stainless steel vessel of diameter approximately 1 m (Fig. 10) which is kept under ultrahigh vacuum ($10^{-11}$ Torr) by a series of pumps [75]. Recent comprehensive reviews and monographs [1,2, 73, 74] give an up - to - date description of the technique and its application to the growth of semiconductor layers for electronic devices.

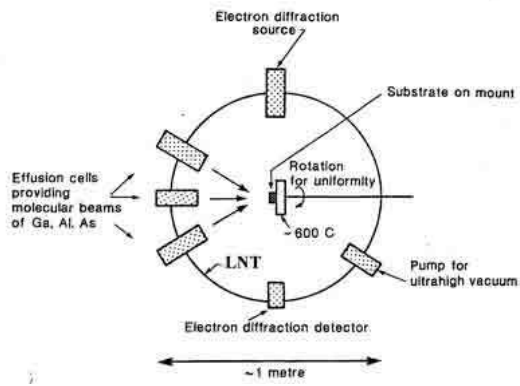

**Fig. 10. The MBE method for the growth of heterostructures.**

On one side (see, Fig. 10) a number of cells (typically eight) are bolted onto the chamber. These Knudsen cells are of some complexity, again to control the processes for which they are responsible. Inside each a refractory material boat contains a charge of one of the elemental species (for example, Ga, Al or As) for growth of the semiconductor, Si (for n - type doping), and Be or B (for p - type doping). Each boat is heated so that a vapor is obtained which leaves the cell for the growth chamber through a small orifice. The vapor is accelerated by the pressure differential at the orifice and forms a beam that crosses the vacuum chamber to impinge on a GaAs substrate which is mounted on a holder controlled from the opposite site of the chamber. The substrate holder contains a heater, as the quality of the grown crystal is a sensitive function of the substrate temperature (about $580^0$C and $630^0$C is best for GaAs and AlAs [1, 2]). The flux rate is controlled by the temperature within the Knudsen cell. The control and monitoring of the fluxes from the different cells ensure approximately a monolayer's worth of molecular beam species impinge on the substrate in 1 s. Thus the growth rates of the layers are typically $10^{-6}$mh$^{-1}$. Shutters in front of the orifices can be opened and closed in less than 0.1 s [74] and so combinations of Ga to Al flux can be varied to produce the species for growing Al$_x$Ga$_{1-x}$As alloys. The opening and closing of different shutters determines the multilayer structure that is grown in terms of both semiconductor composition and doping profile (see, also [9, 76]).

The advantage of MBE are:

1. The use of only high - purity elemental sources rather than less pure compounds may ultimately result in the highest purity.

2. Growth occurs in UHV apparatus where background concentrations of undesirable gases such as $H_2O$, CO and $O_2$ are very low.

3. In situ analytical tools (e.g. high - energy electron diffraction (RHEED), see, Fig. 10) may be used to monitor the crystal structure and composition.

4. Extreme control of growth rate and composition leads to very abrupt ($\langle$ 20 Å) changes in composition and/or doping level.

5. Pattern growth is possible through masks, by focused ion beams and on areas defined by electron beam writing.

BN is found to be an excellent crucible (boat) material for growth of high - purity III - V compounds [73]. A lower limit of the atomic or molecular flux from an diffusion cell (see, e.g. [2, 74])

$$J = 1.11 \cdot 10^{22}[AP(T)]/d^2(mT)^{1/2}, \qquad (24)$$

where the flux J is expressed in molecules cm$^{-2}$s$^{-1}$, A is the area of the opening, P(T) is the equilibrium pressure at the cell temperature, d is the distance to the substrate and m is the mass of the diffusing molecule. For typical systems, A ~ 1 cm$^2$, d $\cong$ 10cm and P(T) ranges from 10$^{-2}$ to 10$^{-3}$ Torr. These values give a flux at the substrate of 10$^{15}$ - 10$^{16}$ molecules cm$^{-2}$s$^{-1}$ and a growth rate of 1 - 10 monolayers per second, the typical growth rate in MBE systems (for details see [1, 73, 74, 75]).

Today, most of the III - V binary, ternary and a few quaternary semiconducting alloys have been grown by the MOCVD technique. GaAs and Al$_x$Ga$_{1-x}$As have been the most fully researched. MOCVD, which is also known as organometallic chemical vapour deposition (OMCVD), takes place in a glass reactor (the photograph see in [75]), typically about 0.3 m long and about 0.1 m in diameter (see Fig. 11).

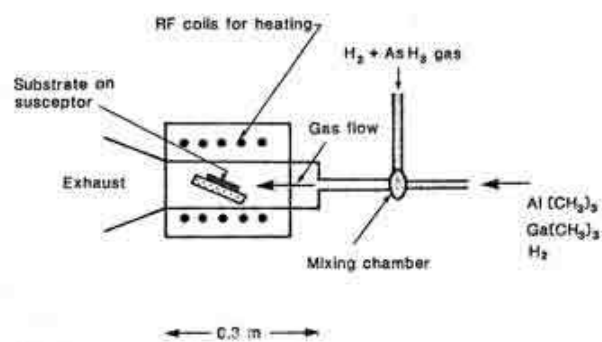

**Fig. 11. A scheme of the growth cell of an MOCVD method.**

In that reactor, a heated substrate site at an angle to a laminar flow of gas. Radiofrequency (rf) inductive heating is used to achieve substrate temperature comparable to those for MBE growth, although research is aimed at being able to use even lower substrate temperature while maintaining high - quality growth.

In the case of growth of Si layers, several different gases containing Si atoms can be used. They include silicon tetrachloride (SiCl$_4$), silane (SiH$_4$), and dichlorosilane (SiH$_2$Cl$_2$). In the case of silicon tetrachloride, the following reaction with hydrogen occurs:

$$SiCl_4 + 2H_2 \rightarrow Si + 4HCl. \qquad (25)$$

The reaction can be conducted at temperatures in the range of 1150 - 1250$^0$ C (see, e.g [2]). In the case of using silane and dichlorosilane, the reaction can be conducted at even lower temperatures (1000 - 1100$^0$ C). These temperatures are well below the melting point of Si (T$_m$ = 1412$^0$ C [73]). Thus, these reactions release atoms of Si, and the relatively low - temperature regimes provide efficient crystal growth onto the seed.

For growth of III - V compounds, the following reactions are used

$$A_{III}(CH_3)_3 + B_V H_3 \rightarrow A_{III}B_V,$$
$$A_{III}(CH_3)_3 + B_V H_3 \rightarrow A_{III}B_V + 3CH_4. \qquad (26)$$

Here A$_{III}$ is B, Al, Ga, IN, Tl and B$_V$ is N, P, AS, Sb, Bi. These reactions take place at temperatures of $\approx$ 600 - 700$^0$ C [2]. Dopants zinc [Zn(CH$_3$)$_2$] or silane are used to provide the dopant species. It is important that epitaxial methods can be applied to produce new materials that are difficult to grow by other methods. Examples are wide - bandgap nitrides of the group III elements. These include INGaN and AlGaN compounds.

In conclusions, both group IV elements (C, Si, Ge, Sn, Pb) and III - V compounds are successfully grown with thickness control of the order of one monolayer. Different types of doping - uniform doping, modulation doping, and delta - doping are realized with high accuracy [10, 11]. Since in the chemical reactor te partial pressures of chemicals are much higher than the pressure in the mlecular beams of the MBE method, the rate of crystal growth realized in the MOCVD method is higher than that of MBE. The former may be used in industrial production, while the latter is rather well suited for research laboratories.

The principles behind both MBE and MOCVD growth were established in the 1970s and refined in the 1980s, since which time further developments have taken place. New methods of growth research whose benefits will be realized during last two decades can be found in next references [1, 2, 73, 74].

2.2. Nanolitography and etching technologies.

The purpose of this paragraph is to describe the nanolitography and etching technologies used in fabrication of semiconductor nanostructures for physics and device studies. These state - of - the - art fabrication techniques are available precisely because they are foreseen as essential in one or other strategy for the fabrication of future devices (see, also [3, 9, 10]). As was shown above MBE and MOCVD grow high - quality single - crystal wafers and crystalline multilayered structures which thicknesses may be on the nanometer scale. However, to produce an individual device or electric circuit scaled down to nanosize in two or three dimensions, one needs to exploit the so - called nanolitography. For these processes, whereby short - wave radiation, for example short - wave ultraviolet (UV), electron beams, X - radiation, and ion beams are used to produce finer structures.

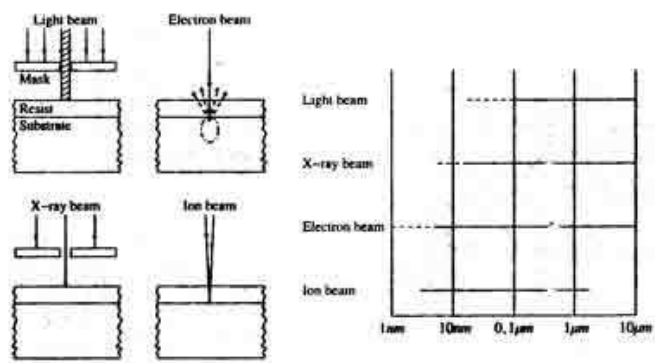

**Fig. 12. Overview of different litography method.**

Figure 12 presents a rough overview of the different nanolitography methods. At first side it is possible to create ever finer structures by using higher - energy radiation. But it must be noted that the material defects also increase proportionally [72]. It is clear that one practical limit to the smallest feature sizes that can be faithfully be reproduced is the wavelength of that light, i.e. the diffraction limit with visible light, in the range 400 - 800 nm. According Grenville et al. [7] the optical litography using visible light is available down to $0.15 \cdot 10^{-6}$m feature sizes in lithography for integrated circuits (see, also [8, 9]).

Electron - beam direct writing is a usual method to produce fine structures. The major disadvantage of this approach is that all individual structures must be written one after the other, which consumes a lot of time. That is why electron - bean writing is only economical for the mask production and not for direct structuring tasks on wafers. The X - ray lithography is very promising, although no usable lens systems and reflectors for wavelength between 0.5 and 5 nm are known. The imaging has to be performed by so - called contact copies with special masks. The mask carrier is a thin silicon film ($\sim 2 \cdot 10^{-6}$m), which is transparent for X - radiation. The actual masking part is a thin gold layer ($10^{-6}$m) structured by electron - beam writing. In order to avoid contact copy image defects on the semiconductor wafer, the X - ray beams should run as parallel as possible. When using a normal X - ray tube, the distance between the radiation source and the silicon wafer is so small that many image defects occur in the peripherical wafer area. This error $\Delta B$ can be simply measured by the following rule (see, e.g. [74])

$$\Delta B = t \frac{B}{S}, \qquad (27)$$

where t represents the mask - wafer distance, B equals the wafer radius, and S is the distance to the radiation source. To keep this deviation small, the distance S has to be increased as much as possible, which can be achieved by extracting X - ray beams from a synchrotron. When using such a synchrotron, the distance S can be chosen to be relatively large, say 10 m, so that the error $\Delta B$ is reduced to less than $\pm$ 10 nm.

The comparison between the different lithography methods, the highest throughput is still achieved by optical lithography [9, 14]. A higher resolution can be attained by using X - ray or electron beams. The single probe methods, whereby single atoms are manipulated, yields the best results. Remarkable results are also produced by a structure printing process the so - called nanoprinting (see, also [8]). Another method should be mentioned in this context: the scanning tunneling microscope (STM) can be utilized to visualize and analyze the fine structure (for details see next paragraph).

There are two principal forms of etching - using wet chemicals or using (dry) plasmas. Each method has advantages and disadvantages. Wet chemistry has been used since the earliest days of integrated circuit manufacture. The range of dilute acid and alkaline materials used is quite wide. For example HF reacts with $SiO_2$ and does not affect the photoresist or silicon. That is, this wet chemical etch is highly selective. However, the rate of etching is the same for any direction, lateral or vertical, so the etching is isotropic (see, however [8]). Using an isotropic etching technique is acceptable only for relative large structures. For nanosize structures, anisotropic etching with faster vertical etching is preferable.

Anisotropic etching generally exploits a physical process, or some combination of both physical and chemical methods. The best - known method of anisotropic etching is reactive - ion etching. Reactive - ion etching is based on the use of plasma reactions. This method works as follows. An appropriate etching gas, for example a chlorofluorocarbon, fills the chamber with the wafers. The pressure is typically reduced, so that a rf voltage can produce a plasma. The wafer to etch is a cathode of this rf discharge, while the walls of the chamber are grounded and act as an anode.

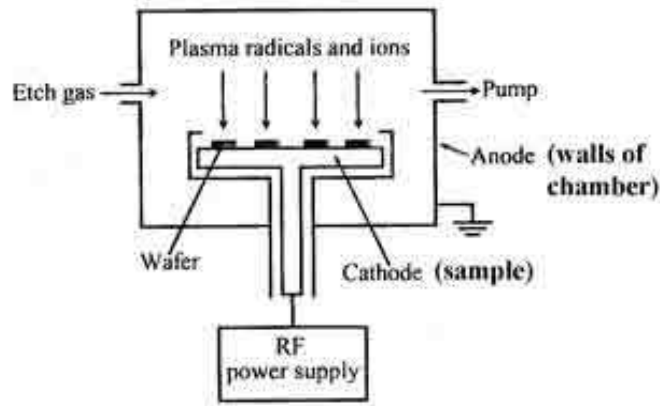

**Fig. 13. Scheme of plasma etching.**

Fig. 13 illustrates a principal scheme for the ion - etching method. The electric voltage heats the light electrons and they ionize gaseous molecules, creating positive ions and molecular fragments (chemical radicals). Being accelerated in the electric field, the ions bombard the wafer normal to the surface. This normal incidence of bombarding ions contributes to the etching and makes the etching highly anisotropic. This process, unfortunately, is not selective. However, the chemical radicals present in the chamber give rise to chemical etching, which, as we pointed above, is selective. From this, we can conclude, that the method combines both isotropic and anisotropic components can give good results for etching on the nanoscale (see, e.g. Fig. 3.4 in [8]). For further details on etching techniques see [78, 79]).

2.3. Techniques for characterization of nanostructures.

Progress in the fabrication, study, and use of nanostructures would not be possible without adequate techniques for the characterization of these structures. These techniques should allow one to determine the shape and geometrical parameters of nanostructures, the distribution of chemical composition, the strain fields, etc. Knowing all of these one can predict the electronic and optical properties which will ultimately be relevant in applications. In this paragraph we briefly describe next techniques:
   1. Hall measurements;
   2. Secondary ion mass spectrometry (SIMS);
   3. Methods for optical characterization;
   4. Scanning tunneling microscopy (STM);
   5. Atomic - force microscopy (AFM);
   6. Transmission electron microscopy (TEM);
   1. The simplest method for checking of doping levels (electrical characterization) in bulk semiconductors is the Hall measurements [24, 26].
   2. This method involves removing material from a multilayer structure using a beam of high - energy ions (i.e. sputtering) and a mass analysis of the speci4es that can from flat center of the crater (for details see [72, 80]).

3. Photoluminescence (PL), electroluminescence, and photoreflectance spectroscopy have all proved useful in the qualification of certain multilayer structures [3, 76]. AS is well - known, the first two techniques involve the examination of the wavelength - or energy - selected emission of light from a structure when the structure has first been excited by light or an electron beam respectively.  The last technique involves changes of reflectivity at different energies as small electric fields are applied to the solid. All three provide provide information about optical transitions and we shall see the way in which  thin layers have optical transitions modified by quantum - size effects including the shape of heterojunction interfaces (atomically abrupt changes of comparison, or changes over two or more atomic layers, with or without lateral steps in the interface plane).

The structure in the PL spectra as a function of energy can be used to infer the position of various energy levels - in particular the bandgap of the different layers - while the linewidths of the PL features can be interpreted in terms of uniformity and absence of fluctuation. PL as a diagnostic technique has a number of advantages: it is nondestructive, fairly simple to implement, able to give a rapid turnaround of information to crystal growers, and capable of use in a wafer - mapping mode (i.e. checking out the uniformity of layers in a manufacturing environment). The disadvantage include complications that arise if one has (as is often the case) an active device structure with heavily doped contact layers on either side; such doped layers  tend to reabsorb and redistribute the luminescent energy. The qualitative results produced are particularly useful for many structures of optical devices (lasers etc.). Changes to the PL spectra under the influence of temperature changes, applied stress, magnetic fields, etc. all provide further insight into the thicknesses, compositions, and uniformity of multilayers. Photoluminescence excitation (PLE) spectroscopy involves the detection of light at a fixed energy (generally near a prominent band - edge feature, e.g. the lowest bandgap) while the excitation energy is monochromatic and is swept in energy. The amplitudes of features  in the PLE spectra provide further information about the cross - section  of the optical absorption processes involved in the relaxation of the excited electrons (excitons [55]). Typical PL and PLE data for a range of thin layers of GaAs between thicker AlGaAs layers, according [16], is shown in Fig. 14.

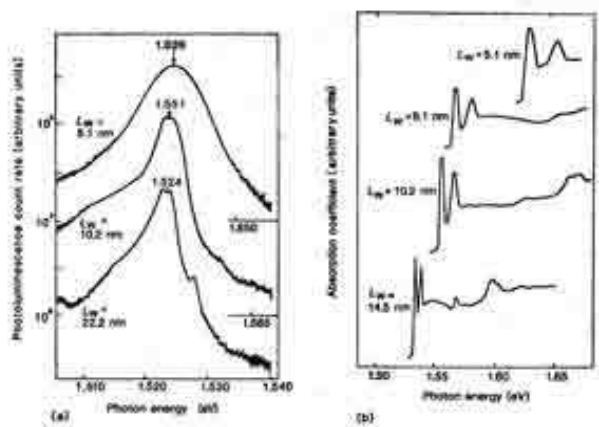

**Fig. 14. (a) Photoplumenescence spectra for wells of different thickness $W_L$**

and (b) the photoluminescence excitation spectra from GaAs quantum wells (after [16]).

4. STM yields surface topographies and work - function profiles on an atomic scale directly in real space. In terms of classical physics, a transfer process of an electron from one solid into another can be taught of as an electron transfer over a vacuum barrier. This process requires additional energy and because of this it has a small probability. On the other hand, according to quantum mechanics, a particle can penetrate a classically forbidden spatial region under a potential barrier. Thus, electron transfer between two solids can occur as a tunneling process through (under) vacuum barrier. The principle of STM, which is based on electron tunneling, is straightforward. It consists essentially of a scanning metal tip (one electrode of the tunnel junction) over the surface to be investigated - the second electrode. At present, the resolution of STM reaches 0.5 Å vertically and well below 2 Å laterally (see, e.g [6]). On the other hand, STM is subject to some restrictions in application: only conductive samples can be investigated, and measurements usually have to be performed in UH vacuum. At the same time, the tunnel current is sensitive to material composition and strain. Atomic resolution in both lateral and vertical directions makes STM an ideal tool for the investigation of growing surfaces at this scale, which can give insight into growth mechanisms. STM systems attached to a growth chamber allow measurements to be made without breaking the vacuum after growth [76]. It is remarkable that, apart from providing structural information, low - temperature STM has been used for wavefunction mapping of single electron states in nanostructures. Applied to the InAs dots (islands) the STM methods directly reveal s -, p -d -, and even f - type states as made visible by an asymmetry of the electronic structure (see, Fig. 15), which can be attributed to a shape symmetry of the islands (see, also [81, 5]).

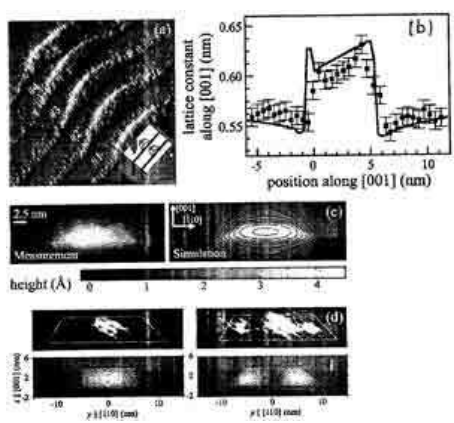

**Fig. 15. Cross - sectional STM: (a) an image of a stack of InAs islands in GaAs; (b) the lattice parameter in the growth direction in an InAs island (the experimental data were obtained from cross - sectional STM; the solid line is from a simulation assuming an In content increasing from island base to island apex);(c) comparison between measured and simulated height profiles for a similar sample; and (d) the electronic wavefunction measured at two different tip biases, compared with simulation for the ground and the first excited states (after [5]).**

5. An atomic force microscope (AFM) measures the force between the sample surface and a very fine tip. The force is measured either by recording the bending of a cantilever on which the tip is mounted - it contact mode - or by measuring the change in resonance frequency due to the force - the trapping mode. With a typical resolution of several nanometers laterally and several Å vertically, AFM is ideally suited to characterize the shapes of nanostructures. With AFM, any surface can be investigated. Furthermore, most semiconductor materials oxidize under ambient conditions, so that, strictly speaking, the AFM images usually show the surface of this oxide. When obtaining quantitative data such as lateral sizes and heights of structures, this has to be kept in mind, as well as the fact that this image is actually a convolution of the sample's surface morphology with the shape of the microscope tip.

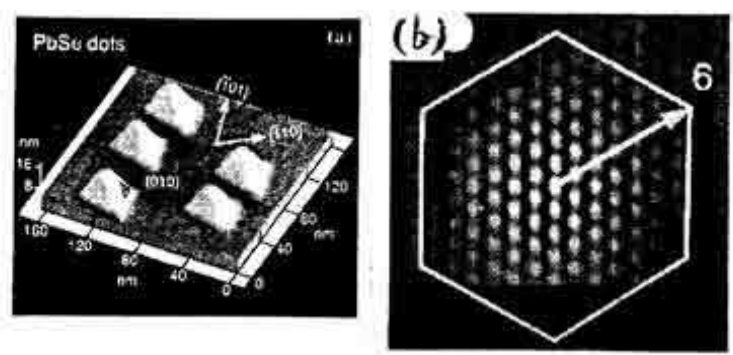

**Fig. 16. PbSe islands with {001} facets; (a) AFM image of the top surface of a PbSe/PbEuTe island multilayer grown on BaF$_2$; (b) the autocarrelation function. Islands are arranged in a regular array up to the sixth - nearest neighbor (after [5]).**

Examples of the quantitative analysis of AFM images are shown in Fig. 16 [5]. There, the top surface of PbSe/PbEuTe multilayers is shown. From Fig. 16$^a$, one can see that PbSe forms triangular pyramids with [001] - type side facets. The lateral ordering can also be analyzed. In Fig. 16$^b$, a hexagonal in - plane arrangement of pyramids is evident (for details see [4, 5]).

6. Among the methods which allow one "to see" things at the nanometer scale, two types of electron microscopy play an important role. Transmission electron microscope (TEM) makes possible the visualization of thin slices of material with nanometer resolution [6]. This technique has subnanometer resolution, and, in principle, can resolve the electron densities of individual atoms (see, also [8, 74]). A TEM operates much like an optical microscope, but uses electrons instead of visible light, since the wavelength of electrons is much smaller than that of visible light. As we have already discussed, the resolution limitation of any microscope is based on the wavelength of the probe radiation. Since electrons are used instead of light, glass lenses are no longer suitable. Instead, a TEM uses magnetic lenses to deflect electrons. In a TEM, the electrons are collimated from the source and passed through the sample, and the resulting pattern of electron transmission and absorption is magnified onto a viewing screen. In scanning electron microscope (SEM), the electron beam is not projected through the whole sample area. Instead, it is raster - scanned across the surface, and the secondary electrons, or X - rays, emitted from the surface are recorded. This generates a low - resolution image, but allows the direct mapping of surface features, and can even be used for elemental analysis. The scheme the both types of electron microscope are depicted in Fig. 17. Thus, the structural and chemical information provided by TEM (SEM) has made it the single most important analytical tool available (for details see [82, 83]).

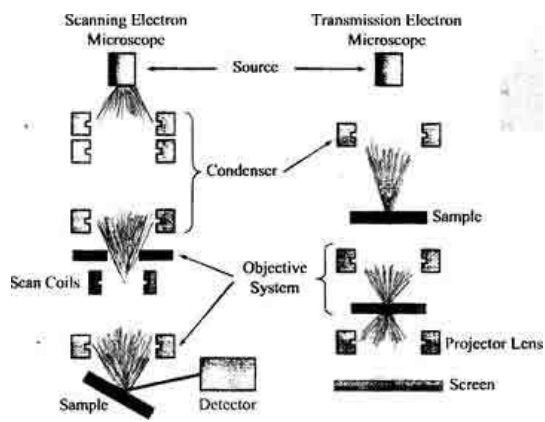

**Fig. 17. Schematic diagrams of SEM and TEM.**

One example of atomically resolved TEM, from which the position of unit cells, strain, and composition information were derived, is depicted on Fig. 18. Remarkably, these techniques make it possible to visualize a detailed map of the strain for an object of size a few tens of nanometers.

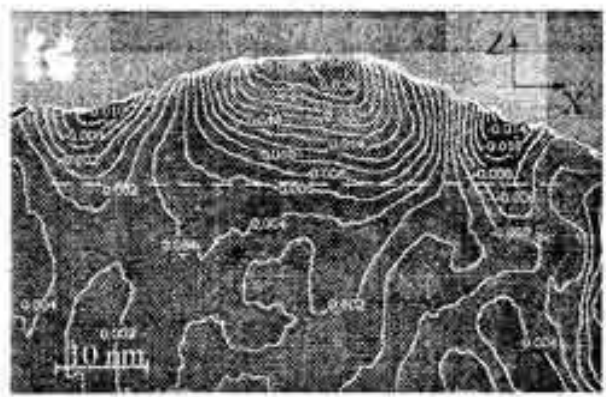

**Fig. 18. Strain distribution obtained from the TEM images of InGaAS islands in GaAs by using the method of digital analysis of lattice images (aftetr [5].**

From Fig. 18 it is seen clearly how the strain increases at the apex of the InGaAs island, while around the island the strain changes its sign.

Concluding this paragraph, we should note, that powerful characterization techniquFig. 19. Schematics of Si isotope superlattices. Thickness of each isotope layer are 1.1; 1.6; and 3.2 for $^{28}Si_8/^{30}Si_8$; $^{28}Si_{12}/^{30}Si_{12}$ and $^{28}Si_{24}/^{30}Si_{24}$ samples, respectively. Low index denotes the thickness of each isotope layer in atomic monolayers, each 0.136 nm thick (after [94]).

2.4. Nuclear technology.

The advances in epitaxial thin film homo and hetero-structures. synthesis, which have been achieved through a variety of epitaxial techniques [1, 2], have led to a vast array of new solid - state structures with many fascinating properties (see, e.g. [8 - 11]). In view of this vast variety of activities and discoveries of isotope hetero-structures. including isotope hetero-structures. has been studied only in two last decades [87 - 95]. In combination with the well established neutron transmutation doping (NTD [12]) technique, isotope hetero-structures. appear to represent a family of solid state structures, which offer new possibilities and numerous advantages over the traditional multilayer structures (see above). The formation of a doped isotope multilayer structure can be broken down into two independent steps: growth of the structure with isotopically pure or deliberately mixed layers and selective doping with the NTD process (see, also [94]). The formation of an isotope multilayer structure differs from the traditional methods only in sofar that isotopically pure and deliberately mixed sources must be used, and, the most importantly, that no dopants are introduced during the growth process. The absence of any dopants during the growth process automatically eliminates all dopant induced effects including autodoping and dopant interdiffusion between adjacent layers [91]. In principle all the established epitaxial techniques can be applied to the growth of isotope multilayer structures. The only requirement is to availability of semiconductor grade pure isotopes. The doping of an isotope hetero-structures. is achieved with the NTD [12] techniques after growth process has been completed. The NTD technique is isotope selective and therefore it can used superlatively for the creation of the low - dimensional structure. The cross -section for thermal neutron capture and the subsequent nuclear processes of practically every stable isotope of all elements have been measured, studied and documented (see also [12] and references therein).

As we all know that breaking the crystal translational symmetry without strongly influencing its electronic band structure can be done by means of a modification in the mass of one or more atoms composing the crustal. Without translational symmetry, the wave vector conservation requirements can be circumvented. Ideal models for most studies of elementary excitations are represented by isotopically pure crystals. A new field offering interesting physical studies is opened with the growth of isotopically tailor - made single crystals. The translational symmetry operations can be removed in part by artificial fabricating isotopic superlattice in which layers of two isotopically enriched materials alternate periodically. MBE of isotopically controlled germanium has enabled studies of low - dimensional phonons in isotope superlattice [89, 90, 91, 92] and quantum dots [93].

In this paragraph we describe the results of Raman measurements on novel kind of hetero-structures., a series of isotopic superlattice' of germanium and silicon [90 - 94]. These samples represent an excellent model system to study the vibrionic properties of superlattice because the electronic structure should be affected only weakly by changes in the isotopic mass (see, e.g. [34, 39]).

Since these changes are the only difference between the superlattice' constituents, Raman spectroscopy is the only nondestructive method to investigate their structural properties. Experimental data are compared with the results of planar force - constant model [89]. Let us consider the case of Ge, with the five isotopes of it (see, also [44]).

The readers will ask themself one should see five phonons (or more if they know that there are two atoms per primitive cell), corresponding to the five different masses, or only one corresponding to the average mass. We are all know that the latter is true. The transition from the average mass vibrations to those localized at all possible pairs is an example of the Anderson localization phenomenon [96], which is observed in Raman experiments on LiH$_x$D$_{1-x}$ system (for details see [40]). In a three - dimensional crystal fluctuations in the parameters of the secular equation lead to localization (measured in units of frequency, i.e. ($\Delta$M/M)$\omega_0$) are larger than the bandwidth of the corresponding excitations [44, 45]. For optical phonons in Ge this bandwidth is 100 cm$^{-1}$ while ($\Delta$M/M)$\omega_0$ ⩽ 0.04 · 300 = 12 cm$^{-1}$ (see e.g. [64, 96]). Hence no phonon localization (with lines corresponding to all pairs of masses) is expected, in agreement with the observation of only one line at 304 cm$^{-1}$ (at 77 K) for natural Ge (see Fig. 31 in [44]). For comparison we indicate that the bandwidth in the LiH$_x$D$_{1-x}$ mixed crystal is more than 500 cm$^{-1}$ therefore the crystal and localized phonons are coexistenced (for details see [40]).

In superlattice composed, for example, of n layers of $^{70}$Ge and m layers of $^{76}$Ge repeated periodically, one would expect to find optical modes localized or nearly localized in each of two constituents. Schematics of Si isotope superlattice depicted on Fig. 19 [94]. Koijma et al. have grown three kinds of silicon isotope superlattice ($^{28}$Si$_n$/$^{30}$Si$_n$, with n = 8, 12 and 24) using the solid - source MBE technique [1, 73]. In this paper n denotes the thickness of each isotope layer in atomic monolayers, each 0.136 nm thick. The periodicities, i.e. the number $^{28}$Si/$^{30}$Si pait layers stacked vertically, are 80, 50 and 30 for n = 8, 12 and 24 samples respectively. The resulting total thickness of the superlattice are 160 - 200 nm (see, Fig. 19). The source for the $^{28}$Si layer is actually $^{nat}$Si which is composed of 92.2% $^{28}$Si. The source for the $^{30}$Si layer is a single Si crystal isotopically enriched to $^{30}$Si(~98.74% [95]).

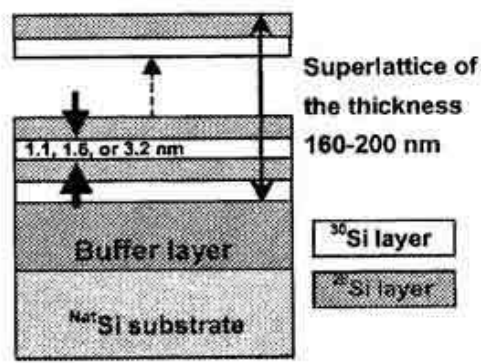

**Fig. 19. Schematics of Si isotope superlattices. Thickness of each isotope layer are 1.1; 1.6; and 3.2 for $^{28}$Si$_8$/$^{30}$Si$_8$; $^{28}$Si$_{12}$/$^{30}$Si$_{12}$ and $^{28}$Si$_{24}$/$^{30}$Si$_{24}$ samples, respectively. Low index denotes the thickness of each isotope layer in atomic monolayers, each 0.136 nm thick (after [94]).**

In MBE in individual effusion cells equipped with crucibles made of high purity tantalum. The crucible temperature is maintained at 1400$^0$ C for a growth rate of ~ 0.01 nm/s. The base pressure of the vacuum is 5 · 10$^{-10}$ torr and the pressure during growth is ~ 10$^{-9}$ Torr.

As was shown above, the E vs k dispersion of phonons in the superlattice is zone - folded due to the new periodicity, na, introduced by the $(^{28}Si)_n$ - $(^{30}Si)_n$ unit where a is the periodicity of the bulk Si. Because Raman spectroscopy, to fist order, probes phonons situated at k ~ 0 in the dispersion relation, while only one longitudinal optical (LO) phonon peak is observed with bulk Si, multiple LO phonon peaks should appear for isotope superlattice due to the zone folding or phonon localization (see, e.g. [98]).

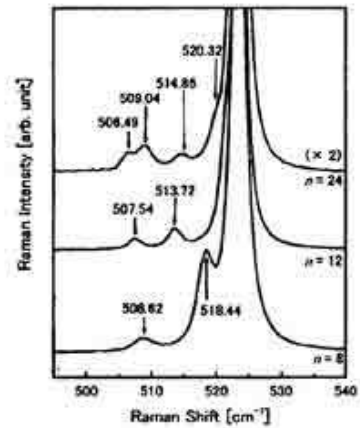

**Fig. 20. Raman spectra of the $^{28}Si_n/^{30}Si_n$ samples with n = 8, 12 and 24 (after [94]).**

Fig. 20 shows the Raman spectra of Si superlattice. As expected, many peaks are observed on the shoulders of the large $^{nat}Si$ substrate LO peak around 523.5 cm$^{-1}$. The wave numbers of the identified peaks are indicated in Fig. 20 for comparison with theoretical predictions fulfilled in the planar bond - charge model for Si (see [97]). As was shown in paper [95] theoretical curves are not smooth due to anticrossings. In general, the agreement between the experimental and theoretical results is excellent, except for the one detail: while $LO_1$ ($^{28}Si$) peaks in n = 12 and 24 samples are hidden in the large substrate peak, the $LO_1$ ($^{28}Si$) peak is observed experimentally for the n = 8 sample and its position deviates from the calculation (for details see Fig. 4 in [94]).

Raman spectra of a seria of isotopic $^{70}(Ge)_n^{74}(Ge)_n$ superlattice with $2 \leq n \leq 32$ ($8 \leq n \leq 24$) was published in papers [90, 91]. Three modes could be observed (see Fig. 21) for the $^{70}(Ge)_{16}^{74}(Ge)_8$ "as - grown" superlattice as theoretically predicted [89]. We should underline that the excellent agreement between results of papers [90, 91].

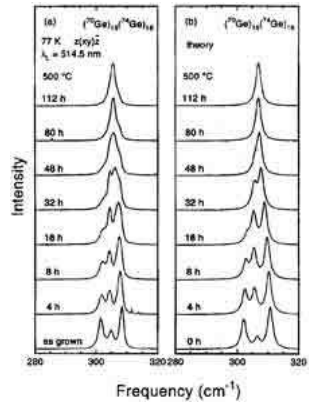

**Fig. 21.** (a) Experimental Raman spectra of a ($^{70}$Ge)$_{16}$($^{74}$Ge)$_{16}$ superlattice for different annealing steps at 500$^0$ C. (b) Calculated Raman spectra for the same superlattice using the same parameters (after [91]).

In concluding of this paragraph we should stressed that isotopic superlattice represent an excellent model system for the investigation of confinement of optical phonons. Both frequencies and relative intensities of the measured spectra are in good agreement with calculations based on a planar bond - charge model and the bond - polarizability approach.

**3. Electron excitations in low - dimensional structures.**

3.1. Wave - like properties of electrons.

In classical physics we deal with two kinds of entities: particles, such as a small mass which obeys Newtoniat's equations , and waves as, for example, electromagnetic waves which behave according to Maxwell's equations. Moreover, classical physical models assume the continuity of quantities and involve no restrictions concerning very small physical structures. The quantum theory shows, however, that values of some measurable variables of a system, can attain only certain discrete meanings. Therefore, in dealing with very small objects, like atoms, the above classification (particles and waves) is not enough to describe their behavior, and we have to turn to quantum mechanics, and to the dual concept of wave - particle. For instance, if light interacts with a material, it is better to think of it as being constituted by particles called photons instead of waves. On the other hand, electrons of which have primary concept of particle, behave like waves, when they move inside a solid of nanometric dimensions.

In third decade of XX century, Davison and Germer showed that electrons impinging against and in fact diffracted, as if they were waves, and in fact followed Bragg's low of diffraction [26]. The more details the electron's waves was described in paper [99]. The beautiful photograph in Fig. 22 clearly shows the wave - particle dualism of the electron by means of the accumulation of many single shots, corresponding to independent electrons, in an interferometry experiment performed by A. Tonomura (for details see [99]).

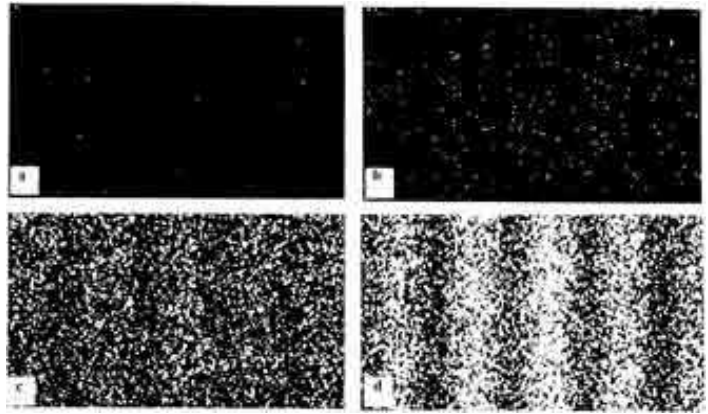

**Fig. 22. Observing the wave - like properties of the electrons in the double slitexperiment (after [99]). The pictures (a) →(c) have been taken at various times: picturees on the monitor after (a) 10 electrons, (b) 200 electrons, (c) 6000 electrons, and (d) 140 000 electrons. Electrons were emitted at a rate of 10 per second. (after A. Tonomura, 2006, Double - slit experiment (http://hqrd.hitachi.co.jp/globaldoubleslit.cfm)).**

In 1924 de Broglie assumed, to every particle of momentum p, a wave of wavelength

$\lambda = \frac{\hbar}{p}$,  (28)

where as usually $\hbar$ is a Planck's constant ($\hbar = \frac{h}{2\pi}$) and

$p = \frac{h}{\lambda} = \frac{h}{2\pi} \cdot \frac{2\pi}{\lambda} = \hbar k$.  (29)

In (29) k is so - called wave number. According quantum mechanics the electron is described by successive quantum - mechanical states, which represent a certain probability that the particle may be located in a specific spatial region. These measures of probability can be calculated from the wave function $\Psi$ that results from the solutions of a partial differential equation called the Schrödinger equation

$[-\frac{\hbar^2}{2m}\nabla^2 + V(\vec{r}, t)]\Psi = i\hbar\frac{\partial \Psi}{\partial t}$,  (30)

where $\nabla^2$ is the operator ($\frac{\partial^2}{\partial x^2} + \frac{\partial^2}{\partial y^2} + \frac{\partial^2}{\partial z^2}$), and $V(\vec{r}, t)$ is the potential energy, which is generally a function of position and time. As is known, the function $\Psi$ does not have a physical meaning, the product of $\Psi$ by its conjugate ($\Psi^*$) is a real quantity, such that the indicated above probability dP of funding a particle in a small volume dV is given by

$dP = |\Psi|^2 dV$.  (31)

If the potential energy V is not time dependent, we can search for a solution to Eq. (30) of the form

$\Psi(\vec{r}, t) = \Psi(\vec{r})e^{-i\omega t}$.  (32)

Substituting Eq. (32) in (30) and writing $E = \hbar\omega$, we can find the time - independent Schrödinger equation

$[-\frac{\hbar^2}{2m}\nabla^2 + V(\vec{r})]\Psi(\vec{r}) = E\Psi(\vec{r})$  (33)

for the time - independent wave function $\Psi(\vec{r})$. Schrödinger's equation can be solved exactly in a few cases (see, e.g. [15, 27]). Probably the simple stone is that of free particle, as for instance a free electron of energy E and mass m. In this case $\Psi(\vec{r}) = 0$ and the solution of Eq. (30) is easily found to be

$$\Psi = Ae^{i(kx - \omega t)} + Be^{i(-kx - \omega t)} \qquad (34)$$

where

$$k = \left(\frac{2mE}{\hbar^2}\right). \qquad (35)$$

Therefore, the free electron is described by a wave, which according to the de Broglie relation has momentum and energy given, respectively, by

$$p = \hbar k, \qquad E = \frac{p^2}{2m}. \qquad (36)$$

In general we will assume that the electron travels in one direction, for example, along the x - axis from left to right, and therefore the coefficient B in Eq. (34) is zero. The wave function for the free electron can simply be written as:

$$\Psi = Ae^{i(kx - \omega t)}. \qquad (37)$$

Another example in which Schrödinger's equation can be solved exactly is that of the hydrogen atom for which the potential is Coulombic, i.e. V varies with distance r between proton and electron in the form 1/r. Solving Schrödinger's equation, one gets the well - known relation for the electron energies [27]:

$$E_n = -\frac{m_r e^4}{2(4\pi\varepsilon_0)^2 \hbar^2 n^2} = -\frac{13.6}{n^2} \text{ eV}, \qquad n = 1, 2, 3, \ldots \qquad (38)$$

In the last expression $m_r$ is the reduced proton - electron mass ($m_r = \frac{M_p m}{M_p + m}$). In solid state physics, the mathematical model of the hydrogen atom is often used< as for example, in the study of the effects of impurities and excitons in crystals [26]. Although the equation giving the values of the energy is very similar to Eq. (38), the values of the binding energy $E_n$ are much smaller since the dielectric constant of the medium has to substitute the value of the permittivity of vacuum $\varepsilon_0$. For instance, in the case of silicon, the value of the dielectric constant is about $12\varepsilon_0$ [100].

Another important relation that derives heuristically from the model described above is Heisenberg's uncertainty principle: in any experiment, the products of the uncertainties, of the particle momentum $\Delta p_x$ and its coordinate $\Delta x$ must be larger than $\hbar/2$, i.e.

$$\Delta p_x \cdot \Delta x \geq \hbar/2. \qquad (39)$$

There are of course corresponding relations for $\Delta p_y \cdot \Delta y$ and $\Delta p_z \cdot \Delta z$. It is important to remark that this indeterminacy principle is inherent to nature, and has nothing to do with errors in instruments that would measure $p_x$ and x simultaneously. The second part of this principle is related to the accuracy in the measurement of the energy and the time interval $\Delta t$ required for the measurement, establishing

$$\Delta E \Delta t \geq \hbar/2. \qquad (40)$$

So, uncertainty principle denotes that the location or the momentum of a particle, and its energy or its time of observation can only be determined imprecisely. This statement is very important is we are considering nanoelectronic applications, because the dimensions of such devices are so small that we can use the uncertainty principle to roughly estimate the relevant nanoelectronic effects, for example, the tunneling effect. In the following sections some important nanoelectronic structures will be discussed. Thereby in is inevitable to apply the wave model of matter to describe the behavior of the electrons involved. The upcoming example of the potential well shows that is not possible to correctly determine the behavior of an electron in such configuration by using the classical - particle model.

In conclusion of this paragraph we should note that the interpretation above of $|\Psi|^2$ suggests the introduction the term information. The information delivered by a measuring process is inversely proportional to the probability of localizing a particle in the observation space [99, 101]. Although this relation to information theory is interesting, the concept was not generally adopted by physicists.

### 3.2. Dimensionality and density of states.

As we know from solid state physics, most physical properties significantly depend on the density of states (DOS) function S. The DOS function, at a given value E of energy, is defined such that $S(E)\Delta E$ is equal to the number of states (i.e. solution of Schrödinger equation) in the interval energy $\Delta E$ around E (see, e.g. [10]). We also know that if the dimensions $L_i$ (i = x, y, z) are macroscopic and if proper boundary condition are chosen, the energy levels can be treated as a quasi - continuous [26]. On the other hand, in the case where any dimensions $L_i$ gets small enough, the DOS function becomes discontinuous. Let us next obtain the DOS function for several low - dimensional solids (see, also [10, 74]).

We also know that every electron state is defined by the set of numbers ($k_x$, $k_y$, $k_z$). According to the Pauli exclusion principle there will be two electrons (spin up and spin down) for each occupied state. Since the electron energy is proportional to $k^2$, the occupied points in k - space, expressed by the set of all combinations of values of $k_x$, $k_y$, $k_z$ will be located inside a sphere of radius $k = k_{max}$. On the other hand, the difference between two consecutive values of each $k_i$ component (i = x, y, z) is $2\pi/L$. Therefore each allowed value of $\vec{k}$ ($k_x$, $k_y$, $k_z$) should occupy a volume in k - space given by

$$(\frac{2\pi}{L})^3 = \frac{(2\pi)^3}{V}, \tag{41}$$

where V is the volume of the crystal. Thus, the number of electron states with values lying between k and k + dk should be

$$2\frac{4\pi k^2 dk}{(2\pi)^3/V} = \frac{V k^2 dk}{\pi^2}, \tag{42}$$

where the factor 2 takes into account the spin. Since the $E = E(\vec{k})$ relation is given by Eq. (35), we have finally for the DOS function in energy the expression

$$\frac{dS(E)}{dE} = \frac{V}{2\pi^2}(\frac{2m^*}{\hbar^2})^{3/2}\sqrt{E}. \tag{43}$$

From the last relation, we see that $\frac{dS(E)}{dE}$ increases as the square root of energy (see, Fig. 23).

The behavior of electrons when their motion is restricted along one direction in the wells of infinite height corresponds to a well - known problem in quantum mechanics, the so - called particle in a box of infinite wells [27].

In two -dimensional nanostructures carrier motion for both electrons and holes is not allowed in the direction perpendicular to the well, usually taken as the z - direction because of the potential wells. However, in the other two spatial directions (x, y) parallel to the crystal interfaces, the motion is not restricted, i.e. the electrons behave as free electrons. It is well - known from quantum mechanics that, in the case of infinite potentials barriers, the wave functions and energy levels of the bound electrons are given by

$$\Psi_n(z) = (\frac{2}{a})^{1/2}\sin(\frac{\pi n z}{a}), \tag{44}$$

$$E_n = \frac{\hbar^2 \pi^2}{2m^* a^2} n^2, \qquad n = 1, 2, 3,.... \qquad (45)$$

Here $m^*$ is the effective mass of the electrons in the well material for the motion along z - direction and a is the width of the well. From last relation (45) we can derive several important consequences:

1. In general, quantum size effects will be more easily observable in quantum structures of very small size a, and for materials for which the electron effective mass is as small as possible. In this case, GaAs nanostructures are very convenient since $m^* \sim$ 0.067 $m_0$ [8, 14], where $m_0$ is the free electron mass. This is equivalent to saying that in materials for which the electron mobility or the free electron path are large (see above), quantum effects are easier to observe.

2. Quantum size effects, which require energy transitions of electrons between levels, are better observed at low temperatures, since the mean thermal energy of carriers is of the order of kT.

As was indicated above, the motion of electrons in the quantum well is confined only in one direction,z, but in the (x, y) planes the electrons behave as in a three - dimensional solid. Therefore the electron wave function is separable as the product of $\Psi_x$, $\Psi_y$ and $\Psi_z$ (see, also [10]) i.e.

$$\Psi = \Psi_x \cdot \Psi_y \cdot \Psi_z, \qquad (46)$$

where, $\Psi_x$ and $\Psi_y$ satisfy the Schrödinger equation for a free electron, i.e. travelling wave, while $\Psi_z$ is given by the Schrödinger equation with a square well potential V(z) and therefore can be expressed in Eq. (44). The expression for the total energy of electrons in the potential well, can then be written as

$$E(k_x, k_y, n) = \frac{\hbar^2}{2m^*}(k_x^2 + k_y^2) + E_n = \frac{\hbar^2}{2m^*}(k_x^2 + k_y^2) + \frac{\hbar^2 \pi^2}{2m^* a^2} n^2, \qquad n = 1, 2, 3, ........ \qquad (47)$$

where the quasi - continuous values of $k_x$, $k_y$ are determined by the periodic boundary conditions as in the case of free electron in the bulk. Using the same algebra, we get for the DOS function in two - dimensional (2D) case:

$$\frac{dS(E)}{dE} = \frac{2m^*}{\pi^2 \hbar^2}. \qquad (48)$$

Note that in the 2D case, the DOS function is a constant, independent of energy and exhibits a staircase shaped energy dependence (see, also Fig. 25 below) in which all the steps are of the same height, but located at energies $E_n$ given by (45). It can be appreciated that interval energy between 0 and $E_1$ is not allowed. For E such that $E_1 \langle E \langle E_2$ the electrons will be located in the subband corresponding to n = 1 and the value will be $\frac{m^*}{\pi^2 \hbar^2}$. For the energy interval between $E_2$ and $E_3$, the electrons can be located either in the n = 1 or in the n = 2 subbands, and consequently the DOS function would be twice the above value, i.e. $\frac{2m^*}{\pi^2 \hbar^2}$ ($\frac{4m^*}{\pi^2 \hbar^2}$; $\frac{4m^*}{\pi^2 \hbar^2}$ etc., see Fig. 23).

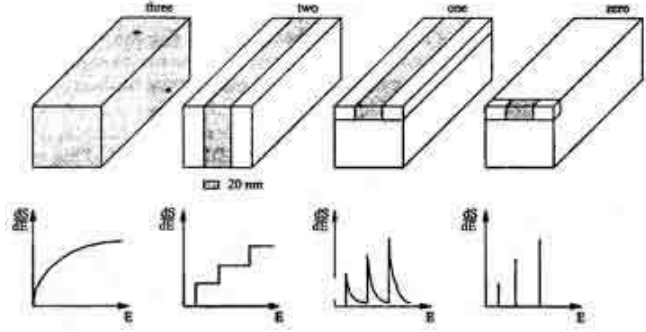

**Fig. 23. Normal structures with different dimensions: normal solid - state body, quantum well, quantum wire, quantum dot. Additionally their DOS are illustrated.**

Such picture directly observed by optical absorption measurements [102]. When an electron is allowed only one - dimensional motion (along. say, the x - direction), the energy is given by

$$E = \frac{\hbar^2 k_x^2}{2m^*}. \qquad (49)$$

A procedure analogous to that used above then yields for the DOS the expression

$$\frac{dS(E)}{dE} = \frac{1}{\pi}\left(\frac{2m^*}{\hbar^2}\right)\frac{1}{\sqrt{E}}. \qquad (50)$$

The Eq. (50) shows that the DOS function of a one - dimensional electron gas (1D) has a square - root singularity at the origin. This result will have important consequences in the physical properties of quantum wires.

Quantum dots (QDs) are often nanocrystals with all three dimensions in the nanometre range ($L_x$, $L_y$, $L_z$). In this case, there is no continuous DOS function, since there is quantization in three spatial directions. To consider the energy spectrum of a zero - dimensional system, we have to study the Schrödinger equation (33) with a confining potential that is a function of all three directions. The simplest case is the quantum box in the form of a parallelepiped with impenetrable wells. The corresponding potential, V (x, y, z) is

$$V(x, y, z) = \begin{cases} 0, \text{ inside of the box,} \\ +\infty, \text{ outside of the box} \end{cases}, \qquad (51)$$

where the box is restricted by the conditions $0 \leq x \leq L_x$, $0 \leq y \leq L_y$, $0 \leq z \leq L_z$. Using the results above analysis discussed previously (see, also [10]), one can write down the solutions of the Schrödinger equation for a box:

$$E_{n_1, n_2, n_3} = \frac{\hbar^2 \pi^2}{2m^*}\left(\frac{n_1^2}{L_x^2} + \frac{n_2^2}{L_y^2} + \frac{n_3^2}{L_z^2}\right), \qquad n_1, n_2, n_3 = 1, 2, 3, \ldots \qquad (52).$$

$$\Psi_{n_1, n_2, n_3}(x, y, z) = \sqrt{\frac{8}{L_x L_y L_z}} \sin\left(\frac{\pi x n_1}{L_x}\right)\sin\left(\frac{\pi y n_2}{L_y}\right)\sin\left(\frac{\pi z n_3}{L_z}\right). \qquad (53)$$

Of fundamental importance is the fact that $E_{n_1, n_2, n_3}$ is the total electron energy, in contrast to the previous cases, where the solution for the bound states in a quantum well and quantum wire gave us only the energy spectrum associated with the transverse confinement (see Eqs. (47) and (49)). Another unique feature is the presence of three discrete quantum numbers $n_1$, $n_2$, $n_3$ resulting straightforwardly from the existence of three directions of quantization. Thus, we obtain three - fold discrete energy levels and wave functions localized in all three directions of the quantum box. In a quantum dot of a parallelepipeds shape we have three quantum numbers $n_1$, $n_2$, $n_3$, that substitute for the three components of the wavevector $\vec{k}$: $k_x$, $k_y$, $k_z$. The discrete spectrum in a quantum box and the lack of free propagation of a particle in any direction are the main features distinguishing quantum dots from quantum wells and quantum wires. As is well - known, these features are typical for atomic systems as well [27].

Since in the case of quantum dots the electrons are totally confined, the energy spectrum is totally discrete and the DOS function is formed by a set peaks (see, Fig. 23) in theory with no width and with infinite height. In practice, the peaks should have a finite width, as a consequence, for instance, of the interaction of electrons with lattice phonons and impurities.

3.3. Electron in quantum dot.

If electron motion is quantized in all three possible directions, we obtain a new physical object, a macroatom. Questions concerning the usefulness of such objects for applications naturally arise from the point of view of their electronic applications. A fundamental question is the following: what is the current through a macroatom? A valid answer is that there exist the possibility of passing an electric current through an artificial atom due to tunneling of electrons through quantum levels of the macroatom (see, e.g. [103, 104]). The field of single electron tunneling (SET) comprises of phenomena where the tunneling of a microscopic charge, usually carried by an electron or a Cooper pair, leads to microscopically observable effects (see, also [105, 106]). The basic principles governing single charge tunneling through QD are briefly outlined in this paragraph.

For the description of our task, let us imagine a semiconductor of nanometric size in the three spatial dimensions, for example a QD (see Fig. 23). Below we will show that even change of one elementary charge (electron) in such small systems has a measurable effect in the electrical and transport properties of the dot [104]. This phenomenon is known as Coulomb blockade, which we will discuss in the simplest possible terms (see, also [100]). Let us imagine a semiconductor dot structure, connected to electron reservoirs (e.g. drain and source)of each side by potential barriers or tunnel junctions (see, Fig 24[a]).

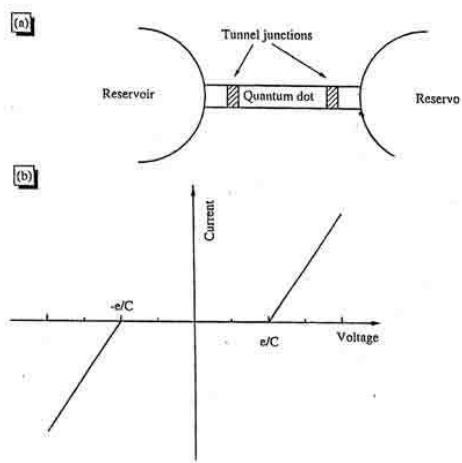

**Fig. 24. (a) Scheme of a quantum system to observe Coulumb blickade effects; (b) I - V characteristics in a quantum dot showing the Coulomb blockade effect.**

In order to allow the transport of electrons to or from reservoirs, the barriers will have to be sufficiently thin, so that the electrons can cross them by the tunnel effect. Further suppose that we wish to change the number N of electrons in the dot by adding just one electron, which will have to tunnel for instance from left reservoir into the dot. For this to happen, we will to provide the potential energy eV to the electron by means of a voltage source. If the charge in the QD is Q and its capacitance C, the potential energy is $Q^2/2C$. Therefore an energy of at least $e^2/2C$ will have to be provided to the electron, which means that for the electron to enter the dot, the voltage will have to be raised to at least $e^2/2C$. Since the electron can either enter the dot or leave (this process is equivalent to a hole entering the dot), we see that electrons cannot tunnel if

$|V| \langle e/2C.$ (54)

Therefore, there this is a voltage range, between -e/2C and e/2C, represented Fig. 24$^b$, in which current cannot go through the dot, hence the name of Coulomb blockade given to this phenomenon (see, also [105] and references therein). Evidently if the above process is continued and we keep adding more electrons, we will have the situation represented in Fig. 25, in which we will observe discontinuities in the current through the QD whenever the voltage acquires the values expressed by:

$V = \left(\frac{1}{2C}\right)(2n+1)e,$   $n = 0, 1, 2, 3, ....$  (55)

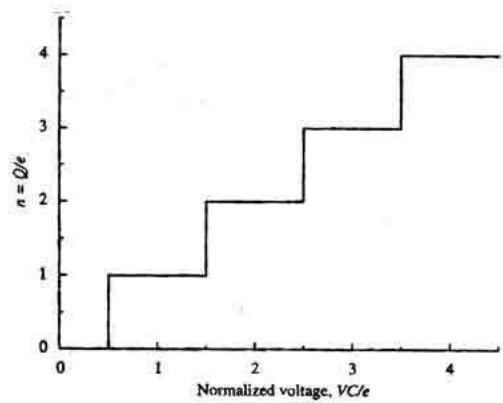

**Fig. 25. Charging of a quantum dot capacitor as a function of voltage, in normalized coordinates.**

Observe that in Fig. 25 we have made use of normalized coordinates, both in horizontal and vertical axes, to better appreciate the effect of the quantification in current and voltage.

It is also interesting to observe from the above equations that as the size of the QD is reduced, and therefore C gets smaller, the value of the energy necessary to change the number of electrons in the dot increases. In this case, it will be easier to observe the Coulomb blockade, since the changes in voltage and electric energy has to be much larger than the thermal energy kT at the working temperature, in order to observe measurable Coulomb blockade effects. Therefore, we should have for the capacitance:

$C \ll \frac{e^2}{kT}.$ (56)

For this condition to be fulfilled, either the capacitance of the dot should be very small (values less than $10^{-16}$F are very difficult to get) or we should work at very low temperatures, usually smaller than 1K.

Another condition to observe SET is that the number of electrons in the dot should not fluctuate in equilibrium. Let us assume that the time taken for an electron to be transferred in or out of dot is of the order of $R_T C$, where $R_T$ is the equivalent resistance of the tunnel barrier and C the capacitance of the dot. Fluctuations in the number of electrons in the dot induce changes in potential energy of the order of $e^2/C$. Therefore we should have, according to the uncertainty principle

$$\Delta E \cdot \Delta t = \frac{e^2}{C} R_T C \rangle h \qquad (57)$$

and consequently for Coulomb blockade effects to be clearly observed we should have

$$R_T \gg \frac{h}{e^2} = 25.8 \text{ k}\Omega. \qquad (58)'$$

In single electron transport experiments, usually the current is measured, which is proportional to the conductance. In terms of the conductance, the above condition can be written as (for details, see, also [100, 104])

$$G \ll \frac{e^2}{h}. \qquad (59)$$

In reality, electrical methods applied to QDs to realize useful devices are not the only method possible. The control of the electric current through the dots can also be realized by means of light, sound, etc. (see, e.g. [10, 11]). Consider here optical control of the dots and optoelectronic functions of zero - dimensional devices. The main peculiarities of the optical properties of QDs arise due to electron and hoe quantization (see, also [107, 108]). In QDs fabricated from semiconductors with different $E_g$, the carrier energies have the form

$$E_e^{QD} = E_g^{QD} + \epsilon_n (n_1, n_2, n_3), \quad E_h^{QD} = -\epsilon_p (n'_1, n'_2, n'_3). \qquad (60)$$

In this expression $E_g^{QD}$ is the fundamental bandgap of the material of the QDs. Very often $E_g^{QD}$ is less than the bandgap $E_g$ of the surrounding material into which the dots are embedded, $\epsilon_n$ and $\epsilon_p$ depend on sets of three discrete quantum numbers for electrons and hole, respectively. The model dependences of $E_g$ (51) can be used for estimation of the energy levels. Owing to these discrete energy spectra, QDs interact primarily with photons of discrete energies:

$$\hbar \omega = \frac{2\pi \hbar c}{\lambda} = E_g^{QD} + \epsilon_n (n_1, n_2, n_3) + \epsilon_p (n'_1, n'_2, n'_3). \qquad (61)$$

In last formula c is the velocity of light and $\lambda$ is wavelength of the light. The different combinations of quantum numbers $(n_1, n_2, n_3)$ and $(n'_1, n'_2, n'_3)$ give a series of optical spectral lines, for which interaction between the dots and light is efficient. Importantly, the fact that $E_g^{QD} \langle E_g$ implies that the light interacting with the dots is not absorbed by the surrounding material (for details see [14, 108]).

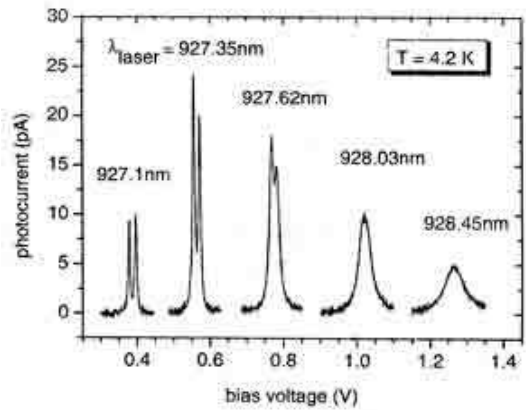

**Fig. 26. Photocurrent resonance for various excitation wavelengeths bias voltage. At low bias the fine structure splitting is fully resolved, at higher bias the linewidth is increased due to fast tunneling (after [109]).**

The optical control of the electric current flowing through a QD can be explained with help of a device that can be called a single - quantum - dot photodiode [107, 108]. In Fig. 26, we present the experimental results obtained [109] from excitation of the ground exciton state ($n_1 = n_2 = n_3 = n'_1 = n'_2 = n'_3$) of a single self - assembled $In_{0.5}Ga_{0.5}As$ QD embedded into a 360 - nm - thick intrinsic GaAs layer. Since $E_g$(InGaAs) $\langle$ $E_g$ (GaAs) the only optically part is the single $In_{0.5}Ga_{0.5}As$ QD. In the paper [109] the experiments were carried out at 4.2K. In Fig.26, the photocurrent is plotted as a function of the electric bias for various wavelengths of illuminating light. The wavelengths are indicated on the photocurrent curves. One can see that photocurrent - voltage dependences have, in fact, a pronounced resonant character. This explained [109] by the fact the quantized electron and hole energies are shorted under an applied electric field, as expected from the so - called Stark effect observable for atoms and molecules. When these energies are such that wavelength given by Eq. (61) corresponds to the illuminating light, the light excites electrons and holes inside the quantum well, which produces the measured photocurrent. As the applied bias increases, the energies a shifted to a smaller values and the resonance wavelength increases. In Fig. 26, spectra for excitation of the same ground exciton state of the dot for different biases are shown. The observed photocurrent spectra are very narrow because a single dot is involved [14]. Spectral broadening becomes visible at high biases when the electron and hole energy levels decay as a result of the increased rate of tunneling from the dot (for details see [109]).

In conclusion we should underline that a very interesting challenge of future nanoelectronics is the control of switching device by just one electron (see, also chapter 4).

3.4. Excitons in nanostructures.

When a semiconductor (insulator) of direct bandgap $E_g$ is shone with near - bandgap light, electron - hole pairs are created. If the electron and the hole were non - interacting only photon energies $\hbar\omega \rangle E_g$ would be absorbed and $E_g$ would be the absorption edge.

The coulombic electron - hole interaction greatly modifies this picture. The electron - hole attraction gives raise to bound states of the relative motion of the exciton [110]. The appearance of intense, narrow absorption lines below the fundamental absorption edge is the manifestation of these bound states.

In the case of confined systems for electrons and holes, such as quantum wells (QWs), quantum wires (QWRs) and quantum dots (QDs), the excitonic effects are much more important than in bulk solids. In effect, as will be shown below, the binding energy of the electron - hole systems forming an excitons is much higher in quantum confined systems than in the case of solids, and, therefore, the excitonic transitions can be observed even at temperatures close to room temperature, as closed to the bulk case for which low temperatures are needed. This makes the role played by excitons in many optoelectronic devices of nanoscale very important (see, also [8 - 11, 15, 74, 109, 111, 112]). It represents the Coulombic binding between the conduction electron and the isotropic part of the $\Gamma_8$ hole. To the zero$^{th}$ approximation in Q, L, M the exciton states formed between the $\Gamma_6$ electron and the $\Gamma_8$ hole [113] are fourfold degenerate and can be calculated like the exciton states in idealized bulk material (see, e.g. [110]).

3.4.1 Excitons in quantum wells.

In diamond - like semiconductors the topmost valence band is fourfold degenerate at the zone centre of the Brillouin zone ($\Gamma_8$ symmetry) (see e.g. [100]). In the spherical approximation the valence Hamiltonian can be written [112, 113]

$$H_v(\vec{k}) = \frac{\hbar^2 \vec{k}^2}{2m_h} 1 + \frac{\hbar^2}{2m_l}\left(\vec{k}\cdot\vec{J}\right)^2, \qquad (62)$$

where 1 is the 4 × 4 identity matrix and $\vec{J}$ a spin 3/2 matrix. The effective masses related to the heavy (h) and light (l) hole masses:

$$\frac{1}{m_h} = \frac{1}{8m_{hh}} - \frac{9}{8m_{lh}}; \qquad \frac{1}{m_l} = -\frac{1}{2m_{hh}} + \frac{1}{2M_{LH}}. \qquad (63)$$

To obtain Eq. (63) we have used the fact that for $\vec{k} \parallel \vec{J}$ heavy hole states correspond to $m_j = \pm 3/2$ and light hole states to $m_j = \pm 1/2$.

The relative motion of the exciton is thus described by Hamiltonian which is also a 4 × 4 matrix:

$$H_{exc} = [\frac{p^2}{2m_c^*} - \frac{e^2}{kr}]1 - H_v(\frac{p}{\hbar}). \qquad (64)$$

Baldereschi and Lipari [114] have shown that $H_{exc}$ can be rewritten

$$H_{exc} = P(\vec{r}, \vec{p})1 + \begin{bmatrix} Q & L & M & 0 \\ L^* & -Q & 0 & M \\ M^* & 0 & -Q & -L \\ 0 & M^* & -L^* & Q \end{bmatrix} \qquad (65)$$

where P, Q, L M are functions of $\vec{r}$ and $\vec{p}$ which have S - symmetry (P) and D - symmetry (Q, L, M) respectively. The important point is that Q, L, M only involve valence band parameters whereas P also involves $m_c^*$ and the Coulombic term -$e^2$/kr. In fact:

$$P(\vec{r}, \vec{p}) = \frac{p^2}{2\mu_0} - \frac{e^2}{kr}, \qquad (66)$$

where:

$$\frac{1}{\mu_0} = \frac{1}{m_c^*} + \frac{\gamma_1}{m_0}. \qquad (67)$$

Here $\gamma_1 is$ one of the Luttinger parameters [112] describing the $\Gamma_8$ hole kinematics.

The diagonal term P($\vec{r}$, $\vec{p}$) is thus much larger than the other ones. It represents the Coulombic binding between the conduction electron and the isotropic part of the $\Gamma_8$ hole (see also [113]). The zero$^{th}$ approximation in Q, L, M the exciton states formed between the $\Gamma_6$ electron and the $\Gamma_8$ hole are fourfold degenerate and can be calculated like the exciton states in idealized bulk materials (see, e.g. [110, 113]) except that the reduced mass of the exciton involves neither the heavy hole nor the light hole masses but an average of the two. As shown by Baldereschi and Lipari [114], the corrections to the zero$^{th}$ approximation are very small under most circumstances (see, also [115]).

To calculate exciton states in quantum well hetero-structures. one should add the barrier potentials for the electrons and the holes to Eq. (65) which have next expression

$$V_{barr}(z_e, z_h) = V_e Y(z_e^2 - \tfrac{L^2}{4}) + V_h Y(z_h^2 - \tfrac{L^2}{4}), \qquad (68)$$

where L is the quantum well thickness and $V_e$, $V_h$ the barrier heights for the electrons and holes. Once again we have neglected some band structure effects, e.g. the effective mass and dielectric mismatches between the host materials (see also [111]).For the L values where the size quantization is important, we expect the $z_e$, $z_h$ motion to be forced by the quantum well effects. We thus need to keep the exact hole masses in order evaluate the hole confinement energies correctly. This precludes the use of Baldareschi and Lipari's type of approach which is based on a suitable averaging of heavy and light hole masses. Up to present time there does not exist any fully satisfactory treatment of the exciton binding in quantum nanostructures (see, also [122]). Miller et al. [116] and Greene and Bajaj [117, 118] have approximated the exciton Hamiltonian (see, also [111]) by

$$H_{ex} = (P + V_{barr})\mathbf{1} + \begin{bmatrix} Q & & & 0 \\ & -Q & & \\ & & Q & \\ 0 & & & Q \end{bmatrix} \qquad (69)$$

Miller at al. took [116] $V_e$, $V_h$ to be infinite, whereas Greene and Bajjaj [111] accounted for the finite barrier effects both groups were specifically interested in GaAs-Ga$_{1-x}$Al$_x$As quantum wells. The advantage of including Q in H$_{ex}$ is that a correct evaluation of the size quantization of the holes can be obtained. The drawback of this approximation is the inclusion of terms (brought by Q) for the in plane exciton motion which are, in principle, as small as the last over terms L, M (although the latter are off diagonal). If we remember that the L, M terms actually give rise to strong anticrossings between the hole subbbands [119, 120], there is some possibility that they can significantly contribute to the excitonic binding itself (see, also [111]).

As the exciton Hamiltonian Eq. (69) is a diagonal matrix, the excitons fall into two categories; the heavy hole (P + Q) and the light hole (P - Q) excitons. The heavy hole exciton Hamiltonian corresponds to m$_j$ = ± 3/2 and is written as

$$H_{ex}^{hh} = \frac{p_{z_e}^2}{2m_c^*} + \frac{p_{z_h}^2}{2m_{hh}^*} - \frac{e^2}{k|\vec{r_e} - \vec{r_h}|} + V_e Y(z_e^2 - \tfrac{L^2}{4}) + V_h Y(z_h^2 - \tfrac{L^2}{4}) + \frac{p_\perp^2}{2\mu_{hh}}, \qquad (70)$$

where m$_{hh}$ and $\mu_{hh}$ are defined by

$$\tfrac{1}{m_{hh}} = \tfrac{1}{m_0}(\gamma_1 - 2\gamma_2) \quad \text{and} \quad \tfrac{1}{\mu_{hh}} = \tfrac{1}{m_c^*} + \tfrac{1}{m_0}(\gamma_1 + 2\gamma_2). \qquad (71)$$

The light hole exciton Hamiltonian corresponds to m$_j$ = ± 1/2 and is written as:

$$H_{ex}^{hh} = \frac{p_{z_e}^2}{2m_c^*} + \frac{p_{z_h}^2}{2m_{lh}^*} - \frac{e^2}{k|\vec{r_e} - \vec{r_h}|} + V_e Y(z_e^2 - \tfrac{L^2}{4}) + V_h Y(z_h^2 - \tfrac{L^2}{4}) + \frac{p_\perp^2}{2\mu_{lh}}, \qquad (72)$$

where m$_{hh}$ and $\mu_{lh}$ are given by the expression:

$$\frac{1}{m_{lh}} = \frac{1}{m_0}(\gamma_1 + 2\gamma_2) \text{ and } \frac{1}{\mu_{lh}} = \frac{1}{m_c^*} + \frac{1}{m_0}(\gamma_1 - 2\gamma_2), \qquad (73)$$

where $\gamma_1$ and $\gamma_2$ are well - known Luttinger parameters [112].

Thus, the heavy hole and light hole excitons again resemble those obtained in idealized quantum well structures. However $\mu_{lh}$ is not necessary smaller than $\mu_{hh}$. In fact in GaAs the opposite is true: $\mu_{lh} = 0.051\ m_0$ and $\mu_{hh} = 0.04\ m_0$ whereas $m_{lh} = 0.08 m_0$ and $m_{hh}\ 0.45 m_0$ (see, e.g. [10]). The inclusion of the Q term in $H_{ex}$ inverts the parts played by the heavy and light masses along and perpendicular to the z axis. The heavy hole exciton is indeed heavy along z but light in the layer plane and vice versa. Thus in GaAs - $Ga_{1-x}Al_xAs$ quantum well, the curves which represent the binding energies versus the GaAs slab thickness of the two kinds of excitons should cross. For large wells the light and heavy hole confinement (governed by $m_{lh}$ and $m_{hh}$, respectively) are almost complete and the light hole exciton is more tightly bound because its effective bulk Rydberg is larger than that of the heavy holes. On the other hand, for narrow GaAs wells, the light holes are less confined than the heavy holes. The Coulombic interaction between the electron and the hole in the light hole exciton is thus weaker than the one in the heavy hole exciton. Consequently the light hole exciton is less bound than the heavy hole exciton. Fig. 27 shows Bajaj [111] results concerning the ground bound exciton states in GaAs - $Ga_{1-x}Al_xAs$ quantum wells for two aluminium mole fractions x = 0.15 and x = 0.3.

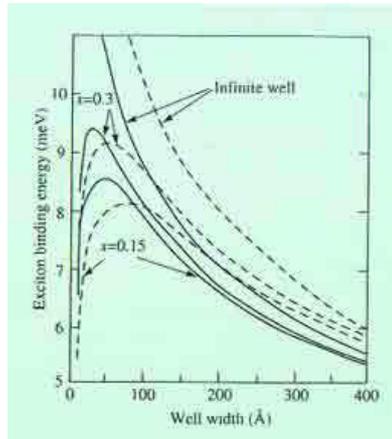

**Fig. 27. Variation of the binding energy of the ground state $E_{1s}$ of the heavy - hole exciton (solid lines) and the light - hole (dashed lines) as a function of the GaAs quantum well size (L) for aluminium concentration x = 0.15 and x = 0.3 and for infinite potential wells (after [111]).**

In these curves, the Dingle's rule [121] which states that the conduction band shares (85%) of the total bandgap difference between GaAs and $Ga_{1-x}Al_xAs$ has been used. Otherwise, the overall shapes of these curves look familiar: the exciton binding energies admit a maximum value versus the GaAs well thickness, whose location and amplitude depend on $V_e$, $V_h$ and $\mu_{hh}$ and $\mu_{lh}$.

To summarize, the Coulombic bound states in hetero-structures. are qualitatively well understood [10] The effect of off - diagonal terms in the exciton Hamiltonian, however, an issue for quantitative understanding.. We should remember that the 2D can only be approached hypothetically in infinitely quantum well. Fig. 28 illustrates the results of calculations of the exciton biding energy as a function of the width of an infinitely deep CdTe quantum well [10]. As is well - known the magnitude of the bulk exciton binding energy for CdTe is 10.1 meV (see, e.g. [113]).

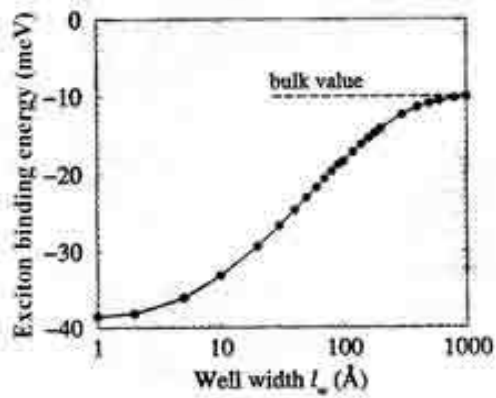

**Fig. 28. Exciton binding energy in an infinitely deep CdTe quantum well (after [10]).**

The negative values on the graph of Fig. 28 illustrate that it is indeed a bound state. From Fig. 28 it can be seen that two - dimensional exciton binding energy is equal four meaning of bulk exciton binding energy. Fig. 29 displays the corresponding Bohr radii $r_{ex}$ for the energies of Fig. 28. Remembering that the the Bohr radius in bulk, $r_{ex}(3D) = 67$ Å, then the 2D limit, i.e.

$$\lim_{l_w \to 0} r_{ex}(2D) = \frac{r_{ex}(3D)}{2} \qquad (74)$$

is satisfied. The 3D limit is obeyed, although the data on the graph show a slight scatter around the bulk radii of 67 Å. According [10], the source of this discrepancy is numerical accuracy. At the lager well widths, the wave function needs to be known at many points in order to calculate the binding energy to very high tolerance (according [10] thus leading to long computational times).

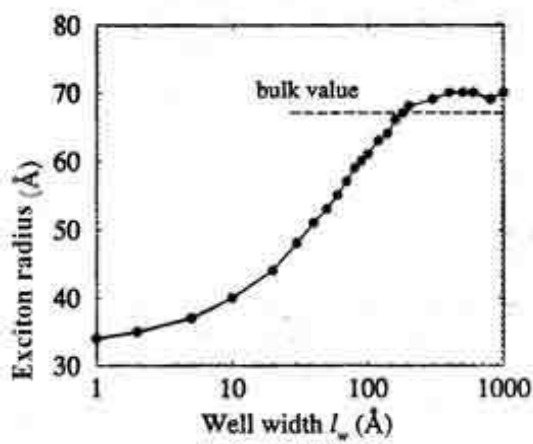

**Fig. 29. Bohr radius of two - dimensional exciton in an infinitely deep CdTe**

**quantum well (after [10]).**

To concluding this paragraph we should note, that the Coulombic interaction produces an excitonic series at every subband [74]. But normally only 1S exciton is visible for each transition. In very good QWs it is, however, sometimes possible to distinguish the 2S exciton [123]. The relative positions offer a very good guide to the actual exciton binding energy. In practice the most important effect of the Coulomb interaction on QW band structure is the appearance of strong, sharp spectral resonances just below the band edge for the first electron - heavy - hole and electron - light - hole transition (for details see [123]).

### 3.4.2. Excitons in quantum wires.

One - dimensional (1D) semiconductor structures have received interest in recent years, and promising advances have been obtained in quantum - wire fabrication and in application, e.g. photodetectors, photodiodes and laser devices [123 - 126]. In analogy to the in - plane dispersion discussed above, in a quantum wire it is possible to decouple the motion along the length of the wire. Taking the axis of the wire along x (see, also Fig. 23) then the total potential V(x, y, z) can always be written as the sum of a two - dimensional confinement potential plus the potential along the wire, i.e.

V(x, y, z) = V(x) + V(y, z).  (75)

The eigenfunction can then be written as a product of two components:

Ψ(x, y, z) = Ψ(x)Ψ(y, z).  (76)

Substituting both (equation (75) and equation (76)) into general three - dimensional Schrödinger equation for constant effective mass, then

$$-\frac{\hbar^2}{2m^*}(\frac{\partial^2}{\partial x^2} + \frac{\partial^2}{\partial y^2} + \frac{\partial^2}{\partial z^2})\Psi(x)\Psi(y, z) + [V(x) + V(y, z)]\Psi(x)\Psi(y, z).$$  (77)

Writing the energy as a sum of terms associated with two components of the motion, we get

$$-\frac{\hbar^2}{2m^*}(\Psi(y, z)\frac{\partial^2\Psi(x)}{\partial x^2} + \Psi(x)\frac{\partial^2\Psi(y, z)}{\partial y^2} + \Psi(x0\frac{\partial^2\Psi(y, z)}{\partial z^2})\Psi(y, z)V(x)\Psi(x) + \Psi(x)V(y, z)\Psi(y, z) =$$
$$= (E_x + E_{y,z})\Psi(x)\Psi y,z).$$  (78)

It is now we can write

$$-\frac{\hbar^2}{2m^*}(\Psi(y, z)\frac{\partial^2\Psi(x)}{\partial x^2} + \Psi(y, z)V(x) = \Psi(y, z)E_x\Psi(x)$$  (79)

$$-\frac{\hbar^2}{2m^*}[\Psi(x)\frac{\partial^2\Psi(y, z)}{\partial y^2} + \Psi(x)\frac{\partial^2\Psi(y, z)}{\partial z^2}] + \Psi(x)V(y, z)\Psi(y, z) = \Psi(x)E_{y,z}\Psi(y, z).$$  (80)

In the above Ψ(y, z) is not acted upon by any operator in the first equation, and similarly for Ψ(x) in the second equation, and thus they can be divided out. In addition, the potential component along axis of the wire V(x) = 0, thus giving the final decoupled equations of motion as follows:

$$-\frac{\hbar^2}{2m^*}\frac{\partial^2\Psi(x)}{\partial x^2} = E_x\Psi(x)$$

(81)

and

$$-\frac{\hbar^2}{2m^*}[\frac{\partial^2\Psi(y, z)}{\partial y^2} + \frac{\partial^2\Psi(y, z)}{\partial z^2}] + V(y, z)\Psi(y, z) = E_{y,z}\Psi(y, z).$$  (82)

The equation (81) is satisfied by a plane wave of the form $\exp(ik_x x)$, thus giving the standard dispersion relation:

$E_x = \frac{\hbar^2 k^2}{2m^*}$. (83)

The second of the last equations of motion, equation (82), is merely the Schrödinger equation for the two-dimensional confinement potential characterizing a quantum wire (QWr).

Further we consider the cylindrical quantum wire and we'll be used the polar coordinate. With definition of the modulus r and angle $\theta$ as in ordinary case, the Cartesian coordinates then follow as:

y = rsin$\theta$ and z = rcos$\theta$ (84).

and r = $\sqrt{y^2 + z^2}$ . (85).

The wave function $\Psi(y, z)$ can clearly be written in terms of the new variables r and $\theta$; however the circular symmetry the wave functions should not have a dependence on the angle $\theta$. Thus, the wave function can actually be written as $\Psi(r)$, and the Schrödinger equation therefore becomes:

$-\frac{\hbar^2}{2m^*}(\frac{\partial^2}{\partial y^2} + \frac{\partial^2}{\partial z^2})\Psi(r) + V(r)\Psi(r) = E_r\Psi(r),$ (86)

where the index on $E_r$ just indicates that this eigenvalue is associated with the confined cross-sectional motion, as opposed to the unconfined motion along the axis of the wire. In addition, the circular symmetry of the potential which defines the wire can be written as V(r). Now:

$\frac{\partial}{\partial y}\Psi(r) = \frac{\partial}{\partial r\Psi(r)} \cdot \frac{\partial r}{\partial y}.$ (87)

Differentiating both sides of equation (85) with respect y, gives:

$\frac{\partial r}{\partial y} = \frac{1}{2}(y^2 + z^2)^{-\frac{1}{2}} \cdot 2y = \frac{y}{r}.$ (88)

Hence:

$\frac{\partial}{\partial y}\Psi(r) = \frac{\partial}{\partial r}\Psi(r) \cdot \frac{y}{r}.$ (89)

The second derivative is then:

$\frac{\partial}{\partial y}\frac{\partial}{\partial y}\Psi(r) = \frac{\partial}{\partial y}[\frac{\partial}{\partial r}\Psi(r) \cdot \frac{y}{r}].$ (90)

and thus :

$\frac{\partial^2}{\partial y^2}\Psi(r) = \frac{y^2}{r^2}\frac{\partial^2}{\partial r^2}\Psi(r) + \frac{\partial}{\partial r}(\frac{1}{r} - \frac{y}{r^2}\frac{\partial r}{\partial y})$ (91)

Finally:

$\frac{\partial^2}{\partial y^2}\Psi(r) = \frac{1}{r}\frac{\partial}{\partial r}\Psi(r) - \frac{y^2}{r^3}\frac{\partial}{\partial r}\Psi(r) + \frac{y^2}{r^2}\frac{\partial^2}{\partial r^2}\Psi(r)$ (92)

and similarly for z, hence:

$(\frac{\partial^2}{\partial y^2} + \frac{\partial^2}{\partial z^2})\Psi(r) = \frac{2}{r}\frac{\partial}{\partial r}\Psi(r) - \frac{(y^2+z^2)}{r^3}\frac{\partial}{\partial r}\Psi(r) + \frac{(y^2+z^2)}{r^2}\frac{\partial^2}{\partial r^2}\Psi(r).$ (93)

Recalling that $y^2 + z^2 = r^2$, then

$(\frac{\partial^2}{\partial y^2} + \frac{\partial^2}{\partial z^2})\Psi(r) = \frac{1}{r}\frac{\partial}{\partial r}\Psi(r) + \frac{\partial^2}{\partial r^2}\Psi(r).$ (94)

Substituting into equation (86) gives the final form for the Schrödinger equation as follows:

$-\frac{\hbar^2}{2m^*}[\frac{1}{r}\frac{\partial}{\partial r} + \frac{\partial^2}{\partial r^2}]\Psi(r) + V(r)\Psi(r) = E_r\Psi(r).$ (95)

In this case, reliance has been made on the specific form of the kinetic energy operator, and hence this Schrödinger equation is only valid for a constant effective mass.

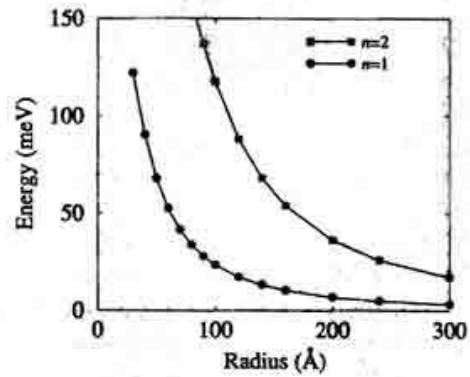

**Fig. 30. The confinement energy in a finite barrier circular cross - section quantum wire (after [10]).**

The numerical solution of the Eq. (95) is shown in Fig. 30 [10] This figure displays the result of calculations of the electron confinement energy versus the wire radius, for GaAs wire surrounded by $Ga_{0.8}Al_{0.2}As$, for constant effective mass. As expected, the confinement energy decreases with increasing radius and the odd - parity eigenstate is of higher energy than for even.

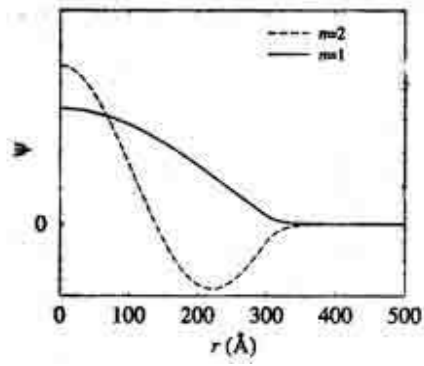

**Fig. 31. The radial component of the wave function $\Psi(r)$ for the lowest two eigenstates in a finite - barrier quantum wire with radius 300 Å of circular cross - section (after [10]).**

This latter point is highlighted in Fig. 31, which plots the radial motion $\Psi(r)$ for the 300Å radius wire. The even - (n = 1) and odd - (n= 2) parity nature of the eigenstates can clearly be seen (for details see [126] and references therein).

3.4.3. Excitons in quantum dots.

It is perhaps easier to deal with a finite barrier quantum dot (QD) with spherical rather than cuboid symmetry. The approach is rather similar to that derived earlier for the circular cross - section quantum wire (QWr). Given the spherical symmetry of the potential, then the wave function would also be expected to have spherical symmetry, hence the Schrödinger equation for a constant effective mass could be written (see, e.g. [14, 74])

$$-\frac{\hbar^2}{2m^*}\left(\frac{\partial^2}{\partial x^2} + \frac{\partial^2}{\partial y^2} + \frac{\partial^2}{\partial z^2}\right)\Psi(r) + V(r)\Psi(r) = E_r\Psi(r), \qquad (96)$$

where the index on $E_r$ has been added just to indicate that this energy is associated with the confinement along the radius. In this case:

$$r = \sqrt{x^2 + y^2 + z^2}. \qquad (97)$$

The transition can be made from Cartesian (x, y, z) to spherical polar coordinates, in effect just r, in the same way above. Using equation (93), each of the three Cartesian axes gives an equation of the following form:

$$\frac{\partial^2}{\partial x^2}\Psi(r) = \frac{1}{r}\frac{\partial}{\partial r}\Psi(r) - \frac{x^2}{r^3}\frac{\partial}{\partial r}\Psi(r) + \frac{x^2}{r^2}\frac{\partial^2}{\partial r^2}\Psi(r). \qquad (98)$$

Therefore, the complete $\nabla^2\Psi(r)$ is given by:

$$\left(\frac{\partial^2}{\partial x^2} + \frac{\partial^2}{\partial y^2} + \frac{\partial^2}{\partial z^2}\right)\Psi(r) = \frac{3}{r}\frac{\partial}{\partial r}\Psi(r) - \frac{(x^2+y^2+z^2)}{r^3}\frac{\partial}{\partial r}\Psi(r) + \frac{(x^2+y^2+z^2)}{r^2}\frac{\partial^2}{\partial r^2}\Psi(r). \qquad (99)$$

and

$$\left(\frac{\partial^2}{\partial x^2} + \frac{\partial^2}{\partial y^2} + \frac{\partial^2}{\partial z^2}\right)\Psi(r) = \frac{2}{r}\frac{\partial}{\partial r}\Psi(r) + \frac{\partial^2}{\partial r^2}\Psi(r). \qquad (100)$$

Substituting into the Schrödinger equation then:

$$-\frac{\hbar^2}{2m^*}\left(\frac{2}{r}\frac{\partial}{\partial r} + \frac{\partial^2}{\partial r^2}\right)\Psi(r) + V(r)\Psi(r) = E_r\Psi(r). \qquad (101)$$

Such spherical symmetric Schrödinger equations have been investigated before (see, e.g. [27]). The last equation as in the previous case, is numerically solved and Fig. 32 shows the results of calculations of the three lowest energy levels of a spherical GaAs QD surrounded by a finite barrier composed of $Ga_{0.8}Al_{0.2}As$, with a sharp boundary. In fact, the formalism above, as that of the circular cross - section QWr, is applicable for any radial potential profile V(r), e.g. it is also valid for diffused interfaces [10].

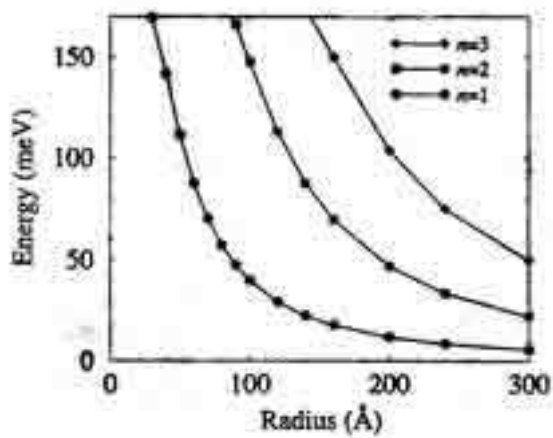

**Fig. 32. The confinement energy in a spherical GaAs quantum dot surrounded by a $Ga_{0.8}Al_{0.2}As$ barrier (after [10]).**

Again, the behavior of the energies as a function of the spatial dimension, as shown in Fig. 32, is as expected in confined systems, namely the confinement energy decreases as the size of the system increases. Fig. 33 displays the corresponding radial components of the wave functions. It can be seen that they all have a maximum at the centre of the potential and that as the principal quantum number n increases, then the number of nodes increases.

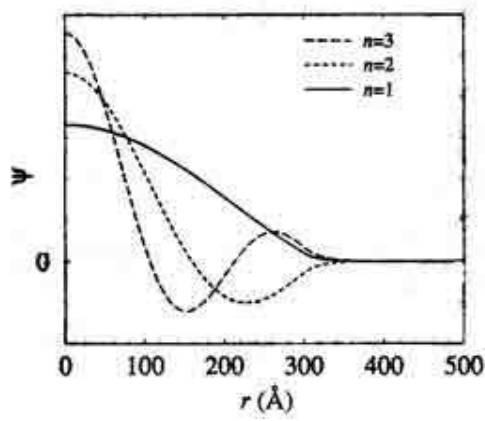

**Fig. 33. The wave functions of the three lowest energy states in the 300 Å spherical quantum dot (after [10]).**

3.5. Biexcitons in quantum dots.

In 1958, Moskalenko [127] and Lampert [128] suggested that in crystals besides excitons more complex electronic quasi - particles might exist, made up of three or four carriers. The latter one, consisting of two electrons and two holes is well known as biexcitons or excitonic molecules [129]. As the density of excitons is increased, biexcitons are formed by increasing the light intensity. Biexcitons can be generated either through ordinary excitation of the crystal or by two - photon absorption each photon having an energy

$$h\nu = E_x - \frac{E_{B_{xx}}}{2}, \qquad (102)$$

where $E_{B_{xx}}$ is the biexciton binding energy and $E_x$ is the exciton energy

$$E_x = E_g - E_{B_x} + \frac{\hbar^2 k^2}{2m_x}. \qquad (103)$$

In the last relation $E_g$ is the band - gap energy, $E_{B_x}$ is the exciton binding energy and $\frac{\hbar^2 k^2}{2m_x}$ is the kinetic energy with which an exciton moves through the crystal (see, also [130]).

Compared to the bulk material, an increased stability of biexcitons due to the two - dimensional carrier confinement is observed for typical III -V structures like GaAs/AlGaAs QWs [116, 131] (see Fig. 34) or for wide bandgap II - VI materials like CdZnSe/ZnSe [132]. AS a consequence of the enhanced biexciton binding energy, a variety of optical properties, like e.g. the photoluminescence(PL) spectrum, the optical gain or the four - wave mixing signal especially in wide bandgap II - VI QWs are strongly influenced by biexcitons (see [132] and references therein.

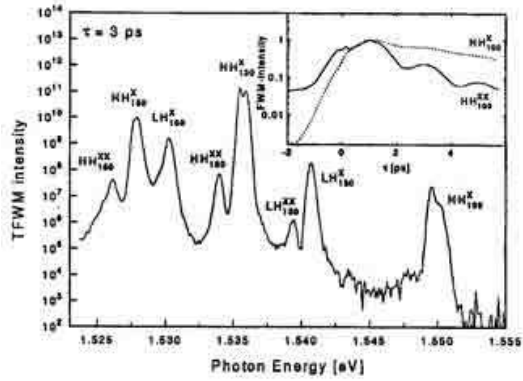

**Fig. 34. Spectrally resolved four - wave mixing at $\tau = 3$ ps showing the heavy hole and light hole biexcitons. Insert shows the four - wave mixing intensity of the heavy hole exciton and biexciton as a function of delay (after [131]).**

Below we briefly review some results obtained from optical spectroscopy on epitaxially grown single SQDs based on II - VI and II - N compounds. As was indicated above the biexciton (XX or $X_2$ is a four - particle state. In its lowest energy state configuration, two electrons and two holes with antiparallel spins occupy the first quantized state of the conduction and the valence band in the SQDs, respectively (see, e.g. [13]) We should add that the QDs in the material systems described here are quite small with diameters in the order of 10 nm and heights of a few nm. The biexciton state is therefore a singlet state with a total spin of $J = 0$. Thus, the exciton state X represents the final state for the biexciton recombination (see, also [133]). In II - VI semiconductors, as in III -V materials with a zincblende crystal lattice, Coulomb interaction leads to positive biexciton binding energies (see Eq. (103)), i.e. the energetic distance between XX ($X_2$) and X smaller than the energy difference between the first exciton state and the ground state. A typical optical fingerprint for the $X_2$ is therefore an additional PL line at the low energy side of the exciton emission X that exhibits a strong (quadratic) dependence on the excitation power [130]. This behavior is clearly visible in left panel of Fig. 35. At low excitation density, the PL spectrum of CdSe/ZnSe SQDs consists of emission peaks stemming from exciton recombination of two individual QDs. With rising excitation density additional lines emerge, red shifted by about 24 meV with respect to the excitonic emission X, and rapidly increasing in intensity, which can be attributed to biexciton emission $X_2$. The biexciton binding energy is obviously much larger than in III -As based QDs where a typical values of a few meV ($\sim 2$ meV [131]) have been determined (see, also [130, 134]). When having a closer look on the PL spectra presented in Fig. 35, some more information can be extracted. One should have in mind that in QDs, the light hole level is shifted to higher energies due to strain and confinement and thus, excitons are formed between electrons and heavy holes. The ground state of a heavy hole exciton in a SQD is a spin quadruplet, which can be by the z - component (= component, according [132] in growth direction) of the total exciton spin $J_z$. If the z - component of the electron spin, $s_z = \pm 1/2$, and the z - component of the total angular momentum of the heavy hole $j_z = \pm 3/2$, are antiparallel, in such case, we get $J_z = s_z + j_z = \pm 2$ (the dark exciton states [133]).

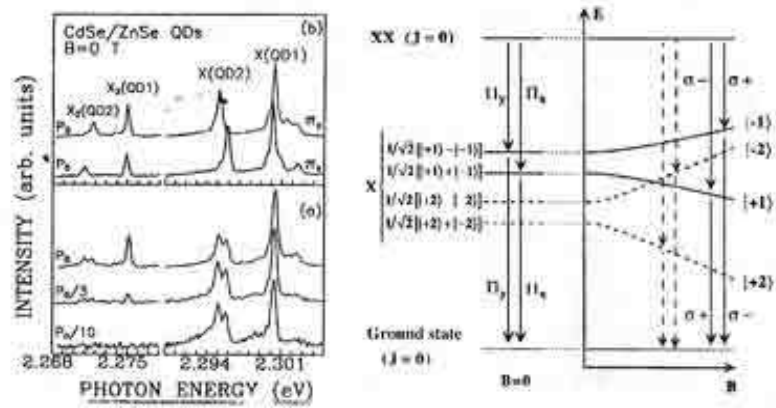

**Fig. 35.** Left side: Excitonic (X) and biexcitonic ($X_2$) emission from two individual CdSe/ZnSe SQDs for different excitation powers. The PL spectra shown in the lower panel are unpolarized, the data presented in the upper panel represent linearly polarized PL spectra ($\pi_x$ and $\pi_y$, respectively). Right side: Energy lvel schem for the biexciton - exciton cascade in a QD (after [132]).

In II - VI QDs the energy difference $\Delta_0$ between bright and dark exciton states that is given by the isotropic electron - hole interaction energy, amounts to about 1 meV and more which is nearly an order of magnitude larger than in InAs/GaAs QDs [14]. As can be seen in Fig. 35, the exciton fine structure is reflected both in the exciton and in the biexciton recombination: SQD1 does not show a significant splitting of the exciton PL signal, while SQD2 exhibits a doublet with an energy separation of almost 1 meV indicating a reduced QD symmetry. Exactly the same behavior is observed in the corresponding biexciton lines. Moreover, the high energy component of the X emission in SQD2 ($\pi_x$ polarized) corresponds to the low energy component of the $X_2$ emission and vice versa, in agreement with energy level scheme (see Fig. 35). All these effects are easily accessible in wide bandgap II - VI QDs because the characteristic energy splitting are significantly enhanced with respect to III - As semiconductor QDs. We may expect more significant value of the exchange splitting for exciton and biexciton states in QD of isotope - mixed crystals (see, also [39, 95]). Thanks to the large biexciton binding energy, II - VI QDs were the first, where the biexciton - exciton cascade could be traced directly in the time domain on SQD level [135]. Fig. 36. depicts transient PL spectra (left) of both emission lines and the time - dependent intensity of the exciton and the biexciton signal (right panel). The biexciton emission shows a monoexponential decay with a time constant of 310 ps.

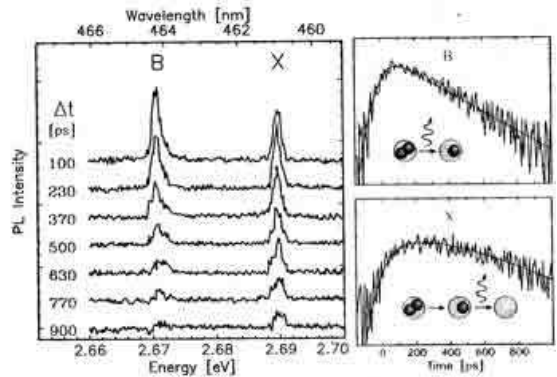

**Fig. 36. Left panel: Transient PL spectra from a single CdSe/ZnSe QD showing the single exciton X and the biexciton transition (here denoted by B = $X_2$). Right panel: Decay curves for the exciton and the biexciton PL signal (for details see text) (after [132]).**

The exciton reveals a more complex behavior: the onset of the exciton line ids delayed, resulting in "plateau - like" characteristics of the exciton decay curve. The excitation density according authors of this experiment was set to a value where an average number of two electron - hole pairs per excitation pulse in the SQD was generated. Model calculations taking into account the biexciton state, the bright and the dark exciton states and the "empty"QD (corresponding to a QD population with 2, 1 and 0 excitons, respectively), confirm that the exciton state is fed by the biexciton recombination causing the delayed onset and the "plateau - like" characteristics of the exciton emission dynamics (for details see [14] and references therein).

**4. Applications of low - dimensional structures.**

The knowledge gained in previous discussion makes it possible to consider and analyze a variety of different nanostructure devices. In this chapter for the first step we consider electronic and optical devices. Some of these mimic well - known microelectronic devices but with small dimensional scales. This approach applications to devices with shorter response times and higher operational frequencies that operate at lower working currents, dissipate less power, and exhibit other useful properties and enhanced characteristics. Such example include, in the first step, the field effect transistors will be consider below. On the other hand, new generations of the devices are based on new physical principles, which can not realized in microscale devices. Among these novel devices are the resonant - tunneling devices described in next section, and single - electron - transistor as well as optoelectronic devices (light - emitting diodes and lasers).

4.1. Resonant tunneling diodes.

As was shown above that electrons in heterojunctions and in QWs can respond with very high mobility to applied electric fields parallel to the interfaces (see, also [74]). In this paragraph, the response to an electrical field perpendicular to the potential barriers at the interfaces will be considered. Under certain circumstances, electrons can tunnel through these potential barriers, constituting the so - called perpendicular transport (see, also [136]).

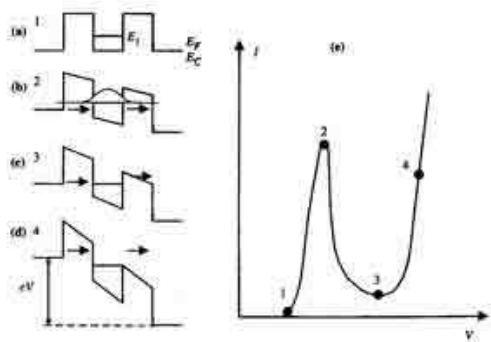

**Fig. 37. Schematic representation of the conduction band of a resonant tunnel diode:(a) with no valtage, (b - d) for increasing applied voltage, (c) - current - voltage characteristic.**

Tunneling currents through hetero-structures. can show zones of negative differential resistance (NDR) (see, Fig. 37) , which arise when the current level decreases for increasing voltage (see, also [ [11]). The operation of NDR QW electronic devices is based on the so - called resonant tunnel effect (RTE), which takes place when the current travels through a structure formed by two thin barriers with a QW between them. The I - V characteristics of RTE devices depicted on the F. 37. This figure also shows the representation of the conduction band of a double heterojunction with a QW between the junctions. The thickness of the QW is supposed to be small enough (5 - 10 nm) as to have only one allowed energy level $E_1$ (resonant level). The well region is made from lightly doped GaAs surrounded by higher gap AlGaAs (see, e.g. [8, 103]). The outer layers are made from heavily doped n - type GaAs ($n^+$ GaAs) to facilitate the electrical contacts. The Fermi level of the $n^+$ GaAs is represented within the conduction band, since it can be considered a degenerated semiconductor [113].

Let us suppose that an external voltage, V, is applied, starting from 0V. It can be expected that some electrons tunnel from the $n^+$ GaAs conduction band through the potential barrier, thus resulting in increasing current for increasing voltage (region 1 - 2 in the I - V curve of the Fig. 37$^c$). When the voltage increases, the electron energy in $n^+$ GaAs increases until the value $2E_1/e$ is reached, for which the energy of the electrons located in the neighborhood of the Fermi level coincides with that of level $E_1$ of the electrons in the well (see, Fig. 37$^b$). In this case, resonance occurs and the coefficient of quantum transmission through the barriers rises very sharply. In effect, when the resonant condition is reached, the electron wave corresponding to the electrons in the well is coherently (see, e.g. Fig. 10.18 in [74]) reflected between two barriers . In this case, the electron wave incident from the left excites the resonant level of the electron in the well, thus increasing the transmission coefficient (and thus the current through the potential barrier (region 2 in Fig. 37$^c$). If the voltage further increased (region 2 - 3), the resonant energy level of the well is located below the cathode lead Fermi level and the current decreases, thus leading to the so - called negative differential resistance (NDR) region (region 2- 3 of the Fig. 37). Finally, for even higher applied voltage , Fig. 37$^d$, the current again rises due to the thermo ionic emission over the barrier (region 4). RTD used in microwave applications are based on this effect. A figure of merit used for RTD is the peak - to - valley current ratio of their I - V characteristic, given by the ratio between the maximum current (point 2) and the minimum current in the valley (point 3). Although the normal values of the figure of merit are about five for AlGaAs - GaAs structures at room temperature, values up to 10 can be reached in devices fabricated from strained InAs layers, surrounded by AlAs barriers and operating at liquid nitrogen temperature[11]. If RTD are simulated by a negative resistance in parallel with a diode capacitance C and a series resistance $R_S$, as is the case of normal diodes, it is relatively easy to demonstrate that the maximum operation frequency increases as C decreases. The resonant tunnel diode is fabricated from relatively low - doped semiconductors, which results in wide depletion regions between the barriers and the collector region, and accordingly, small equivalent capacity. For this reason, RTDs can operate at frequencies up to several THz, much higher than those corresponding earlier tunnel diodes which just reach about 100GHz, with response time under $10^{-13}$ s. Small values of the NDR , i.e. an abrupt fall after the maximum on the I - V curve result in high cut - off frequencies of operation. In fact, RTDs are the only purely electronic devices that can operate up to frequencies close to 1 THz, the highest of any electron transit time device (see, also [137}).

4.2. Field effect transistors.

The previously analyzed diodes are simplest electronic devices, for which the current is controlled by the diode bias and vice versa. A useful function can be performed mainly due to nonlinearity of current - voltage dependences. In contrast, in three - terminal devices known as transistors there exist the possibility of controlling the current through two electrodes by varying the voltage or the current through third electrode. Below we briefly describe the field effect transistors (FETs) on the base of the nanowires. Nanowire FETs can be configured by depositing the nanomaterial onto an insulating substrate surface, and making source and drain on the ends nanowire. Fig. 38 illustrates this approach.

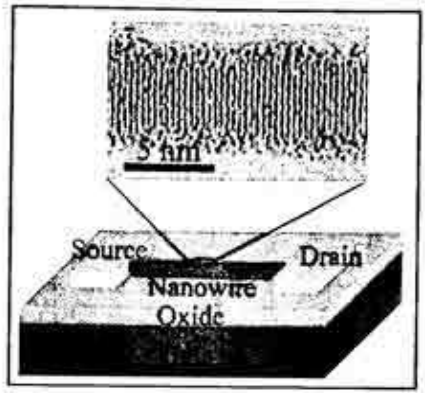

**Fig. 38. A schematic diagram of a Si - FET with nanowire, the metal source, and drain electrodes on the surface of a SiO$_2$/Si substrate (after [100]).**

There, we show a schematic diagram of a Si - nanowire FET with the nanowire, the metal source and drain electrodes on the surface of the SiO$_2$/Si substrate (see, also [100]). This approach may serve as the basis for hybrid electronic systems consisting of nanoscale building blocks integrated with more complex planar silicon circuitry [11]. We should note that an extremely small FET may be built on the basis of carbon nanotube [138]. In conclusion of this part we have noted that the nanowire devices discussed here have great potential for applications in nano - and optoelectronics.

4.3. Single - electron - transistor.

The so - called single electronics [103 - 106] appeared in the late 1980s, is at present time a tremendously expanded research field covering future digital and analog circuits, metrological standards, sensors, and quantum information processing and transfer [11]. The basic device, called a single electron device (SED), literally enables the control of electrons on the level of an elementary charge (see, also [100, 137]). There are rich varieties SEDs (see, e.g. [139 - 141]), but the operation principle of all SED is basically the same. SEDs rely on a phenomenon that occurs when electrons are to enter a tiny conducting material. When the tiny conducting material, or metallic "island", is extremely small, the electrostatic potential of the island significantly increases even when only one electron enters it. For example, for a nanometer scale island having a capacitance C of, say, 1 aF ($10^{-18}$F), the increase in the voltage, which is e/C with e = $1.6 \cdot 10^{-19}$ C, reaches 16 mV. This is much larger than the thermale noise voltage at room temperature, 25.9 mV. Coulomb repulsion prevents additional electrons from entering the island unless the island potential is intentional lowered by an external bias. If the island potential is lowered gradually, the other electrons can enter the island one by one with negligibly small power dissipation (for details see [140] and references therein).

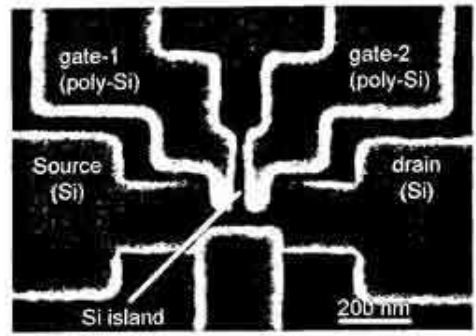

**Fig. 39. A scanning electron microscope image of a single electron transistor (after [141]).**

The single - electron transistor works as follows. The electron transfer is determined by two factors: the Coulomb charging of the dot and the quantized energy levels in the dot (see above). If the drain is biased with respect to the source, an electric current occurs in the regime of single - electron transfer. By applying the voltage to the gate and changing the QD parameters, one can change the conditions of electron tunneling and affect the source - drain current. Examples of modulation of the conductance in single - electron transistors by the gate voltage are presented in Fig. 40.

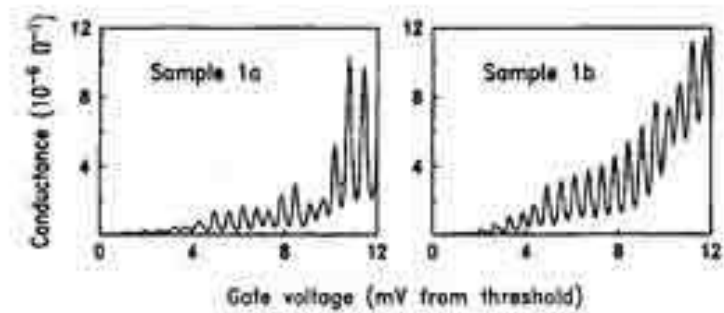

**Fig. 40. Conductance as a function of $V_g$ for two samples with the same geometry (after [139]).**

The devices have almost the same geometry. Their dimensions are large enough to have a number of quantized levels. In Fig. 40 each peak in the conductance corresponds to transfer of one electron, when an energy level enters into resonance with the electron states in the contacts. Though the conductance versus gate - voltage dependences are different, i.e. not reproducible, the peak spacing is the same for both devices. It is determined by the change in the gate voltage required to change the charging energy of the QDs by one electron. The Fig. 40 shows clearly that the electric current is modulated significantly by the gate voltage. Thus, for transistors with single - electron transport, strong control of very small electric current may be possible.

4.4. Light - emitting diodes and lasers.

So far we have studied electronic nanoscale devices, i.e., a class of devices that exploits electrical properties of nanostructures and operates with electric input and output signals.    another class is composed of optoelectronic devices, which are based on both electrical and optical properties of materials and work with optical and electric signals.  In this paragraph we will analyze two very important classes of optoelectronic devices: light - emitting diodes and lasers (diodes as well as photodetectors. As will be shown below, the energy of the electric current flowing through these diodes is transformed into light energy. These optoelectronic devices have a huge number of applications and deserve consideration in details (see, also [142 - 145]).

Although stimulated emission [72] from the injection laser diode is very important (see, below), practically, sub - threshold operation of the diode - when only spontaneous light is emitted - is in many cases advantageous and has a number of applications. Diodes operating with spontaneous light emission are called light - emitting diodes [144]. The important characteristic of the light - emitting diode is the spectral distribution of emission The spectrum of emission  is determined, primarily, by the electron/hole distributions. Thus, the ambient temperature T, defines both spectral maximum and the spectral width of emission. The peak value of the spectral distribution can be estimated as [8, 74]

$\hbar\omega = E_g + \frac{k_B T}{2}.$         (104)

The full width at half maximum of the distribution is $\Delta\omega \approx 2k_B T/\hbar$ and is independent of $\omega$. In terms of the wavelength, $\lambda$, we obtain

$\Delta\lambda = [\lambda_m^2/(2\pi c)]\Delta\omega$

or

$\Delta\lambda = 1.45\lambda^2 k_B T,$         (105)

where $\lambda_m$ corresponds to the maximum of the spectral distribution, $\Delta\lambda$ and $\lambda_m$ are expressed in micrometers, and $k_B T$ is expressed in eV. Fig. 41 shows the spectral density as a function of the wavelength for light - emitting diodes based on various materials.  For these different materials, the spectral linewidth increase in proportion to $\lambda^2$, in accordance with Eq. (105).

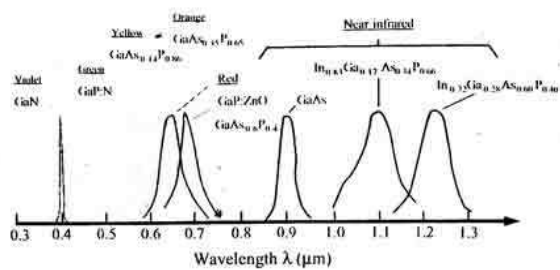

**Fig. 41. The spectra of light - emitting semiconductor diodes with different bandgaps (after [146]).**

From Fig.41, one can see that light - emitting diodes cover a wide spectral region from the infrared - about 8 $\mu$m for INGaAsP alloys - to the near ultraviolet - 0.4 $\mu$m for GaN. Light - emitting diodes are, indeed very universal light sources [14].

Semiconductor lasers incorporating low - dimensional hetero-structures., QWs and QDs, are attracting considerable interest of their potential for improved performance over QW lasers (see, e.g. [144, 145]). This prediction is based, in the single - particle picture, on the sharper density of states resulting from the confinement of the charge carriers in two or three directions. Among other advantages, the ideal QD and QWr lasers would exhibit higher and narrower gain spectrum, low threshold currents, better stability with temperature, lower diffusion of carriers to the device surfaces, and a narrower emission line than double heterostructure or QW lasers (see, also [148]). The observation of lasing from excitons in optically excited V - groove GaAs/AlGaAs QWr laser structures was detail describe in paper [147]. The observable emission is attributed to the recombinations of excitons associated with the lowest energy electron - and hole - subbbands of the QWr. Moreover these authors show that the emission energy remains nearly constant within the inhomogeneously broadened photoluminescence line of the QWrs for both continuous wave (cW) and pulsed optical excitation over a wide range of power densities. These results corroborate the important role played by electron - hole Coulomb correlations [124] in the optical emission from quasi - 1D QWrs in the density regime of the Mott transition.

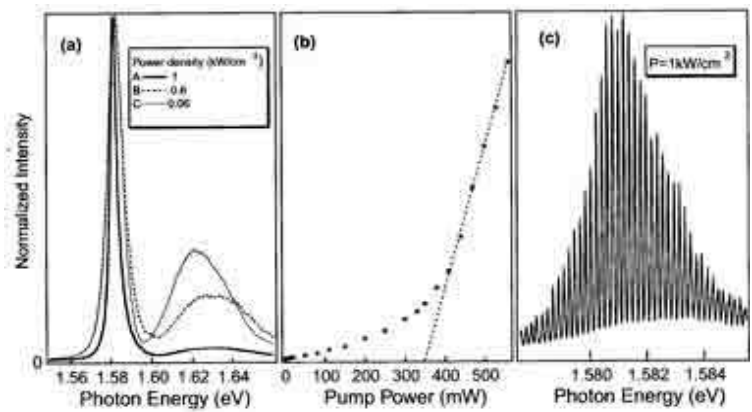

**Fig. 42. (a) Photoluminescence spectra at 10 K of the QWr laser sample above, below and near the lasing threshold in TE - polarization. (b) Dependence on input excitation power of the PL output power;arrows indicate the excitation powers used for the optical spectra depicted in (a). (c) High resolution emission spectrum above the lasing threshold showing the Fabri - Perrot modes of the optical cavity (after [147]).**

Optical emission of the QWr laser structure are displayed in Fig. 42 for different values of the optical power density below, at and above the threshold for lasing in the QWr. Upon increasing the pump power, these authors observe a nearly constant energy of the peak at 1.581 eV that corresponds to the optical transition $e_1$ - $h_1$ associated with the ground electron - hole - subband of the QWrs. A significant spectral narrowing is also found as the power density is increased and crosses the lasing threshold. This evidences the existence of amplified spontaneous emission within this inhomogeneously broadened PL line in this density regime. The observable emission intensity varies linearly at low excitation power over three orders of magnitude (from 0.1 to 100 mW) [147]). Above the lasing threshold (at 350 mW) the intensity variation is again linear (see, Fig. 42$^b$), indicating that the modal gain is saturated. In Fig. 42$^c$, a high - resolution emission spectrum obtained above threshold features well - resolved Fabry - Perot modes that correspond to different longitudinal optical modes within the inhomogeneuous line of the QWr - PL. Detailed investigations of PL and PLE spectra (see, Fig. 43) of the QWr allowed the indicated authors to conclude that the lasing emission originates from the recombination of excitons as it is case for the QWr - peak of the cw - PL spectrum (details see [147]).

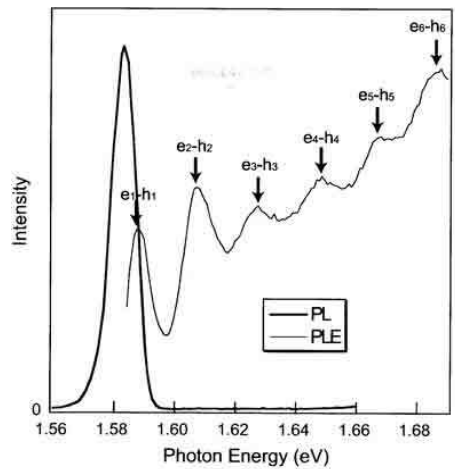

**Fig. 43. Linearly - polarized PLE spectrum and the corresponding PL spectrum of an etched QWr laser sample at 10 K. The polarization of the excitation is parallel tothe wire axis. The different optical transition e$_n$ - h$_n$ are marked by arrows (after [147]).**

In QDs, as indicated above, carriers are confined in the three directions in a very small region of space, producing quantum effects in the electronic properties. As we can see from Fig. 23, the electronic joint density of states for QD shows sharp peaks corresponding to transitions between discrete energy levels of electrons and holes. Outside these levels the DOS vanishes. In many ways, the electronic structure of a QD resembles that a single atom [107]. Lasers based on QDs could have properties similar to those of conventional ion gas lasers, with the advantage that the electronic structure of a QD can be engineered by changing the base material, size and shape. In the next we assume that the QDs are small enough so that the separation between the first two electron energy levels for both electrons and holes is much larger than the thermal energy KT. Then for an undoped system, injected electrons and holes will occupy only the lowest level. Therefore, all injected electrons will contribute to the lasing transitions from the $E_{1e}$ to the $E_{1hh}$ levels, reducing the threshold current with respect to other systems with lower confinement. The evolution of the threshold current density obtained along the years for various laser structures is shown in Fig. 44. The lowest threshold currents have already been reached for QD lasers [149]. As long as the thermal energy is lower than the separation between the fist and second levels, the emission band in an ideal QD laser is very sharp and does not depend on temperature (see, also [145]). Therefore, QD lasers should have a better stability with temperature without the need for cooling. We should add that QDs have the narrowest spectrum and the highest gain (for details see, also [8, 74, 145]).

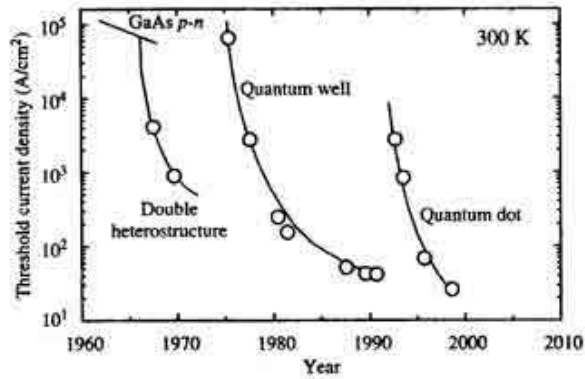

**Fig. 44. Evolution of threshold current density for lasers based on different confinement structures (after [149]).**

4.5. Quantum computation and quantum communication.

4.5.1. Background.

The concept of information is too broad to be captured completely by a single definition (see, e.g. [150]). But we all know that information may be not only created,elaborated, transmitted through space and preserved or stored throughout time, but also may be extracted and use for communication. Before introduced some of definitions of information theory, it is desirable to remove one possible cause of misapprehension. Possible combinations of the letters a, n, and t are tan, ant, nat. These words may have meaning and significance for readers but their impact on individuals will vary, depending on the reader's subjective reaction. Subjective information conveyed in this way is impossible to quantify in general. Therefore the meaning of groups of symbols is excluded from the theory of information; each symbol is treated as an entity in its own right and how any particular grouping is interpreted by an individual is ignored. Information theory is concerned with how symbols are affected by various processes but not with information in its most general sense [151].

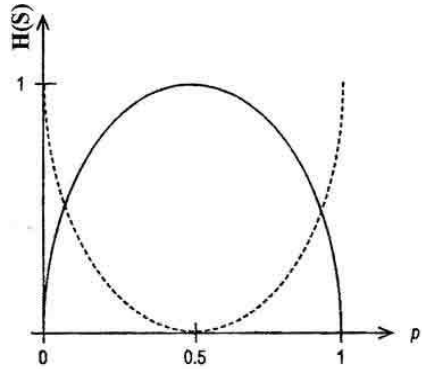

**Fig. 45. Shannon's average information or entropy H as a function of the probability p of one of the final states of a binary (two state) devices. H is measure of the uncertainty before any final state occured and expresses the average amount of information to be gained after the determination of the outcome. A maximum uncertainty of one bit (or maximum gain information, once the result is known) exists when the two final states are equiprobable (p = 0.5). The dotted curve represents (1 - H) (see Eq. 113) an objective measure of the "prior knowledge" before operating the device (after [13]).**

According Shannon [152] we call $I_0$ the information value in bits if state $|0\rangle$ is seen, and $I_1$ the same for the occurrence of $|1\rangle$. We state that in the case of perfect symmetry (see Fig. 45), i.e. for $p_0 = p_1 = 0.5$ we should obtain $I_0 = I_1 = 1$b. And it is reasonable to demand that for $p_i = 1$ (i = 0 or 1) we should get $I_i = 0$, whereas for $p_i \to 0$, $I_i \to \infty$. What is between these limits? In general, the function we are looking for I(p), should be a continuous function (see Fig. 45), monotonically decreasing with p so that $I_k \rangle I_i$ if $p_k \langle p_i$ (naturally the value of the information gained should be greater for the less probable state. In all above expression p is probability [153]. Namely this function fulfilling such conditions was chosen by Shannon [152] for what is usually called the information content of an outcome that has the probability $p_i$ to occur (see, also [101, 151 - 154]):

$I_i = -K \ln p_i$ . (106)

In order to obtain I = 1b for $p_i = 0.5$ we must set K = 1/ln2. The negative sign is needed so that $I \geq 0$ (p always $\leq 1$). Turning to logarithm of base 2 we can write

$I_i = - \log_2 p_i$. (107).

The base of 2 is therefore especially apposite for dealing with binary digits (bits) and can therefore be expected to be important in application to computing and coding. Tables of logarithms to the base 2 are available but if they are not to hand calculations can be carried out by observing that

$\text{Log}_2 x = \frac{\log_{10} x}{\log_{10} 2} = \frac{\ln x}{\ln 2} = \ln x \log_2 e.$ (108)

In general

$\log_a x = \ln x / \ln a$ (109)

and, since the restriction $a \rangle 1$ has been imposed above, $\ln a \rangle 0$ so that the logarithms which arise always be positive multiplies of the natural logarithm. This fact will be used frequently in subsequent consideration.

The most important and useful quantity introduced by Shannon [152] is related to the next question: Given the probability values for each alternative, can we find an expression for the amount of information we expect to gain on the average before we actually determine the outcome? A completely equivalent question is: How much prior uncertainty do we have about outcome? It is reasonable to choose the weighted average of $I_0$ and $I_1$ for the mathematical definition of the a priory average information gain or uncertainty measure H:

$$H = p_0 I_0 + p_1 I_1 = - p_0 \log_2 p_0 \; p_1 \log_2 p_1 \qquad (110)$$

in which as usually $p_0 + p_1 = 1$ (for details see [152]). Since H is a quantitative measure of the uncertainty of the state of a system, Shannon called it the entropy of the source of information. Since $p_i$ may be zero, some term in H could be undetermined in this definition so, when $p_i = 0$, the value zero is assigned to the corresponding term in (110). Let us set, for our case $p = p_0$; then $p_1 = 1 - p$ and:

$$H = - p \log_2 p - (1 - p) \log_2 (1 - p). \qquad (110^a)$$

Fig. 45 shows a plot of H as a function of p (solid line). It reaches the maximum value of 1b (maximum average information gain in one operation or in one toss of a coin) if both probability values are the same ($p = 1/2$). If $p = 1$ or 0, we a ready know the result before we measure, and the expected gain of information will be zero - there is no a priory uncertainty. A measure of the average information available before we actually determine the result would be 1 - H; $p = 1$ or 0 indeed gives 1b of "prior knowledge", and $p = 1/2$ represents zero prior information (broken line in Fg. 45), that is, maximum uncertainty.

We can generalize the definition (106 - 110) for any number N of possible final states, which will then read:

$$H = - \sum_{i=0}^{N-1} p_i \log_2 p_i \text{ with } \sum_i p_i = 1 \qquad (111)$$

The function H has an absolute maximum when all $p_i$ are equal, i.e. when there is no a priory bias about the possible outcome. In that case, by definition of the probability $p_i$, it is easy to verify that ($p_1 = p_2 = \ldots = p_N = 1/N$)

$$H = \log_2 N. \qquad (112)$$

When all the events are equally probably, the most uncertainty prevails as to which event will occur. It therefore satisfactory that the entropy should be a maximum in such situation. The fact that H(S) is a maximum when the events are equally uncertain but zero when there is certainty provides some justification for considering entropy as a measure of uncertainty. To finalize this part we should indicate, as can see from Fig. 45, always fulfilled the relation between information and entropy:

$$I + H(S) = 1. \qquad (113)$$

4.5.2 Introduction in quantum information and quantum computation.

This part of our review is not intended to cover all developments in the quantum information theory and quantum computation. Our aim is rather to provide the necessary insights for an understanding of the field so that various non-experts can judge its fundamental and practical importance. Quantum information theory and quantum computation are an extremely exciting and rapidly growing field of investigation. Before we discuss some fundamental concepts of quantum information we should remind some of the basic quantum physics for the benefits of readers less familiar with subject.

Classical information theory has been around for ever seven years and there are hundreds of well - tested textbooks not only for physics and mathematics students but also for biologists, engineers and chemics (see, e.g. [150 - 153] and references therein). In contrast, quantum information (QI) theory is in its infancy and it involves physics concepts (for more details see below) that are not familiar to everybody. The most fundamental difference between a classical and a quantum system is that the latter cannot be observed (measured) without being perturbed in a fundamental way [154]. Expressed in more precise terms, there is no process that can reveal any information about the state of a quantum system without disturbing it irrevocably. Thus quantum systems cannot be left undisturbed by measurement, no matter how ideal the instruments are: there are intrinsic limitations to the accuracy with which the values of certain magnitudes or observables [27] as they called in quantum mechanics, can be determined in measurements.

The intrinsic limitation to our potential knowledge of a quantum system is most concisely expressed in the form of the Heisenberg uncertainty principle. For a single particle traveling along the x - axis with momentum $p_x$, this principle states [27] that

$$\Delta x \Delta p_x \geq \hbar/2 \qquad (114)$$

where $\Delta x$ and $\Delta p_x$ are the standard deviations of measured values of position and momentum, respectively, obtained for a given type of particle in a series of experiments under strictly identical circumstances of preparation (experimental setup and initial conditions) and measurement (instrumentation and timing). According to the meaning of standard deviation, $\Delta x$ and $\Delta p_x$ represent the approximate ranges within which the values of the position and momentum can be expected to be found with reasonable probability (68 % for a Gaussian distribution [150]) if measured under the specified conditions.

There are many different kinds of experiments show a fundamental property of all quantum systems, valid as long as the system is left undisturbed (free from irreversible interactions with the outside macroscopic world [101]), namely, the possibility of being in a single state made up of the superposition [155] of two or more basis states. By superposition we do not mean that the system is sometimes in one, and sometimes in another state: it is simultaneously in two or more component states. We should underline that there is no classical equivalent to this situation. The principle of superposition tell us that a general state of the photon between vertical and horizontal polarization would be

$$|\Psi\rangle = c_v |\Phi_v\rangle + c_h |\Phi_h\rangle, \qquad (115)$$

where $c_v$ and $c_h$ are two complex numbers [27].

In the quantum formalism the values $c_v c_v^* = |c_v|^2$ and $c_h c_h^* = |c_h|^2$ (the star indicating complex conjugate) are the probabilities of finding the system respectively in the state $|\Phi_v\rangle$ or $|\Phi_h\rangle$ after measurement was made to find out which polarization was taken:

$$p_v = |c_v|^2 \text{ and } p_h = |c_h|^2. \qquad (116)$$

Since their sum must equals one, we require the normalization condition [27]

$$|c_v|^2 + |c_h|^2 = 1 \qquad (117)$$

With this normalization, relation (117) can also be written in polar form $|\Psi\rangle = \cos\alpha |\Phi_v\rangle + e^{i\varphi}\sin\alpha |\Phi_h\rangle$ in which $\cos^2\alpha = p_v$ and $\sin^2\alpha = p_h$. The expression brings out explicitly the phase difference $\varphi$. We will come back to this form later.

### 4.5.3. Information is physical.

As is well - known, information is not a disembodied abstract entity: it always tied to physical representation (see, e.g. [156]). It is represented by engraving on a stone tablet, a spin, a charge, a hole in a punched card, a mark on the sheet of paper, or some other equivalent*[)]. This ties the handling of information to all the possibilities and restrictions of our real physical world, its laws of physics and its storehouse of available parts. This view was implicit in Szilard's discussion [157] of Maxwell demon (see, also [158 - 159] and references therein). the laws of physics are essentially algorithms for calculation (see, also [101, 160 - 162]).

______________________________

*[)]As is well - known [154], in 1961, Landauer had the important insight that there is a fundamental asymmetry in the Nature allows us to process information. Copying classical information can be done reversibly and without wasting any energy, but when information is erased there is always an energy cost of $kT\ln 2$ per classical bit to be paid ( for more details see, also [163]). Furthermore an amount of heat equal to $kT\ln 2$ is damped in the environment at the end of erasing process. Landauer's conjectured that this energy/entropy cost cannot be reduced below this limit irrespective of how the information is encoded and subsequently erased - it is a fundamental limit. Landauer's discovery is important both theoretically and practically as on the one hand it relates the concept of information to physical quantities like thermodynamical entropy and free energy and on the other hand it may force the future designers of quantum devices to take into account the heat production caused by the erasure of information although this effect is tiny and negligible in today's technology. At the same time, Landauer profound insight has led to the resolution of the problem of Maxwell's demon by Bennett [164].

______________________________

Thus, information is something that can be encoded in the state of a physical system, and computation (see also below) is a task that can be performed with a physically realizable device. Therefore, since the physical world is fundamentally quantum mechanical, the foundation of information theory and computation science should be sought in quantum physics. In fact, quantum information has weird properties that contrast sharply with the familiar properties of classical information. Be that as it may, information until recent has largely been thought of in classical terms, with quantum mechanics playing a supporting role in the design of the equipment to process it, and setting limits on the rate at which it could be sent through certain channels. Now we know that a fully quantum theory of information and information processing offers (for details see [160]), among other benefits, a brand of cryptography whose security rests on fundamental physics, and a reasonable hope of constructing quantum computers (see below) could dramatically speed up the solution of certain mathematical problems (see, e.g. [165]) These benefits depend on distinctively quantum properties such as uncertainty, interference and entanglement. Thus quantum information theory generalizes the classical notions of source and channel, and the related techniques of source and channel coding, as well as introducing a new resource, entanglement, which interacts with classical and quantum information in a variety of ways that have no classical parallel (for details, see [164, 166] and references therein).

As was shown for the first time by Schrődinger [155] fundamental properties of quantum systems, which might be include to information processes are [166 - 169]:

1. Superposition: a quantum computer can exist in an arbitrary complex linear combination of classical Boolean states, which evolve in parallel according to a unitary transformation.

2. Interference: parallel computation paths in the superposition, like paths of a particle through an interferometer, can reinforce or cancel one another, depending on their relative phase.

3. Entanglement: some definite states of complete quantum system do not correspond to definite states of its parts.

4. Nonlocality and uncertainty: an unknown quantum state cannot be accurately copied (cloned) nor can it be observed without being disturbed (see, also [167 - 168].

These four elements are very important in quantum mechanics, and as we'll see below in information processing. All (classical) information can be reduced to elementary units, what we call bits. Each bit is a yes or a no, which we may represent it as the number 0 or the number 1. Quantum computation and quantum information are built upon ananalogous concept, the quantum bit [170], or qubit for short. It is a two-dimensional quantum system (for example, a spin 1/2, a photon polarization, an atomic system two relevant states, etc.) with Hilbert space. In mathematical terms, the state of quantum state (which is usually denoted by $|\Psi>$ [155]) is a vector in an abstract Hilbert space of possible states for the system. The space for a single qubit is spanned by a basis consisting of the two possible classical states, denoted, as above, by $|0>$ and $|1>$. This mean that any state of qubit can be decomposed into the superposition

$$|\Psi> = \alpha |0> + \beta |1> \qquad (118)$$

with suitable choices of the complex coefficients *a* and *b*. The value of a qubit in state $|\Psi>$ is uncertain; if we measure such a qubit, we cannot be sure in advance what result we will get. Quantum mechanics just gives the probabilities, from the overlaps between $|\Psi>$ and the possible outcomes, rules due originally by Max Born (see, e.g. [101]). Thus the probability of getting 0 is $|<0|\Psi>|^2 = |a|^2$ and that for 1 is $|<1|\Psi>|^2 = |b|^2$. Quantum states are therefore normalized; $<\Psi|\Psi> =$ (b*a*)·$\begin{pmatrix} b \\ a \end{pmatrix} = 1$ (where $|\Psi>$ is represented by the vector $\begin{pmatrix} b \\ a \end{pmatrix}$) and the probabilities sum to unity (see, also above). Quantum mechanics also tells us that (assuming the system is not absorbed or totally destroyed by the action of measurement) the qubit state of Eq. (118) suffers a projection to $|0>$ ($|1>$) when we get the result 0(1). Because $|\alpha|^2 + |\beta|^2 = 1$ we may rewrite Eq. (118) as (see, e.g. [171])

$$|\Psi> = \cos\theta |0> + e^{i\varphi}\sin\theta |1> \qquad (119)$$

where $\theta, \varphi$ are real numbers. Thus we can apparently encode an arbitrary large amount of classical information into the state of just one qubit (by coding the information into the sequence of digits of $\theta$ and $\varphi$). However in contrast to classical physics, quantum measurement theory places severe limitations on the amount of information we can obtain about the identity of a given quantum state by performing any conceivable measurement on it. Thus most of the quantum information is "inaccessible" but it is still useful - for example it is necessary in its totality to correctly predict any future evolution of the state and to carry out the process of quantum computation.

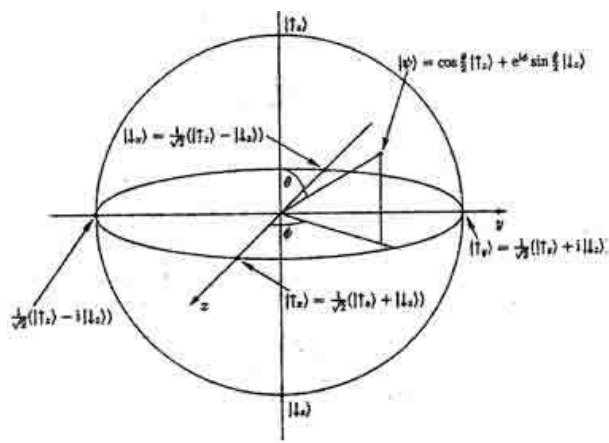

**Fig. 46. The Bloch sphere of the Hilbert space spanned by** $|\uparrow_z\rangle$ **and** $|\downarrow_z\rangle$ **(after [13]).**

The numbers $\theta$ and $\varphi$ define a point on the unit three - dimensional sphere, as shown in Fig. 46. This sphere is often called the Bloch (Poinkare) sphere [171, 172]; it provides a useful means of visualizing the state of a single qubit. A classical bit can only sit at the north or the south pole, whereas a qubit is allowed to reside at any point on the surface of the sphere (for details see [161.]).

4.5.4. Quantum computation.

The theory of computation has been long considered a completely theoretical field, detached from physics (see, e.g. [162, 173, 174]). Nevertheless, pioneers such as Turing, Church, Post and Gődel were able [101, 154], by intuition alone, to capture the correct physical picture, but since their work did not refer explicitly to physics , it has been for a long time falsely assumed that the foundations of the theory of classical computation were self-evident and purely abstract. Only in the last three decades were questions about the physics of computation asked and consistently answered [175, 176]. Subsequently in the development of the subject of quantum computation - which represents a hybrid of quantum physics and theoretical computer science - it was realized that quantum systems could be harnessed to perform useful computations more efficiently them any classical device.

We should stress that the perspective of information theory also provides further new insight into the relationship between entanglement (see above) and non - locality (see, e.g. [177, 178]), beyond the well-studied mediation of non-local correlations between local measurement outcomes. The theory of computation and computational complexity [153] is normally as an entirely mathematical theory with no references to considerations of physics. However, as we know, any actual computation is a physical process involving the physical evolution of selected properties of a physical system. Consequently the issues of "what is computable" and "what is the complexity of a computation" must depend essentially on the laws of physics and cannot be characterized by mathematical alone [156]. This fundamental point was emphasized by Landauer, Deutsch and it is dramatically confirmed by the recent discovers (see, e.g. [160 - 162]) that the formalism of quantum physics allows one to transgress some of boundaries of the classical theory of computational complexity, whose formulation was based on classical intuitions.

As is well-known that a fundamental notion of the theory of computational complexity is the distinction between polynomial and exponential use of resources in a computation (see, also [179]). This will provide a quantitative measure of our essential distinction between quantum and classical computation. Consider a computational task such as following: given an integer $N$, decide whether $N$ is a prime number or not. We wish to assess the resources required for this task as a function of the size of the input which is measured by $n = \log_2 N$, the numbers of bits needed to store $N$. If $T(n)$ denotes the number of steps (on a standard universal computer) needed to solve the problem, we ask whether $T(n)$ can be bounded by some polynomial function in n or whether $T(n)$ grows faster than any polynomial (e.g. exponential). More generally it may consider any language L - a language being a subset of the set of all finite strings of 0's and 1's - and consider the computational task of recognizing the language, i.e. given a string $\sigma$ of length n the computations outputs 0 if $\sigma \in L$ and outputs 1 if $\sigma \notin L$. The language L is said to be in complexity class P (it is mean "polynomial time") if there are exists an algorithm which recognizes L and runs in time $T(n)$ bounded by polynomial function. Otherwise the recognition of L is said to require exponential time.

Thus, the standard mathematical theory of computational complexity assesses the complexity of a computation in terms of the resources of time (number of steps needed) and space (amount of memory required). In the quantum computation we have been led to consider the accounting of other physical resources such as energy and precision (for details see [180]). The algorithm;of quantum computation such as Shor's algorithm [180] depend critically for their efficiency and validity on effects of increasingly large scale entanglements with increasing input size (see, also [181]).

Further evidence for the power of quantum powers came in 1995 when Grover [182] showed that another important problem - the problem of conducting a search through some unstructured search space - could also be speed up on a quantum computer. While Grover's algorithms did not provide as spectacular a speed up as Shor's algorithms, the widespread applicability of search - based methodologies has excited considerable interest in Grover's algorithm (for details see, also [183]).

4.5.5. Quantum teleportation.

The role of entanglement in quantum information processing is fundamental. Motivated by paper [177] Schrődinger in his famous paper [155] wrote" Maximal knowledge of a total system does not necessary include total knowledge of all its parts, not even when these are fully separated from each other and at the momentary not influencing each other at all" and he coined the term "entanglement of our knowledge" to describe this situation [171].

A composite system is a system which consists of two or more parts and the simplest one is a system consisting of two qubits (carried by two particles of the same kind, or other appropriate quantum registers (see [13]). We call the two systems A (Alice) and B (Bob). Any states of each of the systems can be written as

$$|\Psi\rangle_A = \alpha |0\rangle_A + \beta |1\rangle_A \quad \text{and} \quad |\Phi\rangle_B = \gamma |0\rangle_B + \delta |1\rangle_B \tag{120}$$

with $|\alpha|^2 + |\beta|^2 = 1$ and $|\gamma|^2 + |\delta|^2 = 1$. The subindices A and B refer to two physical entities (the qubits) and the vectors $|0\rangle$ and $|1\rangle$ refer to their basis states (in the case of a pair particles, to some binary internal variable like spin, polarization, pair of energy levels, etc.). Each pair of coefficients in (120) satisfies the normalization condition [27]. (The composite state of the two systems is then simply the tensor product (or direct product) of the two states.

$$|\Psi_{prod}\rangle = |\Psi\rangle_A \otimes |\Phi\rangle_B. \tag{121}$$

Such a state is called a product state, but product states are not only physically realizable states. If we let the two systems interact with each other, any superposition of product states is realizable. Hence a general composite state can be written as

$$|\Psi\rangle = \sum_{i,j} \alpha_{ij} |\Psi_i\rangle_A \otimes |\Phi_j\rangle_B \tag{122}$$

where $\sum |\alpha_{ij}|^2 = 1$ and the sets $\{|\Psi_i\rangle\}$ and $\{|\Phi_j\rangle\}$ are orthonormal bases for the two subsystems. **Any composite state that is not a product state is called an entangled state**. A composite quantum state consisting of two parts only, is called a bipartite state [129], as opposed to multipartite states which consist of mor than two parts. For bipartite qubit states, four entangled states play a major role [178], namely the singlet state

$$|\Psi^-\rangle \equiv \tfrac{1}{\sqrt{2}}(|01\rangle - |10\rangle) \tag{123$^a$}$$

and three triplet states

$$|\Psi^+\rangle \equiv \tfrac{1}{\sqrt{2}}(|01\rangle + |10\rangle) \tag{123$^b$}$$

$$|\Phi^-\rangle \equiv \tfrac{1}{\sqrt{2}}(|00\rangle - |11\rangle) \tag{123$^c$}$$

$$|\Phi^+\rangle \equiv \tfrac{1}{\sqrt{2}}(|00\rangle + |11\rangle) \tag{123$^d$}$$

where we have used $|ij\rangle$ as a shorthand notation for $|i\rangle \otimes |j\rangle$. They are called Bell states [178] or EPR [177] pairs. Together they form an orthogonal basis for the state space of two qubits, called the Bell basis. The Bell states are maximally entangled and one can be converted into another by applying a unitary transform locally on any one of the subsystems. Note that if we measure the state of one qubit in a Bell state (that is, measure the Z operator which has eigenvalues $\pm 1$), we immediately know the state of the other particle. In the singlet Bell state, a measurement of qubit A will yield one of the eigenstates $|0\rangle$ and $|1\rangle$, each with probability of 1/2. These results leave qubit B in state $|1\rangle$ or $|0\rangle$, respectively. For a single qubit we could always change to another basis where the outcome of a Z measurement would be given. For a spin - 1/2 particle this means that the spin is always pointing in some direction, even though the state will show up as a superposition in a basis where the state is not one of the basis states. If the particle is entangled with another particle, though, the direction of the spin of that particle alone is not well defined. Actually, for particles in one of the Bell states, the probability for measuring the spin of the particle to "up" (while ignoring the other particle) is 1/2 for any direction (for details see [183]).

Further we briefly describe entangled states of two polarized exciton states in a single dot created and detected optically. As was noted above quantum information, quantum computation, quantum cryptography and quantum teleportation intrinsic quantum mechanical correlations (see [172] and references therein). A fundamental requirement for the experimental realization of such proposal is the successful generation of highly entangled quantum states. In particular, as will be shown below, coherent evolution of two qubits in an entangled state of the Bell type is fundamental to both quantum cryptography and quantum teleportation. Maximally entangled states of three qubits, such as the so - called Greenberger - Horne - Zeilinger (GHZ) states [184], are not only of intrinsic interest but also of great practical importance in such proposals [185]. New systems and methods for the preparation and measurement of such maximally entangled states are therefore being sought intensively (see, e.g. 186 - 188 71]). We should add in this connection, that recent experimental work of Gammon et al. (see, e.g. reviews [189, 190] and references therein)suggests that optically generated excitons in QDs represent ideal candidates for achieving coherent wavefunction control on the nanometer and femtosecond scales.

When two quantum dots are sufficiently, close, there is a resonant energy transfer process originating from the Coulomb interaction whereby an exciton can hop between dots [184]. The Coulomb exchange interaction in QD molecules give rise to a non radiative resonant energy transfer (i.e. Főrster process [181]) which correspond to the exchange of a virtual photon, thereby destroying an exciton in a dot and then re - creating it in a close by dot. As it is well - known that the presence and absence of an exciton in a dot (for example in isotope - mixed crystals serve as a qubit).

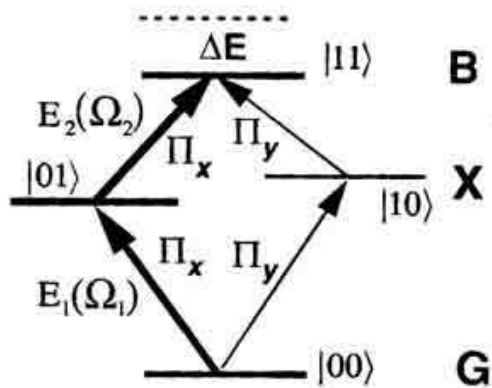

**Fig. 47. Model for a single QD. $|11\rangle$, $|01\rangle$, $|10\rangle$ and $|00\rangle$ denote the biexciton, the exciton and the ground states, respectively. $\Delta E$ is the biexciton binding energy. The optical rules for various transition are indicated.**

The basic quantum operations can be performed on a sequence of pairs of physically distinguishable quantum bits and, therefore, can be illustrated by a simple four - level system shown in Fig. 47. In an optically driven system where the $|01\rangle$ and $|10\rangle$ states can be directly excited, direct excitation of the upper $|11\rangle$ level from the ground state $|00\rangle$ is usually forbidden (see, e.g.[39] and references therein) and the most efficient alternative is coherent nondegenerate two - photon excitation, using $|01\rangle$ and $|10\rangle$ as an intermediate states. The temporal evolution of the non - radiative Raman coherence between states $|01\rangle$ and $|10\rangle$ was directly resolved in qauntum beats measured in differential transmission (DT) geometry as shown in Fig.48.

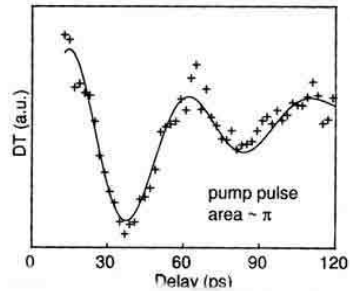

**Fig. 48. An entangled state involving two polarized excitons confined in a single dots was created and detected optically as evidenced by quantum beats between states $|01\rangle$ and $|10\rangle$ shown. The quantum coherence time ($\simeq$ 40 psec.) between these two states is directly extracted from the decay of the envelope (after [189]).**

In order to increase the quantum operations beyond one dot, interdot exciton interaction is required. One proposal is to use an electric field to increase the dipole - dipole interaction between to excitons in separate dots [13]. The procedure we will analyze below is called quantum teleportation and can be understood as follows. The naive idea of teleportation involves a protocol [186, 154] whereby an object positioned at a place A and time t first " dematerializes" and then reappears at a distant place B at some later time t + T. Quantum teleportation implies that we wish to apply this procedure to a quantum object. However, a genuine quantum teleportation differs from this idea, because we are not teleporting the **whole object** but just its state from particle A to particle B. As quantum particles are indistinguishable anyway, this amounts to 'real' teleportation. One way of performing teleportation is first to learn all the properties of that object (thereby possibly destroying it). We then send this information as a classical string of data to B where another object with the same properties is recreated (see Fig. 49).

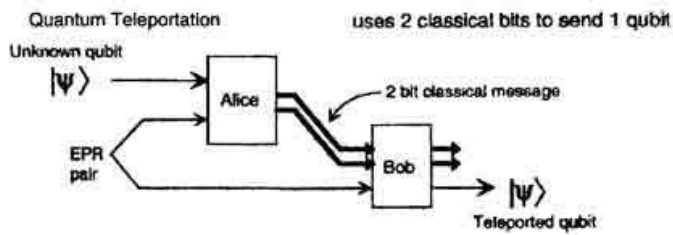

**Fig. 49. Quantum teleportation requires quantum entanglement as a resource. In this case, Alice receives a qubit in an unknown state, and destroys it by performing a Bell measurement on that qubit and a member of an entangled pair of qubits that she shares with Bob. She sends a two-bit classical message (her mesurement outcome) to Bob, who then performs a unitary transformation on his member of the pair to reconstruct a perfect replica of the unknown state. We could see one qubit suffices to carry two classical bits of information.**

One problem with this picture is that, if we have a single quantum system in an unknown state, we cannot determine its state completely because of the uncertainty principle [155, 167]. More precisely, we need an infinite ensemble of identically prepared quantum systems to be able completely to determine its quantum states. So it would seem that the laws of quantum mechanics prohibit teleportation of single quantum systems. However, as we can see above, the very feature of quantum mechanics that leads to the uncertainty principle (the superposition principle [27]) also allow the existence of entangled states [155]. These entangled states will provide a form of quantum channel to conduct a teleportation protocol. We should remind once more, after the teleportation is completed, the original state of the particle at A is destroyed (although the particle itself remains intact) and it is the entanglement in the quantum channel.

As will be show below, coherent evolution of two qubits in an entangled states of the Bell type [178] is fundamental to both cryptography and teleportation.

Consider a system consisting of two subsystems. Quantum mechanics associates to each subsystem a Hilbert space. Let $H_A$ and $H_B$ denote these two Hilbert spaces: let $|i>_A$ (where i = 1, 2, 3,.........) represent a complete orthogonal basis for $H_A$, and $|j>_B$ (where j = 1, 2, 3, ............) a complete orthogonal basis for $H_B$. Quantum mechanics associates to the system, i.e. the two subsystems taken together, the Hilbert space $H_A \otimes H_B$, namely the Hilbert space spanned by the states $|i>_A \otimes |j>$. Further, we will drop the tensor product symbol $\otimes$ and write $|i>_A \otimes |j>$ as $|i>_A |j>_B$ and so on.

Any linear combination of the basis states $|i>_A |j>_B$ is a state of the system, and any state $|\Psi>_{AB}$ of the system can be written as

$$|\Psi>_{AB} = \sum_{i,j} C_{ij} |i>_A |j>_B, \qquad (124)$$

where the $C_{ij}$ are complex coefficients; below we take $|\Psi>_{AB}$ to be normalized, hence

$$\sum_{i,j} |C_{ij}|^2 = 1. \qquad (125)$$

1. A special case of Eq. (124) is a direct product in which $|\Psi>_{AB}$ factors into (a tensor product of) a normalized $|\Psi^{(A)}>_A = \sum_i C_i^{(A)} |i>_A$ in $H_A$ and a normalized state $|\Psi^{(B)}>_B = \sum_j C_j^{(B)} |j>_B$ in $H_B$:

$$|\Psi>_{AB} = |\Psi^{(A)}>_A |\Psi^{(B)}>_B = \left(\sum_i C_i^{(A)} |i>_A\right)\left(\sum_j C_j^{(B)} |j>_B\right).$$
(126)

Note every state in $H_A \otimes H_B$ is a product state. Take, for example, the state $\frac{(|1>_A |1>_B + |2>_A |2>_B)}{\sqrt{2}}$; if we try to write it as a direct product of states of $H_A$ and $H_B$, we will find that they cannot.

2. If $|\Psi>_{AB}$ is not a product state, we say that it is entangled (for details see [187, 188]).

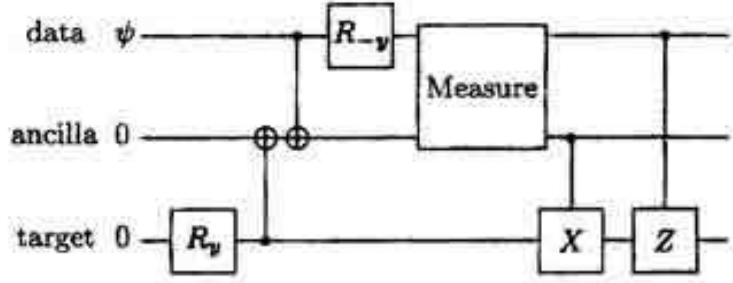

**Fig. 50. Circuit for quantum teleportation. The measurement is in the computational basis, leaving the measurement result stored in the data and ancila qubits. $R_y$ and $R_{-y}$ denote rotations of 90 degrees about the y and -y axes on the Bloch sphere (after [129]).**

Quantum teleportation is a method for moving quantum states from one location to another (Fig. 50) which suffers from none of these problems. Suppose Alice and Bob share a pair of qubits which are initially in the entangled state $(|00\rangle + |11\rangle)/\sqrt{2}$. In addition, Alice has a system which is in some potentially unknown state $|\Psi\rangle$. The total state of the system is therefore

$$|\Psi\rangle \left( \frac{(|00\rangle + |11\rangle)}{\sqrt{2}} \right). \quad (127)$$

By writing the state $|\Psi\rangle$ as $\alpha|0\rangle + \beta|1\rangle$ and doing some simple algebra, we see that the initial state can be rewritten as

$(|00\rangle + |11\rangle)|\Psi\rangle + (|00\rangle - |11\rangle)Z|\Psi\rangle + (|01\rangle + |10\rangle)X|\Psi\rangle + (|01\rangle - |10\rangle)XZ|\Psi\rangle.$
(128)

Here and below we omit normalization factors from the description of quantum states. Suppose Alice performs a measurement on the two qubits in her possession, in the Bell basis consisting of the four orthogonal vectors: $|00\rangle + |11\rangle$; $|00\rangle - |11\rangle$; $|01\rangle + |10\rangle$; $|01\rangle - |10\rangle$, with corresponding measurement outcomes which we label 00, 01, 10 and 11 [187]. From the previous equation, we see that Bob's state, conditioned on the respective measurement outcomes, is given by

$00 : |\Psi\rangle; 01: X|\Psi\rangle; 10 : Z|\Psi\rangle; 11 : XZ|\Psi\rangle.$ \quad (129)

Therefore, if Alice transmits the two classical bits of information, she obtains from the measurement to Bob, it is possible for Bob to recover the original state $|\Psi\rangle$ by applying unitary operators inverse to the identity X, Z and XZ, respectively. More explicitly, if Bob receives 00, he knows his state is $|\Psi\rangle$, if he receives 01 then applying an X gate (see below) will cause him to recover $|\Psi\rangle$, if he receives 10 then applying a Z gate cause him to recover $|\Psi\rangle$, and if he receives 11 then applying an X gate followed by a Z gate will enable him to recover $|\Psi\rangle$ (see Fig. 51). This completes the teleportation process.

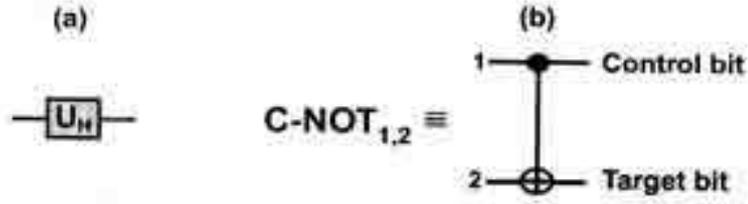

**Fig. 51. Schematic representation of (a) the Hadamard gate, and (b) the C - NOT gate.**

Further we describe a practical scheme capable of demonstrating quantum teleportation which exploits entangled states of excitons in coupled QDs [190]. As we saw above, the general scheme of teleportation [155], which is based on EPR pairs [177] and Bell measurements [178] using classical and purely non - classical correlations, enables the transportation of an arbitrary quantum state from one location to another without knowledge [168] or movement of the state itself through space. In order to in implement the quantum operations for the description of the teleportation scheme, we employ two elements: the Hadamard transformation and the quantum controlled NOT gate (C - NOT gate). In the orthonormal computation basis of single qubits $\{|0\rangle, |1\rangle\}$, the C - NOT gate acts on two qubits $|\varphi_i\rangle$ and $|\varphi_j\rangle$ simultaneously as follows C - NOT$_{ij}(|\varphi_i\rangle|\varphi_j\rangle) \rightarrow |\varphi_i\rangle|\varphi_i \oplus |\varphi_j\rangle\rangle$. Here $\oplus$ denotes addition modulo 2. The indices i and j refer to the control bit and the target bit respectively (see Fig. 52).

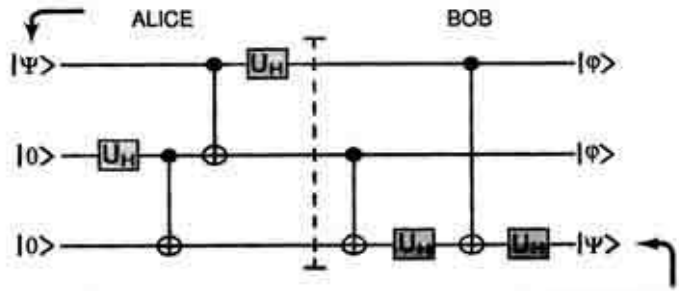

**Fig. 52. Cicuit scheme to teleport un unknown quantum state of exciton from Alice to Bob using arrangement of 3 qubits (coupled quantum dots).**

The Hadamard gate $U_H$ acts only on single qubits by performing the rotations $U_H(|0\rangle) \rightarrow \frac{1}{\sqrt{2}}(|0\rangle + |1\rangle)$ and $U_H(|1\rangle) \rightarrow \frac{1}{\sqrt{2}}(|0\rangle - |1\rangle)$. The above transformation can be written as

$$U_H = \begin{pmatrix} 1 & 1 \\ 1 & -1 \end{pmatrix}, \qquad C\text{ - }NOT = \begin{pmatrix} 1 & 0 & 0 & 0 \\ 0 & 1 & 0 & 0 \\ 0 & 0 & 0 & 1 \\ 0 & 0 & 1 & 0 \end{pmatrix} \qquad (130)$$

and represented of quantum circuits as in Fig. 53. We also introduce a pure state $|\Psi\rangle$ (see Eq. 53).

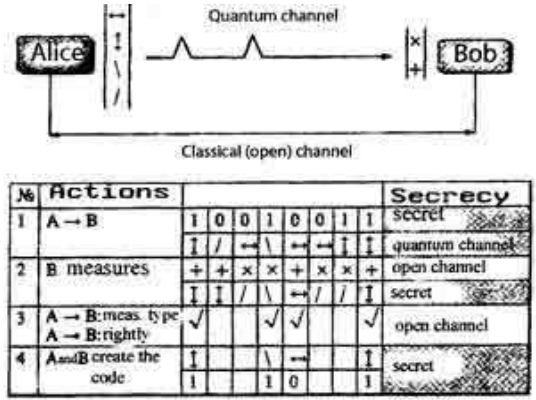

**Fig. 53. One example of the sequence actions for quantum cryptography using different polarized states of photons (after [154]).**

As discussed above, $|0\rangle$ represents the vacuum state for exciton while $|1\rangle$ represents a single exciton. As usual, we refer to two parties, Alice and Bob. Alice wants to teleport an arbitrary, unknown qubit state $|\Psi\rangle$ to Bob. Alice prepares two QDs (b and c) in the state $|0\rangle$ and then gives the state $|\Psi 00\rangle$ as input to the system. By performing the series of transformation, Bob receives as the output of the circuit the state $\frac{1}{\sqrt{2}}(|0\rangle + |1\rangle)_a \frac{1}{\sqrt{2}}(|0\rangle + |1\rangle)_b |\Psi\rangle_c$. Consider a system of three identical and equispaced QDs containing no net charge, which initially prepared inn the state $|\Psi\rangle_a |0\rangle_b |0\rangle_c$. Following this initialization, we illuminate QDs b and c with the radiation pulse $\xi(t) = A\exp(-i\omega t)$ with defining $\tau$. For a 0 or $2\pi$ – pulse, the density of probability for finding the QDs b and c in the Bell state $\frac{1}{\sqrt{2}}(|0\rangle + |1\rangle)$ requires indicated $\tau$ (see e.g. [189]). Hence, this time $\tau_{Bell}$ corresponds to the realization of the first two gates of the circuit in Fg. 53, i.e. the Hadamard transformation over QD b followed by the C - NOT gate between QDs b and c. After this, the information in qubit c is sent to Bob and Alice keeps in her memory the state of QS b. Next, we need to perform a C - Not operation between QDs a and b and, following that, a Hadamard transform over the QD a: this procedure then leaves the system in the state

$$\frac{1}{\sqrt{2}}\{|00\rangle(\alpha|0\rangle + |1\rangle) + |01\rangle(\beta|0\rangle + \alpha|1\rangle + |10\rangle(\alpha|0\rangle - |1\rangle) + |11\rangle(-\beta|0\rangle + |1\rangle)\}$$

. (131)

As we can seen from Eq. (27), we are proposing the realization of the Bell basis measurement in two steps: first we have rotated the Bell basis into the computational basis ($|00\rangle$, $|01\rangle$, $|10\rangle$, $|11\rangle$) by performing the unitary operations shown before dashed line in Fig. 52. Hence, the second step is to perform a measurement in this computational basis. The result of this measurement provides us with two classical bits of information, conditional the states measured by nanoprobing on QDs a and b. These classical bits are essential for completing the teleportation process: rewriting Eq. (131) as

$$\frac{1}{2}\{|00\rangle|\Psi\rangle + |01\rangle\sigma_x|\Psi\rangle + |10\rangle\sigma_z|\Psi\rangle + |11\rangle(-i\sigma_y|\Psi\rangle$$

(132)

we see that if, instead of performing the set of operations shown after the dashed line in Fig. 52., Bob performs one of the conditional unitary operations $I$, $\sigma_x$, $\sigma_z$ or ($-i\sigma_y$) over the QD c, the teleportation process is finished since the excitonic state $|\Psi\rangle$ has been teleported from dot a to dot c. For this reason only two unitary exclusive - or transformations are needed in order to teleport the state $|\Psi\rangle$. This final step can be verified by measuring directly the excitonic luminescence from dot c, which must correspond to the initial state of dot a. For instance, if the state to be teleported is $|\Psi\rangle \equiv |1\rangle$, the final measurement of the near - field luminescence spectrum of dot c must give an excitonic emission line of the same wavelength and intensity as the initial one for dot a.

### 4.5.6. Quantum cryptography.

Cryptology, the mathematical science of secret communications, has a long and distinguished history of military and diplomatic uses dating back to the ancient Greeks (see, e.g. [192 - 194]). It consists of cryptography, the art of codemaking and cryptoanalysis, the art of code-breaking. With the proliferation of the Internet and electronic mail, the importance of achieving secrecy in communication by cryptography [194, 195] - the art of using coded messages - is growing each day.

The two main goals of cryptography are for a sender and intended recipient to be able to communicate in a form that is unintelligible to third parties, and - second - for the authenication [193, 197] of messages to prove that they were not altered in transit. Both of these goals can be accomplished with provable security if sender and recipient are in possession of shared, secret "key" material. Thus key material, which is trust random number sequence, is a very valuable commodity even through it conveys no useful information itself. One of the principal problems of cryptography is therefore the so - called "key distribution problem". How do the sender and intended recipient come into possession of secret key material while being sure that third parties ("eavesdroppers") cannot acquire even partial information about it? It is provably impossible to establish a secret key with conventional communications, and so key distribution has relied on the establishment of a physically secure channel or the conditional security of "difficult" mathematical problems (see, e.g. [187, 195]) in public key cryptography. Amazingly, quantum mechanics has now provided the foundation [188, 194] stone to a new approach to cryptography - quantum cryptography. Namely, the quantum cryptography (QC) can solve many problems that are impossible from the perspective of conventional cryptography (for details see [196]).

QC was born in the late sixties when S. Wiesner [198] wrote "Conjugate Coding". Unfortunately, this highly innovative paper was unpublished at the time and it went mostly unnoticed. There, Wiesner explained how quantum physics could be used in principle to produce bank notes that would be impossible to counterfeit and how to implement what he called a "multiplexing channel" a notion strikingly similar to what Rabin [199] was to put forward more than ten years later under the name of "oblivious transfer" [200].

Later, Bennett and Brassard [201, 202] realized that instead of using single quanta for information storage they could be used for information transmission. In 1984 they published the first quantum cryptography protocol now known as "BB84" [201]. A further advance in theoretical quantum cryptography took place in 1991 when Ekert [203] proposed that EPR [177] entangled two - particle states could be used to implement a quantum cryptography protocol whose security was based on Bell's inequalities [178]. Also in 1991, Bennett and coauthors demonstrated the quantum key distribution (QKD) was potentially practical by constructing a working prototype system for the BB84 protocol, using polarized photons [195, 204, 205].

In 1992 Bennett published a "minimal" QKD scheme ("B92") and proposed that it could be implemented using single - photon interference with photons propagating for long distances over optical fibers [206]. After that, other QKD protocols have been published [129] and experiments were done in different countries (for details see [195, 204 - 207]).

QKD is a method in which quantum states are used to establish a random secret key for cryptography. The essential ideas are as follows: Alice and Bob are, as usual widely separated and wish to communicate (see also Fig. 53 ). Alice send to Bob 2n qubits, each prepared in one of the states $|0>, |1>, |+>, |->$, randomly chosen. As is well-known, many other methods are possible (see, e.g. [167, 188, 195]) we consider here this one merely to illustrate the concept of QC. Bob measures his received bits, choosing the measurement basis randomly between $[|0>, |1>]$ and $[|+>, |->]$. Next Alice and Bob inform each other publicly (i.e. anyone can listen in) of the base they used to prepare or measure each qubit. they find out on which occasions they by chance used the same basis, which happens on average half the time, and retain just those results. In the absence of errors or interference, they now share the same random string of n classical bits (they agree for example to associate $|0>$ and $|+>$ with 0; $|1>$ and $|->$ with 1. This classical bit string is often called the raw quantum transmission, RQT (for details see [188]).

So far nothing has been gained by using qubits. The important feature is, however, that it is impossible for anyone to learn Bob's measurement results by observing the qubits during performance, without leaving evidence of their presence [195]. The crudest way for an eavesdropper Eve to attempt to discover the key would be for her to intercept the qubits and measure them, then pass them on to Bob. On average half the time Eve guesses Alice's basis correctly and thus does not disturb the qubit. However, Eve's correct guesses do not coincide with Bob's, so Eve learns the state of half of the n qubits which Alice and Bob later to trust, and disturbs the other half, for example sending to Bob $|+>$ for Alice's $|0>$. Half of those disturb will be projected by Bob's measurement back onto the original state sent by Alice, so overall Eve corrupts n/4 bits of the RQT. Alice and Bob can now detect Eve's presence simply by randomly choosing n/2 bits of the RQT and announcing publicly the values they have. If they agree on all these bits, then they can trust no eavesdropper was present, since the probability that Eve was present and they happened to choose n/2 uncorrupted bits is $(3/4)^{n/2} \simeq 10^{-125}$ for n = 1000. The n/2 bits form the secret key. From this picture, we see, that Alice and Bob do not use the quantum channel (Fig. 53) to transmit information, but only to transmit a random sequence of bits, i.e. key. Now if the key is unperturbed, then quantum physics guarantees that no one has gotten any information about this key by eavesdropping, i.e. measuring, the quantum communication channel. In this case, Alice and Bob can safely use this key to encode messages. In conclusion of this part we note that the authors of paper [205] performed successfully quantum key exchange over different installed cables, the longest connecting the cities of Lausanne and Geneva (see Fig. 54).

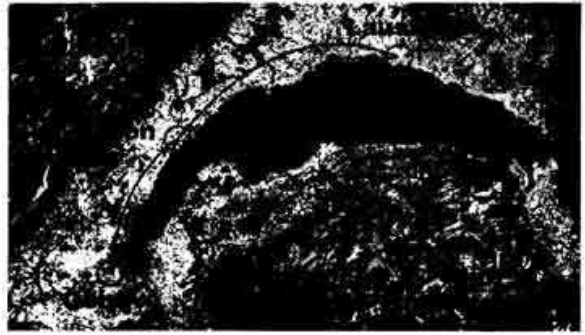

**Fig. 54. Sattelite view of Lake Geneva with the cities of Geneva and Nyon and Lausanne (after [205]).**

**4.6. Quantum computers.**

4.6.1. Introduction.

In quantum mechanics there are some basic principles: such as the correspondence principle, Heisenberg's uncertainty principle, or Pauli's principle, that encode the fundamentals of that theory. The knowledge of those principles provides us with the essential understanding of a quantum mechanics at a glance, without going into the complete formalism of that subject (see, e.g. [208]. A similar thing happens with other areas n physics. In computer science there are guiding principles for the architecture of a computer (hardware) and the programs to be run (software). Likewise, in quantum computing we have seen that there are basic principles associated with the ideas of quantum parallelism (superposition principle) and quantum programming (constructive interference). By principles of quantum computation we mean those rules that are specific to the act of computing according to the laws of quantum mechanics. As was mentioned above, that the quantum version of parallelism is realized through the superposition principle of quantum mechanical amplitudes, likewise the act of programming a quantum computer should be closely related to a constructive interference of those amplitudes involved in the superposition of quantum states in the register of the computer (for details see [209]).

A key step towards the realization of the practical quantum computer is to decouple its functioning into the simplest possible primitive operations or gates (see, also [210]). A universal gate such as NAND (in classical computers) operates locally on a very reduced number of bits (actually two). However, by combining NAND gates in the appropriate number a sequence we can carry out arbitrary computations on arbitrary many bits. This was very useful in practice for it allowed device, leaving the rest to the circuit designer. The same rationale applies to quantum circuits. When a quantum computer is working it is a unitary evolution operator that is effecting a predetermined action on a series of qubits. These qubits form the memory register of the machine, or a quantum register. A quantum register is a string of qubits with a predetermined finite length. The space of all the possible register states makes up the Hilbert space of states associates with the quantum computer. A quantum memory register can store multiple sequences of classical bits in superposition. this is a manifestation of quantum parallelism. A quantum logic gate is a unitary operator acting on the states of a certain set of qubits (see, also Fig. 55A). If the number of such qubits is n, the quantum gate is represented by a $2^n \times 2^n$ matrix in the unitary group $U(2^n)$. It is thus a reversible gate: we can reverse the action, thereby receiving the initial quantum state from final one [163, 210]. One-qubits are the simplest possible gates because they take one input qubit and transform it into one output qubit. The quantum NOT gate is a one-qubit gate (Fig. 55A). Its unitary evolution operator $U_{NOT}$ is [210]

$$U_{NOT} = \begin{bmatrix} 0 & 1 \\ 1 & 0 \end{bmatrix}. \qquad (133)$$

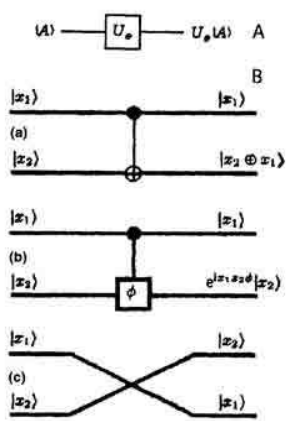

**Fig. 55. A - Example of unitary gate; B - quantum binary gates (a) CNOT gate, (b) CPHASE gate, (c) SWAP gate (after [154]).**

The truth table representing this gate can be found in [154, 188, 209]. It can be see that this quantum NOT gate coincides with the classical counterpart. However, there is a basic underlying difference:the quantum gate acts on qubits while the classical gate operates on bits. This difference allows us to introduce a truly one-qubit gate: the $\sqrt{NOT}$ gate. Its matrix representation is

$$U_{\sqrt{NOT}} = \frac{1}{\sqrt{2}} e^{i\pi/4} \left(1 - i\sigma_x\right). \qquad (134)$$

$$U_{\sqrt{NOT}} U_{\sqrt{NOT}} = \begin{bmatrix} \frac{1+i}{2} & \frac{1-i}{2} \\ \frac{1-i}{2} & \frac{1+i}{2} \end{bmatrix} \begin{bmatrix} \frac{1+i}{2} & \frac{1-i}{2} \\ \frac{1-i}{2} & \frac{1+i}{2} \end{bmatrix} = \begin{bmatrix} 0 & 1 \\ 1 & 0 \end{bmatrix} = U_{NOT}.$$

(135)

This gate has no counterpart in classical computers since it implements nontrivial superposition of basic set. Another one-qubit gate without analog in classical circuity and heavily used in quantum computers is the so-called Hadamard (H) gate [181, 211]. This gate is defined as:

$$U_H = \frac{1}{\sqrt{2}} \begin{bmatrix} 1 & 1 \\ 1 & -1 \end{bmatrix}. \qquad (136)$$

the XOR (exclusive - OR), or CNOT (controlled - NOT) gate is an example of a quantum logic gate on two qubits (see, also [212]). It is instructive to give the unitary action $U_{XOR,\ CNOT}$ of this gate in several form [165]. Its action on the two - qubit basis states is

$U_{CNOT} |00> = |00>; U_{CNOT} |10> = |11>; U_{CNOT} |01> = |01>; U_{CNOT} |11> = |10>;$ (137)

From this definition we can see that the name of this gate is quite apparent, for it means that it executes a NOT operation on the second qubit conditioned to have the first qubit in the state $|1>$. Its matrix representation is

$$U_{CNOT} = U_{XOR} = \begin{bmatrix} 1 & 0 & 0 & 0 \\ 0 & 1 & 0 & 0 \\ 0 & 0 & 0 & 1 \\ 0 & 0 & 1 & 0 \end{bmatrix}. \qquad (138)$$

The action of the CNOT operator (137) immediately translates into a corresponding truth table. The diagrammatic representation of the CNOT gate is shown in Fig. 55B. We shall see how this quantum CNOT gate plays a paramount role in both the theory and experimental realization of quantum computers. It allows the implementation of conditional logic at a quantum level.

Unlike the CNOT gate, there two-qubit gates with no classical analog (see, also [174]). One example is the controlled - phase gate or CPHASE:

$$U_{CPHASE} = \begin{bmatrix} 1 & 0 & 0 & 0 \\ 0 & 1 & 0 & 0 \\ 0 & 0 & 1 & 0 \\ 0 & 0 & 0 & e^{i\Phi} \end{bmatrix}. \qquad (139)$$

It implements a conditional phase shift on the second qubit [209]. Other interesting two-qubit gates are the SWAP gate, which interchanges the states of the two-qubits, and the $\sqrt{SWAP}$ gate, whose matrix representation are

$$U_{SWAP} = \begin{bmatrix} 1 & 0 & 0 & 0 \\ 0 & 1 & 0 & 0 \\ 0 & 0 & 1 & 0 \\ 0 & 0 & 0 & 1 \end{bmatrix}, \qquad (140)$$

$$U_{\sqrt{SWAP}} = \begin{bmatrix} 1 & 0 & 0 & 0 \\ 0 & \frac{1+i}{2} & \frac{1-i}{2} & 0 \\ 0 & \frac{1-i}{2} & \frac{1+i}{2} & 0 \\ 0 & 0 & 0 & 0 \end{bmatrix}. \qquad (141)$$

An immediate extension of the CNOT construction to three qubits yields the CCNOT gate (or $C^2$NOT - controlled - controlled - not gate) which is also called Toffoli gate [215,] (see, also [213, 214]). The Deutsch gate $D(\theta)$ is also an important three - qubit gate. It is a controlled - controlled - S or $C^2$S operation, where

$$\mathsf{U}_{D(\theta)} = \mathsf{i}e^{-\frac{\theta\sigma_x}{2}} = \mathsf{i}\cos\frac{\theta}{2} + \sigma_x\sin\frac{\theta}{2} \qquad (142)$$

is a unitary operation that rotates a qubit about the X axis by an angle $\theta$ and then multiplies it by a factor i and $\sigma_x$. Here $\sigma_x$ is the Pauli matrix

$$\sigma_x = \begin{bmatrix} 0 & 1 \\ 1 & 0 \end{bmatrix}. \qquad (143)$$

Examples of multi- qubit gates can be found in the references [211 - 217].

6.2. Isotope - Based Quantum Computers.

The development of efficient quantum algorithms for classically hard problems has generated interest in the construction of a quantum computer. A quantum computer uses superpositions of all possible input states. By exploiting this quantum parallelism, certain algorithms allow one to factorize [180] large integers with astounding speed, and rapidly search through large databases [182], and efficiently simulate quantum systems [176]. In the nearer term such devices could facilitate secure communication and distributed computing. In any physical system, bit errors will occur during the computation. In quantum computing this is particularly catastrophic, because the errors cause decoherence [154] and can destroy the delicate superposition that needs to be preserved throughout the computation. With the discovery of quantum error correction [180, 218] and fualt-tolerant computing, in which these errors are continuously corrected without destroying the quantum information, the construction of a real computer has became a distinct possibility. The task that lie ahead to create an actual quantum computer are formidable: Preskill [213] has estimated that a quantum computer operating on $10^6$ qubits with a $10^{-6}$ probability of error in each operation would exceed the capabilities of contemporary conventional computers on the prime factorization problem. To make use of error-correcting codes, logical operations and measurement must be able to proceed in parallel on qubits throughout the computer.

Phosphorous donors in silicon present a unique opportunity for solid - state quantum computation [219]. Electrons spins on isolated Si:P donors have very long decoherence times of ~ 60 ms in isotopically purified $^{28}$Si at 7 K [220]. By contrast, electron spin dephasing times in GaAs (for example) are orders - of - magnitude shorter due spin - orbit interaction; and the background nuclear spins of the III - V host lattice cannot be eliminated by isotope selection. Finally, the Si:P donor is a self - confined, perfectly uniform single - electron quantum dot with a non - degenerate ground state. A strong Coulomb potential breaks the 6 - valley degeneracy of the silicon conduction band near donor site, yielding a substantial energy gap of ~ 15 meV to the lowest excited [221] as needed for quantum computation. As we all know, the Si:$^{31}$P system was exhaustively studied more than 40 years ago in the first electron - nuclear double - resonance experiments. At sufficiently low $^{31}$P concentrations at temperature T = 1.5 K, the electron spin relaxation time is thousands of seconds and the $^{31}$P nuclear spin relaxation time exceeds 10 hours. It is likely that at millikelvin temperatures the phonon limited $^{31}$P relaxation time is of the order of $10^{18}$ seconds [222], making, as we said above, this system ideal for quantum computation.

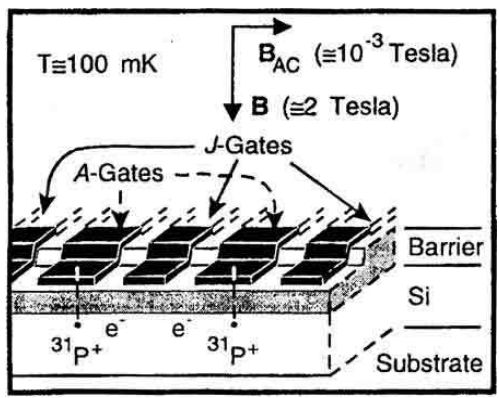

**Fig. 56. Illustration of two cells in a one-dimensional array containing $^{31}$P donors and electrons in a Si host, separated by a barrier from metal gates on the surface. "A gates" control the resonance frequency of the nuclear spin qubits; "J gates" control the electron-mediated coupling between adjacent nuclear spins. The ledge over which the gates cross localizes the gate electric field in the vicinity of the donors (after [219]).**

Kane's original proposal [99] envisions encoding quantum information onto the nuclear spin 1/2 states of $^{31}$P qubits in a spinless I = 0 $^{28}$Si lattice. The Kane architecture employs an array of top - gates (see Fig. 56). to manipulate the ground state wavefunctions of the spin - polarized electrons at each donor site in a high magnetic field B ~ 2 T, at very low temperature (T ≃ 100 mK). "A -gates" above each donor turn single - qubit NMR rotations via the contact hyperfine interaction; and "J - gates" between them induce an indirect two - qubit nuclear exchange interaction via overlap of the spin - polarized electron wavefunctions. In other words, spin - 1/2 $^{31}$P donor nuclei are qubits, while donor electrons together with external A - gates provide single - qubit (using external magnetic field) and two - qubit operations (using hyperfine and electron exchange interactions). Specifically, the single. donor nuclear spin splitting is given by [219]

$$\hbar\omega_A = 2g_n\mu_n B + 2A + \frac{2A^2}{\mu_B B}, \qquad (144)$$

where $g_n$ is the nuclear spin g - factor (= 1.13 for $^{31}$P [219]), $\mu_n$ is the nuclear magneton, A is the strength of the hyperfine coupling between the $^{31}$P nucleus and the donor electron spin, and B is the applied magnetic field. It's clear that by changing A one can effectively change the nuclear spin splitting, thus allow resonant manipulations of individual nuclear spins (Fig. 57).

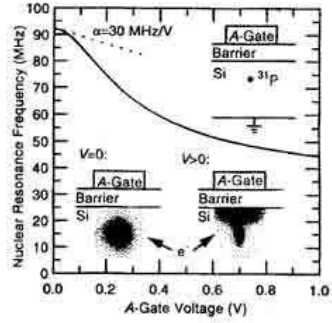

**Fig. 57. An electric field applied to an A gate pulls the electron wavefunction away from the donor and towards the barrier, reducing the hyperfine interaction and the resonance frequency of the nucleus. The donor nucleus-electron system is a voltage-controlled oscillator with a tuning parameter $\alpha$ of the order 30 MHz (after [219]).**

If the donor electrons of two nearby donors are allowed to overlap, the interaction part of the spin Hamiltonian for the two electrons and the two nuclei include electron - nuclear hyperfine coupling and electron - electron exchange coupling [219](see also [223, 224]

$$H = H_{Zeeman} + H_{int} = H_{Zeeman} + A_1 \vec{S_1} \cdot \vec{I_1} + A_2 \vec{S_2} \cdot \vec{I_2} + J\vec{S_1}\vec{S_2}, \qquad (145)$$

where $\vec{S_1}$ and $\vec{S_2}$ represent the two electron spins, $\vec{I_1}$ and $\vec{I_2}$ are the two nuclear spins, $A_1$ and $A_2$ represent the hyperfine coupling strength at the two donor sites, and J is the exchange coupling strength between the two donor electrons, which is determined by the overlap of the donor electron wavefunctions. The lowest order perturbation calculation (assuming $A_1 = A_2 = A$ and J is much smaller than the electron Zeeman splitting) results in an effective exchange coupling between the two nuclei and the coupling strength is (see [219])

$$J_{nn} = \frac{4A^2 J}{\mu_B B (\mu_B B - 2J)}. \qquad (146)$$

Now the two donor electrons are essentially shuttles different nuclear spin qubits and are controlled by external gate voltages. The final measurement is done by first transferring nuclear spin information into electron spins using hyperfine interaction, then converting electron spin information into charge states such as charge locations [223]. A significant advantage of silicon is that its most abundant isotope $^{28}$Si is spinless, thus providing a "quiet" environment for the donor nuclear spin qubits. In addition, Si has also smaller intrinsic spin - orbit coupling than other popular semiconductors such as GaAs. In general, nuclear spins have very long coherence times because they do not strongly couple with their environment, and are thus good candidates for qubits (see, also [129, 224, 222]).

Although the nuclear spin offers unlimited decoherence times for quantum information processing, the technical problems of dealing with nuclear spins through the electrons are exceedingly difficult. A modified versions of the Kane architecture was soon proposed using the spin of the donor electron as the qubit [225, 225[a], 226]. In the first scheme [225], A - gates would modulate the electron g - factor by polarizing its ground state into Ge - rich regions of a SiGe hetero - structure for selective ESR rotations, while two - qubit electron exchange is induced through wavefunction overlap. In the studies of Shlimak et al. [226] was used the new technology to growth of SIGE

hetero - -structures. Recent achievement in Si/Ge technology allow one to obtain high quality heterojunctions with a mobility of about $(1 - 5) \cdot 10^5$ cm$^2$V$^{-1}$s$^{-1}$ [227]. Using Si/Ge hetero - structures has several advantages concerning semiconductor based nuclear spin quantum computers (S - NSQCs). First, the concentration of nuclear spins in Ge and Si crystals is much lower, because only one isotope ($^{73}$Ge and $^{29}$Si [39]) has a nuclear spin, and the natural abundance of this isotope is small (see, also [21]). Second the variation of isotopic composition for Ge and Si will lead to the creation of a material with a controlled concentration of nuclear spin, and even without nuclear spins. Utilization of isotopically engineered Ge and Si elements in the growth of the active Si/Ge layers could help realize an almost zero nuclear spin layer that is coplanar with the 2DEG. Then, one might deliberately vary the isotopic composition to produce layers, wires and dots that could serve as nuclear spin qubits with a controlled number of nuclear spins (see also [129]).

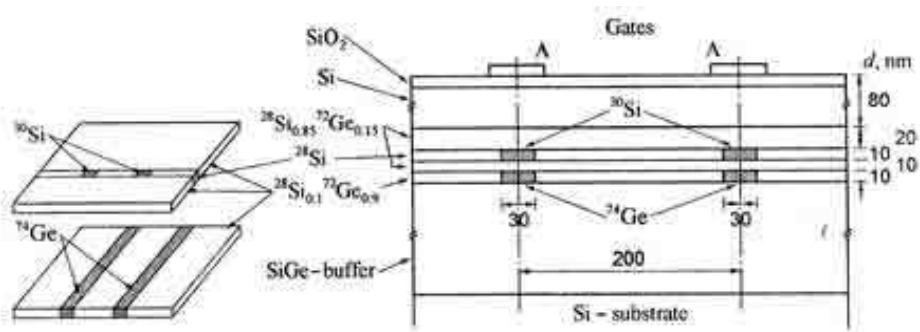

**Fig. 58. Schematics of the proposed device. After NTD, $^{31}$P donors appear only inside the $^{30}$Si - spots and underlying $^{74}$Ge strips will be heavily doped with $^{75}$As donors. All sizes are shown in nm (after [226]).**

The key point of a novel technology is the growth of the central Si and barrier Si$_{0.85}$Ge$_{0.15}$ layers from different isotopes: the Si$_{0.85}$Ge$_{0.15}$ layers from isotope $^{28}$Si and $^{72}$Ge and the central Si layer from isotope $^{28}$Si with $^{30}$Si spots introduced by means of the nano - litography (see Fig. 58) (see [228]). the formation of quasi - one - dimensional Si wires will be achieved in a subsequent operation by the etching of Si layer between wires and the filling of the resulting gaps by the Si$_{0.85}$Ge$_{0.15}$ barrier composed from isotopes $^{28}$Si and $^{72}$Ge. Because different isotopes of Si and Ge are chemically identical, this technology guarantees the high quality of the grown structures [107]. After preparation, these structures will be irradiated with a neutron flux in a nuclear reactor by the fast annealing of radiation damage (see [129] and references cited therein).

As was shown by Di Vincenzo [229] that two - bit gates applied to a pair of electron or nuclear spins are universal for the verification of all principles of quantum computation. Because direct overlap of wavefunctions for electrons localized on P donors is negligible for distant pairs, the authors of paper [226] proposed another principle of coupling based on the placement of qubits at fixed positions in a quasi - one - dimensional Si nanowire and using the indirect interaction of $^{31}$P nuclear spins with spins of electrons localized in the nanowire which they called as "1D - electrons". This interaction depends on the amplitude of the wavefunction of the "1D - electron" estimated at the position of the given donor nucleus $\Psi_n(r_i)$ and can be controlled by the change in the number of "1D - electrons" N in the wire. At N = 0, the interqubit coupling is totally suppressed, each $^{31}$P nuclear spin interact only with its own donor electron. This situation is analogous to that one suggested in the Kane proposal [223, 224] and therefore all single - qubit operations and estimates of the decoherence time are valid also in the model by Shlimak et al. [226].

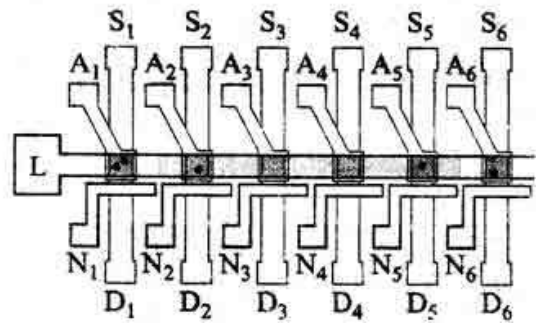

**Fig. 59. Schematics of a $^{28}$Si nanovire L with an array of $^{30}$Si spots (qubits and non - qubits after NTD). Each spot is supplied by overlying A - gate, underlying source - drain - channel and lateral N - gate. This device architecture allows to realize an indirect coupling between any distant qubits (for details see text) (after [226]).**

Below we briefly analyze the schematics of the device architecture which satisfy the scalability requirements of the quantum computer suggested in paper [226]. Fig. 59 shows schematics of the device architecture which allows one to vary l (length of quantum wire) and N. The device consists of a $^{28}$Si nanowire with an array of $^{30}$Si spots. Each spot is supplied by the overlying A - gate, the underlying Source - drain - channel and the lateral N - gate. After NTD, P donors will appear in most of the spots (which transforms these spots into qubits) and not appear in other spots (non - qubits). On Fig. 4 it is assumed that the spots 3 and 4 are non - qubits (0 - spots) and one need to provide coupling between qubits 2 and 5. For this purpose, it is necessary to connect the gates $N_2$, $N_3$, $N_4$ and $N_5$. The negative voltage applied between other N - gates and the wire contact L will lead to pressing - out "1D - electrons" from all corresponding areas and formation of the nanowire with l = 800 nm between the sites 2 and 5 only (shown in grey in Fig. 59). The coupling between qubits 2 and 5 will be realized via injection in the wire of the necessary number of electrons N, using the positive voltage applied to the gates $N_2$ - $N_5$. According [226], the maximal coupling will be realized at N = 7, while at N = 0, the coupling will be totally suppressed.

### 7. Conclusions.

To summarize, we have shown that semiconductor nanostructures (quantum dots in a different materials, different size and shape, and different kind of preparation) can be exploited in order realize all - optical quantum entanglement schemes, even in the presence of noisy environments. A scheme for quantum teleportation as well as computation of excitonic states in isotope - mixed crystals has also been proposed.

Concluding our paper we simply have listed the main parts of new scientific direction in the nanotechnology - isotoptronics. This direction is the further step of nanotechnology, because the size of the devices is $10^{-10}$ m (see, also [10]). The main parts of isotoptronics are:

1. Human health.
2. Neutron transmutation doping semiconductor and other materials.
3. Optical fiber.
4. Quantum low - sized structures (wells, wires, dots) in different materials including isotope - mixed ones.
5. Processors for quantum computers.
6. Isotope memory (including organic world).
7. Problem of the mass (elementary particles and cosmology).
8. Geochronology.

As we could see from this list, such wide applications of isotopes (stables and radioactive), made isotoptronics not only popular in scientific society but also very useful in different branches of human life.

**8. Acknowledgements.**

I would like to express my deep thanks to many authors and publishers whose Figures and Tables used in my review. Many thanks are due to Prof. W. Neder for carefully reading of my manuscript as well as Dr. P. Knight for improving my English. I wish to express my deep gratitude my family for a patience during long preparation of this review.

**Table 1**. Important times for various two - level systems in quantum mechanics that might be used as qubits, including prospective qubits ranging from nuclear physics, through atomic, electronic, and photonic systems, to electron and nuclear spins. The time $t_{switch}$ is the minimum time required to execute one quantum gate; it is estimated as $\hbar/\Delta E$, where $\Delta E$ is the typical energy splitting in the two level system; the duration of a $\pi$ tipping pulse cannot be shorter than this uncertainty time for each system. The phase coherence time as seen experimentally, $t_\Phi$, is the upper bound on the length of time over which a complete quantum computation can be executed accurately. The ratio of these two times gives the largest number of steps permitted in a quantum computation using these quantum bits (after [154]).

| Quantum system | $t_{switch}$, s | $t_\Phi$, s | Ratio |
|---|---|---|---|
| Mössbauer nucleus | $10^{-19}$ | $10^{-10}$ | $10^9$ |
| Electrons: GaAs | $10^{-13}$ | $10^{-10}$ | $10^3$ |
| Electrons: Au | $10^{-14}$ | $10^{-8}$ | $10^6$ |
| Trapped ions: In | $10^{14}$ | $10^{-1}$ | $10^{13}$ |
| Optical microcavity | $10^{-14}$ | $10^{-5}$ | $10^9$ |
| Electron spin | $10^{-7}$ | $10^{-3}$ | $10^4$ |
| Electron quantum dot | $10^{-6}$ | $10^{-3}$ | $10^3$ |
| Nuclear spin | $10^{-3}$ | $10^4$ | $10^7$ |

**Table 2.** Truth table defining the operation of some simple logic gates. Each row shows two input values A and B and the corresponding output values for gates AND, OR, and XOR. The output for for the NOT is shown only for input B.

| A | B | AND | OR | XOR | NOT B |
|---|---|-----|----|----|-------|
| 0 | 0 | 0   | 0  | 0  | 1     |
| 0 | 1 | 0   | 1  | 1  | 0     |
| 1 | 0 | 0   | 1  | 1  | 1     |
| 1 | 1 | 1   | 1  | 0  | 0     |

**6. References.**

**Figure captions.**

Fig. 1. Optical and acoustic modes. The optical modes lie at higher frequencies and show less dispersion than the acoustic modes (for details see text).

Fig. 2. a - First - order Raman scattering spectra Ge with different isotope contents [34] and b - First - order Raman scattering in isotopically mixed diamond crystals $^{12}C_x^{13}C_{1-x}$. The peaks A, B, C, D, E and F correspond to x = 0.989; 0.90; 0.60; 0.50; 0.30 and 0.001 (after [38]).

Fig. 3. Second - order Raman scattering spectra in synthetic diamond with different isotope concentration at room temperature (after [42]).

Fig. 4. Second - order Raman scattering spectra in the isotopically mixed crystals $LiH_xD_{1-x}$ at room temperature. 1 - x = 0; 2 - 0.42; 3 - 0.76; 4 - 1. The arrows point out a shift of LO(Γ) phonons in the mixed crystals  (after [40]).

Fig. 5. The dependence of $\ln(\delta\%) \sim f[\ln(\frac{\partial\omega}{\partial M})]$: points are experimental values and continuous line - calculation on the formulas (23) (after [51]).

Fig. 6. Mirror reflection spectra of crystals: 1 - LiH; 2 - $LiH_xD_{1-x}$; 3 - LiD; at 4.2 K. 4 - source of light without crystal. Spectral resolution of the instrument is indicated on the diagram (after [55]).

Fig. 7. a - Reflection spectra in the A and B excitonic polaritons region of $Cd^{nat}S$ and $Cd^{34}S$ at 1.3K with incident light in the $\vec{E} \perp \vec{C}$. The broken vertical lines connecting peaks indicate measured enrgy shifts reported in Table 18. In this polarization, the n = 2 and 3 excited  states of the A exciton, and the n = 2 excited state of the B exciton, can be observed. b - Polarized photoluminescence spectra in the region of the $A_{n=2}$ and $A_{n=3}$ free exciton recombination lines of $Cd^{nat}S$ and $Cd^{34}S$ taken at 1.3K with the $\vec{E} \perp \vec{C}$. The broken vertical lines connecting peaks indicate measured enrgy shifts reported in Table 18 (after [63]).

Fig. 8. a -Signatures of the $E_0$' and $E_1$ excitonic band gaps of $^{28}Si$ observed (dots) in photomodulated reflectivity. The solid line is a theoretical fit using the excitonic line shape. b - Photomodulated reflectivity spectra of isotopically enriched Si exhibiting isotopic shifts of the $E_0$' and $E_1$ gaps (after [67]).

Fig. 9. A - Photoluminescence (PL) and wavelength - modulated transmission (WMT) spectra of isotopically enriched $^{30}Si$ recorded at 20K ; B - The excitonic indirect band gap and the associated phonon energies as a function of M (after [67]).

Fig. 10. The MBE method for the growth of heterostructures.

Fig. 11. A scheme of the growth cell of an MOCVD method.

Fig. 12. Overview of different litography method.

Fig. 13. Scheme of plasma etching.

Fig. 14. (a) Photoplumenescence spectra for wells of different thickness $W_L$ and (b) the photoluminescence excitation spectra from GaAs quantum wells (after [16]).

Fig. 15. Cross - sectional STM: (a) an image of a stack of InAs islands in GaAs; (b) the lattice parameter in the growth direction in an InAs island (the experimental data were obtained from cross - sectional STM; the solid line is from a simulation assuming an In content increasing from island base to island apex);(c) comparison between measured and simulated height profiles for a similar sample; and (d) the electronic wavefunction measured at two different tip biases, compared with simulation for the ground and the first excited states (after [5]).

Fig. 16. PbSe islands with {001} facets; (a) AFM image of the top surface of a PbSe/PbEuTe island multilayer grown on $BaF_2$; (b) the autocarrelation function. Islands are arranged in a regular array up to the sixth - nearest neighbor (after [5]).

Fig. 17. Schematic diagrams of SEM and TEM.

Fig. 18. Strain distribution obtained from the TEM images of InGaAS islands in GaAs by using the method of digital analysis of lattice images (aftetr [5].

Fig. 19. Schematics of Si isotope superlattices. Thickness of each isotope layer are 1.1; 1.6; and 3.2 for $^{28}Si_8/^{30}Si_8$; $^{28}Si_{12}/^{30}Si_{12}$ and $^{28}Si_{24}/^{30}Si_{24}$ samples, respectively. Low index denotes the thickness of each isotope layer in atomic monolayers, each 0.136 nm thick (after [94]).

Fig. 20. Raman spectra of the $^{28}Si_n/^{30}Si_n$ samples with n = 8, 12 and 24 (after [94]).

Fig. 21. (a) Experimental Raman spectra of a $(^{70}Ge)_{16}(^{74}Ge)_{16}$ superlattice for different annealing steps at $500^0$ C. (b) Calculated Raman spectra for the same superlattice using the same parameters (after [91]).

Fig. 22. Observing the wave - like properties of the electrons in the double slitexperiment (after [99]). The pictures (a) →(c) have been taken at various times: picturees on the monitor after (a) 10 electrons, (b) 200 electrons, (c) 6000 electrons, and (d) 140 000 electrons. Electrons were emitted at a rate of 10 per second. (after A. Tonomura, 2006, Double - slit experiment (http://hqrd.hitachi.co.jp/globaldoubleslit.cfm)).

Fig. 23. Normal structures with different dimensions: normal solid - state body, quantum well, quantum wire, quantum dot. Additionally their DOS are illustrated.

Fig. 24. (a) Scheme of a quantum system to observe Coulumb blickade effects; (b) I - V characteristics in a quantum dot showing the Coulomb blockade effect.

Fig. 25. Charging of a quantum dot capacitor as a function of voltage, in normalized coordinates.

Fig. 26. Photocurrent resonance for various excitation wavelengeths bias voltage. At low bias the fine structure splitting is fully resolved, at higher bias the linewidth is increased due to fast tunneling (after [109]).

Fig. 27. Variation of the binding energy of the ground state $E_{1s}$ of the heavy - hole exciton (solid lines) and the light - hole (dashed lines) as a function of the GaAs quantum well size (L) for aluminium concentration x = 0.15 and x = 0.3 and for infinite potential wells (after [111]).

Fig. 28. Exciton binding energy in an infinitely deep CdTe quantum well (after [10]).

Fig. 29. Bohr radius of two - dimensional exciton in an infinitely deep CdTe quantum well (after [10]).

Fig. 30. The confinement energy in a finite barrier circular cross - section quantum wire (after [10]).

Fig. 31. The radial component of the wave function $\Psi(r)$ for the lowest two eigenstates in a finite - barrier quantum wire with radius 300 Å of circular cross - section (after [10]).

Fig. 32. The confinement energy in a spherical GaAs quantum dot surrounded by a $Ga_{0.8}Al_{0.2}As$ barrier (after [10]).

Fig. 33. The wave functions of the three lowest energy states in the 300 Å spherical quantum dot (after [10]).

Fig. 34. Spectrally resolved four - wave mixing at $\tau = 3$ ps showing the heavy hole and light hole biexcitons. Insert shows the four - wave mixing intensity of the heavy hole exciton and biexciton as a function of delay (after [131]).

Fig. 35. Left side: Excitonic (X) and biexcitonic ($X_2$) emission from two individual CdSe/ZnSe SQDs for different excitation powers. The PL spectra shown in the lower panel are unpolarized, the data presented in the upper panel represent linearly polarized PL spectra ($\pi_x$ and $\pi_y$, respectively). Right side: Energy lvel schem for the biexciton - exciton cascade in a QD (after [132]).

Fig. 36. Left panel: Transient PL spectra from a single CdSe/ZnSe QD showing the single exciton X and the biexciton transition (here denoted by B = $X_2$). Right panel: Decay curves for the exciton and the biexciton PL signal (for details see text) (after [132]).

Fig. 37. Schematic representation of the conduction band of a resonant tunnel diode:(a) with no valtage, (b - d) for increasing applied voltage, (c) - current - voltage characteristic.

Fig. 38. A schematic diagram of a Si - FET with nanowire, the metal source, and drain electrodes on the surface of a $SiO_2$/Si substrate (after [100]).

Fig. 39. A scanning electron microscope image of a single electron transistor (after [141]).

Fig. 40. Conductance as a function of $V_g$ for two samples with the same geometry (after [139]).

Fig. 41. The spectra of light - emitting semiconductor diodes with different bandgaps (after [146]).

Fig. 42. (a) Photoluminescence spectra at 10 K of the QWr laser sample above, below and near the lasing threshold in TE - polarization. (b) Dependence on input excitation power of the PL output power;arrows indicate the excitation powers used for the optical spectra depicted in (a). (c) High resolution emission spectrum above the lasing threshold showing the Fabri - Perrot modes of the optical cavity (after [147]).

Fig. 43. Linearly - polarized PLE spectrum and the corresponding PL spectrum of an etched QWr laser sample at 10 K. The polarization of the excitation is parallel tothe wire axis. The different optical transition $e_n$ - $h_n$ are marked by arrows (after [147]).

Fig. 44. Evolution of threshold current density for lasers based on different confinement structures (after [149]).

Fig. 45. Shannon's average information or entropy H as a function of the probability p of one of the final states of a binary (two state) devices. H is measure of the uncertainty

before any final state occured and expresses the average amount of information to be gained after the determination of the outcome. A maximum uncertainty of one bit (or maximum gain information, once the result is known) exists when the two final states are equiprobable (p = 0.5). The dotted curve represents (1 - H) (see Eq. 113) an objective measure of the "prior knowledge" before operating the device (after [13]).

Fig. 46. The Bloch sphere of the Hilbert space spanned by $|\uparrow_z\rangle$ and $|\downarrow_z\rangle$ (after [13]).

Fig. 47. Model for a single QD. $|11\rangle$, $|01\rangle$, $|10\rangle$ and $|00\rangle$ denote the biexciton, the exciton and the ground states, respectively. $\Delta E$ is the biexciton binding energy. The optical rules for various transition are indicated.

Fig. 48. An entangled state involving two polarized excitons confined in a single dots was created and detected optically as evidenced by quantum beats between states $|01\rangle$ and $|10\rangle$ shown. The quantum coherence time ($\simeq$ 40 psec.) between these two states is directly extracted from the decay of the envelope (after [189]).

Fig. 49. Quantum teleportation requires quantum entanglement as a resource. In this case, Alice receives a qubit in an unknown state, and destroys it by performing a Bell measurement on that qubit and a member of an entangled pair of qubits that she shares with Bob. She sends a two-bit classical message (her mesurement outcome) to Bob, who then performs a unitary transformation on his member of the pair to reconstruct a perfect replica of the unknown state. We could see one qubit suffices to carry two classical bits of information.

Fig. 50. Circuit for quantum teleportation. The measurement is in the computational basis, leaving the measurement result stored in the data and ancila qubits. $R_y$ and $R_{-y}$ denote rotations of 90 degrees about the y and -y axes on the Bloch sphere (after [129]).

Fig. 51. Schematic representation of (a) the Hadamard gate, and (b) the C - NOT gate.

Fig. 52. Cicuit scheme to teleport un unknown quantum state of exciton from Alice to Bob using arrangement of 3 qubits (coupled quantum dots).

Fig. 53. One example of the sequence actions for quantum cryptography using different polarized states of photons (after [154]).

Fig. 54. Sattelite view of Lake Geneva with the cities of Geneva and Nyon and Lausanne (after [205]).

Fig. 55. A - Example of unitary gate; B - quantum binary gates (a) CNOT gate, (b) CPHASE gate, (c) SWAP gate (after [154]).

Fig. 56. Illustration of two cells in a one-dimensional array containing $^{31}$P donors and electrons in a Si host, separated by a barrier from metal gates on the surface. "A gates" control the resonance frequency of the nuclear spin qubits; "J gates" control the electron-mediated coupling between adjacent nuclear spins. The ledge over which the gates cross localizes the gate electric field in the vicinity of the donors (after [219]).

Fig. 57. An electric field applied to an A gate pulls the electron wavefunction away from the donor and towards the barrier, reducing the hyperfine interaction and the resonance frequency of the nucleus. The donor nucleus-electron system is a voltage-controlled oscillator with a tuning parameter $\alpha$ of the order 30 MHz (after [219]).

Fig. 58. Schematics of the proposed device. After NTD, $^{31}$P donors appear only inside the $^{30}$Si - spots and underlying $^{74}$Ge strips will be heavily doped with $^{75}$As donors. All sizes are shown in nm (after [226]).

Fig. 59. Schematics of a $^{28}$Si nanovire L with an array of $^{30}$Si spots (qubits and non - qubits after NTD). Each spot is supplied by overlying A - gate, underlying source - drain - channel and lateral N - gate. This device architecture allows to realize an indirect coupling between any distant qubits (for details see text) (after [226]).